\documentclass[3p]{elsarticle}

\usepackage[hidelinks]{hyperref}
\hypersetup{pdfauthor=author}
\usepackage[table,xcdraw]{xcolor}
\usepackage{pgfplots}
\usepgfplotslibrary{statistics}
\pgfplotsset{compat=1.17}
\usepackage{amssymb}
\usepackage{hologo}
\usepackage{longtable,booktabs}
\usepackage{moreverb}
\usepackage{subcaption}
\usepackage{textcomp}
\usepackage[utf8]{inputenc}
\usepackage{multirow}
\usepackage{multicol}
\usepackage{url}
\usepackage{booktabs}
\usepackage{framed}
\usepackage{xspace}
\usepackage{listings}
\usepackage{pdflscape}
\usepackage{array}
\usepackage{colortbl} 
\usepackage{listingcustom}
\usepackage{float}
\usepackage{graphicx}
\usepackage[misc,geometry]{ifsym} 
\usepackage{fontspec}
\usepackage{academicons}
\usepackage{color}
\usepackage{amsmath}
\usepackage{hyperref} 
\usepackage{aas_macros}
\usepackage[bottom]{footmisc}
\usepackage{soul}
\usepackage{tcolorbox}
\usepackage[title]{appendix}
\usepackage[T1]{fontenc}
\usepackage[leftcaption]{sidecap}
\usepackage{rotating}
\usepackage{pdflscape}
\usepackage{afterpage}
\usepackage{capt-of}% or use the larger `caption` package
\usepackage{multirow,bigdelim,dcolumn,booktabs}
\usepackage{caption}
\usepackage{xcolor}
\usepackage{listings}
\usepackage{parcolumns}
\usepackage{mdframed}
\usepackage{longtable}
\usepackage{booktabs}
\usepackage{makecell}
\tcbuselibrary{breakable}
\hypersetup{colorlinks=true, linkcolor=blue, urlcolor=blue}
\usepackage{totcount}
\newtotcounter{IconsCounter}
\usepackage{balance}

\newcommand{\proof}[1]{{\color{blue} #1}}

\begin{document}

\begin{frontmatter}

\title{
Refactoring for Novices in Java: An Eye Tracking Study on the Extract vs. Inline Methods}

\author[1]{José Aldo Silva da Costa\corref{cor1}}
\ead{jose.aldo@servidor.uepb.edu.br}

\author[2]{Rohit Gheyi}
\ead{rohit@dsc.ufcg.edu.br}

\author[2]{José Júnior Silva da Costa}
\ead{josejunior@copin.ufcg.edu.br}

\author[3]{Márcio Ribeiro}
\ead{marcio@ic.ufal.br}

\author[4]{Rodrigo Bonifácio}
\ead{rbonifacio@cic.unb.br}

\author[2]{Hyggo Almeida}
\ead{hyggo@computacao.ufcg.edu.br}

\author[5]{Ana Carla Bibiano}
\ead{abibiano@inf.puc-rio.br}

\author[5]{Alessandro Garcia}
\ead{afgarcia@inf.puc-rio.br}

% Author notes
\cortext[cor2]{Corresponding author}

\affiliation[1]{
    organization = {State University of Paraíba (UEPB)},  
    country      = {Brazil}
}

\affiliation[2]{
    organization = {Federal University of Campina Grande (UFCG)},  
    country      = {Brazil}
}

\affiliation[3]{
    organization = {Federal University of Alagoas (UFAL)},  
    country      = {Brazil}
}

\affiliation[4]{
    organization = {University of Bras\'{i}lia (UnB)},  
    country      = {Brazil}
}

\affiliation[5]{
    organization = {Pontifical Catholic University of Rio de Janeiro},  
    country      = {Brazil}
}

\begin{abstract}

Developers extract methods to make the code easier to read, understand, and reuse. Inlining keeps logic in a single block, while method extraction isolates functionality into a named unit, which is a common practice. Although prior studies using static metrics have not found clear differences between these practices, the human factor in code comprehension and navigation remains underexplored.
Code metrics might not capture other dynamic aspects such as the impact of method extraction on the visual effort of developers. 
{Therefore, we aim to investigate the impact of Inline Method and Extract Method refactorings on code comprehension using a dynamic approach based on tracking participants' eye movements as they interact with the code. Through eye tracking, we examine how small structural changes such as introducing or removing short methods affect code comprehensibility by analyzing key code areas and comparing visual effort, reading behavior, fixation durations, and revisit patterns.}
We conduct a controlled experiment with 32 novices in Java, {followed by short post-task interviews to capture their difficulties, strategies, and impressions.} Each participant solved eight simple tasks comprising four programs with the method inlined and four with the method extracted. We compare them by measuring time, number of attempts, and visual effort with fixation duration, fixations count, and regressions count. {Complementarily, we survey 58 additional novices to obtain both quantitative and qualitative data.} We found that while traditional metrics such as time revealed significant percentage changes, eye tracking metrics varied even more; in two programming tasks, method extraction improved comprehension and reduced visual effort, with time decreasing by up to 78.8\% and regressions by 84.6\%. {However, for simpler tasks, such as calculating the area of a square, the opposite occurred}: time increased by up to 166.9\% and regressions by 200\%. This suggests that even well-named methods can disrupt novices' linear reasoning and require costly visual navigation. We found evidence that meaningful method names (beacons) do not always improve code comprehension for Java novices in the context of refactoring. Despite the presence of meaningful names, participants still navigated back and forth between the method call and its extracted method, resulting in increased visual effort and cognitive processing time.
To complement our analysis, we surveyed 58 additional novices about their preferences and impressions. While many preferred extracted versions for their readability and reuse, some of their preferences did not always align with their performance, highlighting a gap between perceived and actual comprehension.
The integration of these metrics with participant feedback and their preferences suggests that the modularization introduced by an extracted method may slow down problem-solving for novices due to the increased navigation effort it requires. We observed an increased number of gaze transitions, which can disrupt novices' mental flow. 
Our findings complement previous studies and have important implications for educators suggesting caution when applying premature modularization to novices in Java. They also highlight the need to better understand when and for whom each refactoring should be applied. For researchers, our study underscores the value of eye tracking in revealing effects that static metrics may not capture.

\end{abstract}


\end{frontmatter}

\section{Introduction} 
\label{section: introduction}

Developers often deal with structural problems in code that make it difficult to understand. Refactoring is a disciplined technique for restructuring source code that aims to improve its internal structure without changing its external behavior~\cite{fowler2018refactoring}. According to recent studies, 40\% of developers refactor code nearly every day~\cite{golubev-fse-2021}. Refactorings such as the Extract Method and Inline Method are among the most commonly used~\cite{murphy2006java,golubev-fse-2021}. 
% O que são Extract e Inline refacts
While the Extract Method takes a code fragment, extracts it into a separate method, and gives it a new name, the Inline Method is essentially the opposite, taking a method call and replacing it with the body of the code. 
% Por que devs usam Extract e Inline refacts
Developers apply extraction or inlining to make code easier to understand~\cite{silva2016why}. 
From a cognitive perspective, the Extract Method is expected to support comprehension by introducing named abstractions that serve as beacons---semantic cues that help programmers quickly grasp the purpose of a method without having to inspect its internal details~\cite{siegmund2017measuring}. This beacon effect is especially relevant when method names are meaningful and descriptive, providing high-level insight into the code's intent~\cite{fowler2018refactoring}.

Previous studies have investigated the impact of refactoring practices, often relying on static code metrics to evaluate their effects. For instance, Cedrim et al.~\cite{cedrim2017understanding} found that the Extract Method increased the number of code smells, while the Inline Method left the number unchanged. Similarly, Bavota et al.~\cite{bavota2015experimental} reported that most code smells were not removed by the evaluated refactorings, and Tufano et al.~\cite{tufano2015when} observed that the majority of smells are actually introduced during activities aimed at improving the code. While these studies provide insights into the {negative} effects of refactorings {on the code}, they do not capture the human perspective on how such transformations affect {developers'} code comprehension. Siegmund et al.~\cite{siegmund2017measuring}, on the other hand, examined the neural efficiency of program comprehension with code snippets in Java; however, they did not investigate structural transformations such as refactorings, nor did they employ eye tracking. As a result, the effects of refactoring on code comprehension remain underexplored. To the best of our knowledge, no prior work has investigated the Extract Method and Inline Method refactorings with novices in Java using eye tracking.

% solucao: estudo com eye tracking
Eye tracking has emerged as a promising approach to investigate the impact of {small and localized} changes. With eye-tracking metrics, we can infer how developers allocate their cognitive resources when understanding different code structures, such as inlined versus extracted methods. While traditional metrics (e.g., task completion time, accuracy) reflect overall performance, eye tracking complements this by revealing the underlying processes involved in code navigation and comprehension. 

By analyzing code using eye tracking, we aim to capture the impact of {the Extract Method and Inline Method} on developers' visual effort. {We understand visual effort as the cognitive demand reflected in eye-movement behavior, which we measure with fixation duration, fixation count, and regressions count. These metrics serve as established proxies for the mental effort required to understand the code~\cite{sharafi2015eye}.}
% metodo: estudo controlado
We conduct a controlled experiment with 32 novices in Java to 
measure their objective performance in terms of time, {number of attempts until providing the correct output}, and visual effort in both the entire code and the Area of Interest (AOI). We define our AOIs as the line(s) of code where the logic was affected by inlining or extracting the method.
{In the reading domain, fixations refer to the eyes focusing on certain spots when we read, while regressions refer to skipping back in the text to re-read a word or sentence.}
% quais os programming tasks que seleconamos
We select eight tasks from introductory programming courses: \textit{Sum Numbers}, \textit{Calculate Next Prime}, \textit{Return Highest Grade}, \textit{Calculate Factorial}, \textit{Count Multiples of Three}, \textit{Calculate Area of a Square}, \textit{Check If Even}, and \textit{Count Number of Digits}. For each task, we compare two functionally equivalent versions of programs, one with a method inlined, and the other with the method extracted.
{To complement the visual and performance data}, we interview the experiment participants and apply grounded theory to understand their perceived difficulties. {To complement the interview insights from experiment participants, we also conducted a broader online survey, which aimed to provide a wider perspective on developers' preferences and reasoning about inline and extracted versions. This helps contextualize and generalize the qualitative patterns observed in the experiment.}

% results
{We found reductions in the time in the AOI with the Extract Method for two tasks, ranging from 70--78.8\%.} 
{For these tasks, all three eye-tracking metrics consistently indicated lower visual effort: fixation duration decreased by approximately 74--79\%, fixation count by 68--76\%, and regressions count by 70--79\%.}
For three tasks, the reduction varied from 20--34.4\% in the number of attempts. These results indicate higher productivity and less visual effort. {Interestingly, our eye tracking data revealed that even in the presence of meaningful method names, participants often revisited method calls and bodies repeatedly, indicating confusion, and a breakdown in linear reasoning, particularly in simpler tasks such as calculating the area of a square.} We found increases in the time in the AOI for three tasks that varied from {approximately} 108--167\% and from 100--200\% in the regressions count, which can be associated with more visual effort. We observed differences between the performance of the novices in the experiment and the perceptions of novices in the survey. Novices preferred the extracted method version mainly due to decomposition, which was perceived as improving readability, method reuse, and easier extension, while they preferred the inlining method due to improved readability, removal of unnecessary methods.

Our study demonstrates that small, {localized changes} in code organization, such as applying the Extract Method or the Inline Method, can significantly affect how novices understand and visually navigate code. Through eye tracking, we revealed cognitive patterns and confusion not observable through static metrics alone. These findings complement previous studies by showing that even well-intentioned refactorings, such as extracting a method with a meaningful name, may hinder comprehension by forcing more gaze transitions particularly for Java novices.  
Although previous studies found that semantic cues such as beacons ease bottom-up comprehension by enabling semantic chunking~\cite{siegmund2017measuring}, our results indicate that meaningful method names are not always effective in improving novices' code understanding in Java in the context of refactorings. Our results are consistent with Crosby et al.~\cite{crosby2002theroles}, who observed that novices tend not to use beacons effectively. 
%These effects were observed even in simple programming tasks.
We triangulated completion time, number of answer attempts, and eye-tracking metrics with participant feedback on perceived difficulties and preferences, and gaze transitions. For instance, they needed to direct their attention differently when reading and understanding extracted code compared to inline code, going back and forth and making more regressions. However, these differences can vary depending on the specific task{, the complexity of the logic,} and the characteristics of the code. For simpler tasks such as calculating the area of the square or checking an even number, inlining a method can be more desirable, while for more complex tasks such as calculating the factorial or returning the highest element of a list, the extracted method was more effective. These nuances encourage educators to be more critical of the Extract Method and {Inline Method, and their} potential to hinder code comprehension for novices in Java. It also encourages other researchers to consider eye tracking to evaluate the potential of visual metrics to reveal effects of refactorings that cannot be captured by static code metrics. 
% Contribuições em Bullets
The main contributions of this study are:    
\begin{itemize}
    \item We perform a controlled experiment using eye tracking with 32 novices in Java to objectively investigate their performance-solving programming tasks with the Extract Method and Inline Method (Sections~\ref{sec:methodology} and~\ref{sec:results});
    \item We conduct an online survey with 58 novices in Java to qualitatively investigate their preferences for the inlined or extracted method versions and their motivations (Sections~\ref{sec:methodology} and~\ref{sec:discussion});
    \item We discuss the interviews with the 32 novices and use grounded theory to analyze the data to subjectively investigate their perceptions of the difficulties of the programming tasks (Section~\ref{sec:discussion});
    {\item We provide a replication package with the experimental material available on our website~\cite{our-artifacts}.}
\end{itemize}

This article is organized as follows. In Section~\ref{sec:motivating-example}, we present a motivating example to contextualize our research problem. In Section~\ref{sec:study-definition}, we describe the study definition, followed by the research methodology in Section~\ref{sec:methodology}. In Section~\ref{sec:results}, we report the results, which are then discussed in Section~\ref{sec:discussion}. We examine the threats to validity in Section~\ref{sec:threats}. We present the related work in Section~\ref{sec:related-work}, and finally, in Section~\ref{sec:conclusion}, we present our conclusions.

\section{Motivating Example} 
\label{sec:motivating-example}

\begin{figure}[h]
    \centering
    \includegraphics[width=0.9\textwidth]{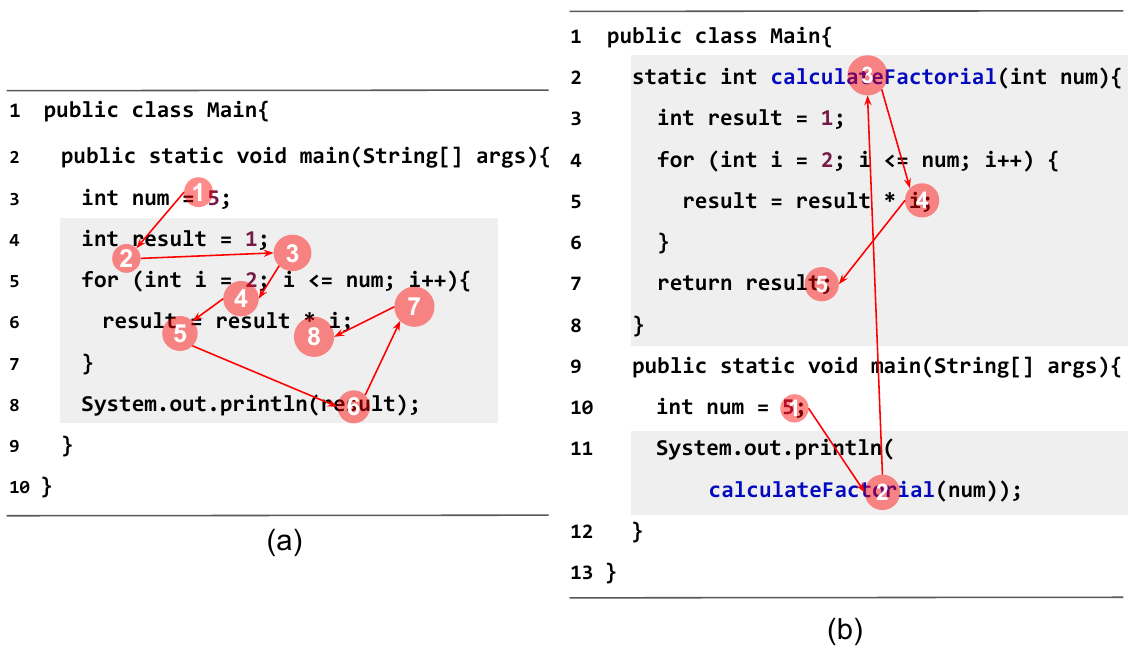}
    \caption{{Gaze patterns for (a) Inline Method version of the code that calculates the factorial of a number, and (b) Extract Method version.}}
    \label{fig: motivating-example-gaze-transitions}
\end{figure}

%Motivating Example
Refactorings can influence how novice developers navigate and comprehend code. To illustrate, we designed two examples, one with moderately complex logic and another with simpler logic, each implemented in two versions: Inline Method and Extract Method. In Figure~\ref{fig: motivating-example-gaze-transitions}, an example we designed, we depict two iterative programs to find the factorial of a number adapted from GeeksforGeeks\footnote{\url{https://www.geeksforgeeks.org/program-for-factorial-of-a-number/}}, a platform with programming learning resources. In Figure~\ref{fig: motivating-example-gaze-transitions}(a), we have the inlined version and in Figure~\ref{fig: motivating-example-gaze-transitions}(b), the extracted version. While both print the same output, the extracted version adds three lines of code and uses a method name that conveys its intention. 
Figure~\ref{fig: motivating-example-gaze-transitions} illustrates gaze patterns for the factorial task in both code styles. In Figure~\ref{fig: motivating-example-gaze-transitions}(a), a subject fixated eight locations, most of which were concentrated on Lines 4--8, with several regressions indicating higher visual effort. In Figure~\ref{fig: motivating-example-gaze-transitions}(b), the subject fixated five locations, with fewer visual regressions and a broader distribution of attention, suggesting reduced visual effort.

\begin{figure}[h]
    \centering
    \includegraphics[width=1\textwidth]{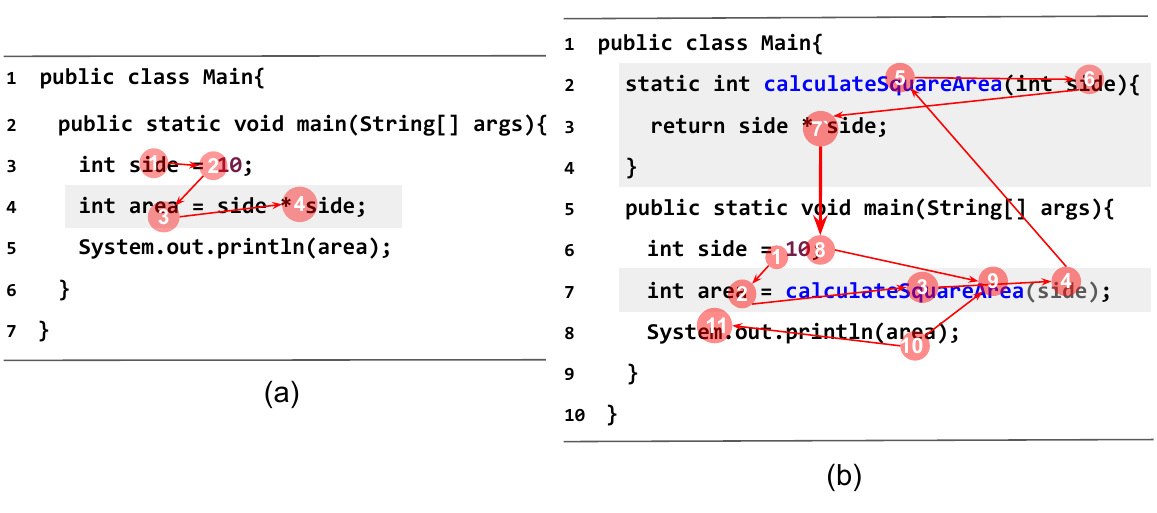}
    \caption{{Gaze patterns for (a) Inline Method version of the code that calculates the area of a square, and (b) Extract Method version of the code.}}
    \label{fig: motivating-example-gaze-transitions-2}
\end{figure}

However, in Figure~\ref{fig: motivating-example-gaze-transitions-2}, another example we designed, we depict two iterative programs that calculate the area of a square. Figure~\ref{fig: motivating-example-gaze-transitions-2} illustrates gaze patterns for the task of calculating the area of a square in both code styles. In the inlined version (Figure~\ref{fig: motivating-example-gaze-transitions-2}(a)), a subject fixated on four locations, concentrated on Lines 3 and 4, and did not make regressions in the code, indicating low visual effort. Conversely, in the extracted version (Figure~\ref{fig: motivating-example-gaze-transitions-2}(b)), the subject fixated on eleven locations, with two visual regressions and a wider distribution of attention. This suggests that, when instructions and code logic are simple, the Inline Method may reduce visual effort compared to the Extract Method. Although method extraction is generally intended to simplify understanding, there are scenarios where it may actually decrease readability, making the inlined version a more suitable alternative~\cite{fowler2018refactoring}. Furthermore, while this example involves a simple piece of code, in more complex scenarios, the distance between a method declaration and its call may be greater, increasing the visual navigation cost.

From a static analysis viewpoint, both the Extract Method and Inline Method versions preserve the same functionality and exhibit similar complexity. {However, prior studies have shown that refactorings may still introduce undesired effects on static properties, suggesting that structural changes can have negative effects on code~\cite{cedrim2017understanding,bavota2015experimental}. Yet, these studies do not address how such changes affect comprehension from a human perspective.} For instance, when an extracted method is named descriptively and encapsulates meaningful logic, it can serve as a beacon~\cite{siegmund2017measuring}, providing a mental shortcut to its purpose and reducing the need for repeated inspection of its body. On the other hand, if the extracted method contains trivial logic, the added indirection may increase visual effort by forcing developers to switch repeatedly between the call site and the method body. {Prior cognitive research has shown that meaningful method names can act as beacons that facilitate comprehension~\cite{siegmund2017measuring}, but this work did not examine the cognitive impact of structural changes such as refactoring and did not employ eye tracking.} Similarly, Fowler~\cite{fowler2018refactoring} provides practical recommendations on when to apply these techniques but does not present empirical evidence, nor does prior work explore these effects in the context of novice developers with eye tracking.

We aim to complement previous studies by combining traditional comprehension measures (e.g., correctness and time) with eye tracking data to analyze how novice developers engage with different code structures {using detailed eye movement data}. Eye tracking can reveal patterns such as visual regressions, fixations, and attention flow, offering a window into the cognitive processes behind code comprehension.
This granularity of data provides insights into the trade-offs between readability and visual effort in each style, which are not captured by static code metrics alone. By triangulating these observations with the subjective perceptions and preferences, we aim to better understand how small changes in code structure affect novice comprehension.

\section{Study Definition}
\label{sec:study-definition}

{In this section,} we define our study according to the Goal-Question-Metrics approach~\cite{basili1994thegqm}. {Our goal is to 
\textbf{analyze} the Extract Method and Inline Method \textbf{for the purpose of} understanding their impact \textbf{with respect to} novices' code comprehension, task performance, and visual attention behavior during code reading \textbf{from the point of view of} novice programmers \textbf{in the context of} a controlled experiment involving code comprehension tasks.}

\begin{table}[t]
\centering
\caption{Summary of the study goal, research questions, and metrics.}
\label{tab:gqm}
\begin{tabular}{p{4.3cm} p{5.6cm} p{5.3cm}}
\toprule
\textbf{Goal} & \textbf{Research Questions (RQs)} & \textbf{Metrics} \\
\midrule

\multirow{12}{=}{\raggedright
Analyze the Extract Method and Inline Method\\ \textbf{for the purpose of} understanding their impact \\\textbf{with respect to} novices' code comprehension, task performance, and visual attention behavior \\\textbf{from the point of view of} novice programmers \\\textbf{in the context of} a controlled experiment involving code comprehension tasks.
}
& RQ$_1$: To what extent does Extract Method impact task completion time?
& Total time; time per AOI \\

& RQ$_2$: To what extent does Extract Method impact the number of answer attempts?
& Number of answer attempts \\

& RQ$_3$: To what extent does Extract Method impact fixation duration?
& Fixation duration (total; per AOI) \\

& RQ$_4$: To what extent does Extract Method impact fixation count?
& Fixation count (total; per AOI) \\

& RQ$_5$: To what extent does Extract Method impact eye regressions?
& Number of regressive saccades \\

& RQ$_6$: How do novices perceive Extract vs.\ Inline Method?
& Interview codes; survey responses \\

\bottomrule
\end{tabular}
\end{table}

The tasks selected for this study are adapted from introductory programming courses to ensure that they are aligned with the knowledge level of novice developers and reflect realistic scenarios. By comparing functionally equivalent programs that implement the Extract Method and Inline Method, we explore whether refactoring methods lead to measurable differences in how novices understand code. 

{We address six Research Questions (RQs).} For each RQ, our null hypothesis is that there is no difference between the Inline Method and the Extract Method versions of the programs with respect to the collected metrics. {For each question, we specify the metrics used to assess the observed effects. In Table~\ref{tab:gqm}, we summarize our goal, RQs, and corresponding metrics.} 

\textbf{RQ$_1$: To what extent {does the Extract Method refactoring} affect task completion time?} Following prior studies~\cite{gopstein2017understanding,oliveira2020atoms}, to answer this question, we measure how much time the subject spends in the whole program to specify the correct output, in addition to the time in specific areas of the code. Comparing the time between the inlined and extracted versions allows us to assess whether refactoring impacts the speed to process and understand the code and whether it provides any benefit in terms of task efficiency.

\textbf{RQ$_2$: To what extent {does the Extract Method refactoring} affect the number of attempts?} To answer this question, we measure the number of attempts made by the subject until specifying the correct output of the program. This metric reflects the difficulty in specifying the correct output of the program. Fewer attempts can suggest that method extraction may have improved code understanding.

\textbf{RQ$_{3}$: To what extent {does the Extract Method refactoring} affect fixation duration?} By measuring the fixation duration, we can assess the effort involved in reading the code, mentally processing it, and comprehending specific code elements. Longer fixations indicate greater effort or attention required by a certain visual stimulus~\cite{busjahn2011analysis}, and comparing the fixation duration in the inlined versus extracted versions helps us understand whether refactoring increases or decreases the visual effort during code reading.

\textbf{RQ$_{4}$: To what extent {does the Extract Method refactoring} affect fixation count?} An increased number of fixations has been associated with more time to understand code phrases~\cite{binkley2013impact}, more attention to complicated code~\cite{crosby2002theroles}, and more visual effort to recall identifiers' names~\cite{sharafi2012women}.  By comparing fixation counts between the two versions, we can evaluate whether refactoring improves or worsens the ease of code reading. This provides insights into the impact of refactoring on code comprehension.

\textbf{RQ$_{5}$: To what extent {does the Extract Method refactoring} affect regressions count?} When readers do not understand what they read in natural language, they make eye regressions~\cite{rayner1998eye}. Likewise, for programming tasks, the regression rate has been used to measure the linearity of code reading~\cite{busjahn2015eye}.
In imperative programming languages, developers may read code lines following a left-to-right, top-to-bottom fashion, similarly to natural language, except for loops, which require the reader to read bottom-to-top at some points. 
To answer this question, we compute the number of regressive eye movements with a direction opposite to the writing system, right-to-left and bottom-to-top. Comparing these values across inlined and extracted versions allows us to assess whether refactoring improves the linearity and flow of comprehension. To make the comparison fair, both versions contain loops with equivalent iteration behavior.

\textbf{RQ$_{6}$: How does the Extract Method refactoring affect the perceptions and motivations of the subjects?}
In addition to the controlled study, we conducted post-study interviews with the 32 novices to gather qualitative feedback about their perceptions, as well as their preferences. Furthermore, we surveyed another 58 novices in Java to investigate their underlying reasons for choosing between the Extract Method and Inline Method. These methods aim to complement the quantitative measures. This mixed-method approach allows us to perform data triangulation by combining quantitative metrics with qualitative data obtained from a survey and post-study interviews.

\section{Methodology} 
\label{sec:methodology}

In this section, we present the pilot study (Section~\ref{pilot}), experiment {sections} (Section~\ref{phases}), evaluated refactorings (Section~\ref{refacts}), programs (Section~\ref{programs}), subjects (Section~\ref{subjects}), treatments (Section~\ref{treatments}), eye tracking system (Section~\ref{eye-tracking-system}), \proof{fixations} and saccades instrumentation (Section~\ref{instrumentation}), and data analysis (Section~\ref{analysis}).

\subsection{Pilot Study} \label{pilot}
% proposito do piloto
Prior to the main experiment, we conducted pilot studies with six {students whose experience levels matched the target novice population} to refine the experimental materials and evaluate the overall setup and design. These subjects were not included in the final data analysis.
% piloto para avaliar materiais, programas
Our materials included programming tasks, a subject characterization form, and a questionnaire for a semi-structured interview. To evaluate our programming tasks, we tested complete code snippets from introductory programming courses with distinct levels of difficulty. We tested the code font size, font style, line spacing, and indentation. {We also piloted the questions included in both the characterization form and the interview questionnaire, refining wording, clarity, and ordering based on feedback from the pilot sessions.}

% vocabulario dos programas em pt
Our participants are Brazilian natives. We designed the vocabulary of the programming tasks to be in the Brazilian Portuguese language to avoid obstacles in language comprehension. By aligning the terms with participants' native language in the programming scenarios, we aim to enable them to focus on the programming challenges rather than being hindered by unfamiliar terminology. The identifiers and methods' names were carefully selected and discussed by the researchers. We used names such as \texttt{result} for storing operation outcomes, \texttt{counter} for counting, and the abbreviation \texttt{num} to represent a number. 

During the pilot studies, some identifiers were refined due to confusion and withdrawals. {We aimed to use names that were easy to understand without being unnecessarily long or overly short, so that fixations on variable names could be meaningfully captured in the eye-tracking analysis. Before running the experiment, two authors reviewed all code snippets and agreed on the identifiers to ensure they were readable and consistent across tasks. We also reused common names across programs (e.g., \texttt{i} for loop variables, \texttt{num}, \texttt{result}, \texttt{count}) to avoid introducing differences in naming that were not related to the experimental conditions.} For example, the variable \texttt{n} to represent a number was renamed to \texttt{num}, and we also clarified some method names such as ``sum numbers from one to \proof{\texttt{n}}'' to ``sum numbers from one to \proof{\texttt{num}},'' and ``count multiples of three'' to ``count multiples of three from the list.'' Being more specific reduced confusion. Additionally, we replaced one of the programs that had a method ``sum the digits'' to ``count the number of digits.'' These refinements aimed to reduce obstacles in the comprehension process and better align with the method extraction mechanics, which suggest the use of intuitive names on the new methods. In addition, meaningful variable and method names reflect more realistic scenarios that students encounter during their studies. More details on the limitations of this approach can be found in Section~\ref{internal-validity}.

% nome dos metodos, que diretries usamos
In addition, for method names, we used lessons learned from previous studies and guidelines~\cite{fowler2018refactoring,arnaoudova2016linguistic,butler2010exploring,lawrie2006s,avidan2017effects,hermans2017peter}. They systematize the use of capitalization, size of identifiers, number of words, the name meaningfulness, use of verbs, among other aspects. We chose method names that convey their intent which were selected to the best of our abilities. This approach is arguably closer to a practical scenario. {In addition, we refined the names of the methods in the pilot study, testing how well the names conveyed the method's intent. We discussed the methods' names among the authors to find the most appropriate ones.} 
%piloto para avaliar design do experiment
After experiment refinement, we organized it into five {sections}: (1) Characterization, (2) Tutorial, (3) Warm-up, (4) Task, and (5) Qualitative Interview. We estimated an average of 60 minutes for each subject to complete all {sections}. Next, we describe these {sections} in detail.

\subsection{Experiment {Sections}} \label{phases}

\begin{figure}[]
\centering
  \includegraphics[width=0.7\textwidth]{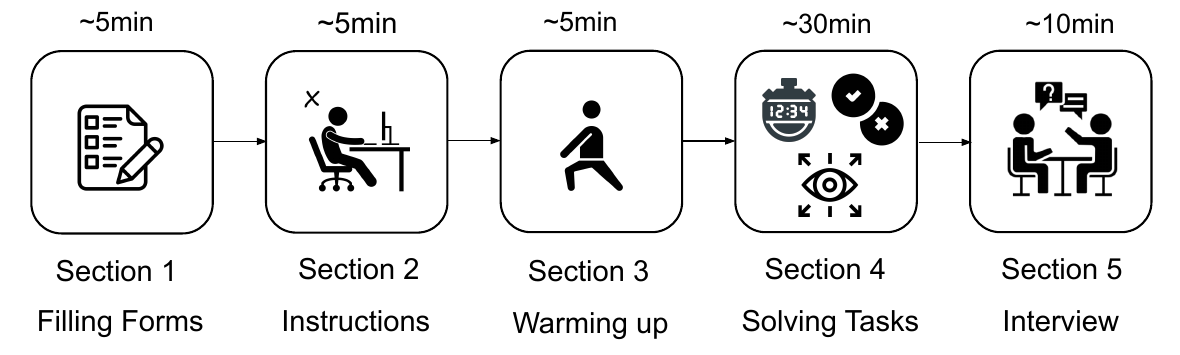}
   \caption{{Experiment sections divided into five sequential steps.}}
   \label{fig:experiment-sections}
\end{figure}

%fase 1 - forms
{We divided the experiment into five sections, as depicted in Figure}~\ref{fig:experiment-sections}. {In Section~1, the subjects fill the consent and characterization forms}. As the subject enters the room, we explain the study, what data we are going to capture, and how we aim to use the data. 
% formulários de cosentimento e caracterização
Each subject voluntarily fills out a consent form, agreeing to participate and being aware that their identity would remain anonymous. Then they fill out a characterization form with questions related to programming experience.

% pre fase 1 - tutorial e explicações
In {Section~2, we provide instructions, delivered as a tutorial, on how to perform the experimental tasks and on how to properly behave in front of the eye-tracking device during the experiment}. {The tutorial consisted of a brief verbal explanation given by one researcher. Its purpose was to guide participants through the experiment procedure without revealing details that could bias their behavior. We explained how the eye-tracking session works, how to sit at the appropriate distance, how to avoid unnecessary movements that could introduce noise, and how to report the output of each program.} %We instruct the subjects on how to sit properly in front of the eye tracker and how to perform the tasks. 
We then proceed with a calibration of the camera on the subject's eyes. During the camera calibration, each subject should look at specific locations on the screen that the camera software indicates. The camera software also reports when the calibration is successful. 

% fase 2 - tarefa de aquecimento
In {Section~3}, each subject warms up for the experiment by solving a simpler program. During the warm-up, we demonstrate how to specify the output out loud instructing them to close their eyes for two seconds before and after solving the program and how we signal the correct and incorrect answer. {The participant verbally communicated the output only to the researcher present in the room. There was only one participant in the room and one or at most two researchers to avoid distraction or social pressure.} After warm-up, the subjects should be more comfortable with the experiment setup and equipment. {We did not observe any evidence that verbalizing the output affected participants' performance.}

% fase 3 - doze tarefas para cada
In {Section~4}, we run the main experiment with eight programs, four with the Inline Method version and four with the Extract Method version. To avoid learning effects, we use a Latin Square design~\cite{box2005stats}, which is detailed in Section~\ref{treatments}.
% fase 4 - entrevista
In {Section~5}, we end the experiment with a semi-structured interview aimed at exploring how participants approached the programs and their subjective impressions. We go through each program and ask three questions: (1) How difficult was it to find the output: very easy, easy, neutral, difficult, or very difficult? (2) How did you find the output? (3) What were the difficulties you had, if any?

% cuidados no ambiente por causa da saude dos \castorsec{subject}e por causa do covid
We started running the experiment after the end of the social distancing measures in the country, when the coronavirus infections were decreasing, and the number of vaccinated people was increasing. For the safety of everyone involved, all the researchers had hand sanitizers and face masks. We limited the number of people in the environment to only one subject at a time. 

% cuidados com o ambiente para evitar ruído nos dados e correção dos dados
We arranged the environment of the experiment to reduce noise in the data. We used a fixed chair which increased the precision of the eye tracker equipment in pilot studies. However, given the camera limitations, obtaining perfect data is virtually impossible. To mitigate it, we as researchers plotted, discussed, and performed data correction by slightly shifting chunks of fixations in the \textit{y}-axis. We discuss in detail this strategy in the threats to validity section (see Section~\ref{internal-validity}). 

\subsection{Evaluated Refactorings} \label{refacts}

% por que escolhemos
We selected and evaluated two refactorings, namely the Extract Method and Inline Method. These refactorings are among the most common refactorings used in practice~\cite{cedrim2017understanding,murphy2011we,silva2016why}.
% qual a mecanica de extract
To perform the Extract Method refactoring, the following mechanics can be used: the developer creates a new method, and names it after the intention of the method. Then the developer copies the extracted code from the source method into the new target method~\cite{fowler2018refactoring}.

% qual a mecanica de inline
To perform the Inline Method refactoring, the following mechanics can be used: the developer finds a call to the method and replaces it with the content of the method, and then deletes the method~\cite{fowler2018refactoring}. The Inline Method is essentially the opposite of the Extract Method. It may vary from one line to multiple lines of code. The main motivation reported by developers to apply the Inline Method is to eliminate unnecessary or too trivial methods~\cite{silva2016why}.

\subsection{Programs} \label{programs}

\begin{figure*}
\centering
  \includegraphics[width=0.62\textwidth]{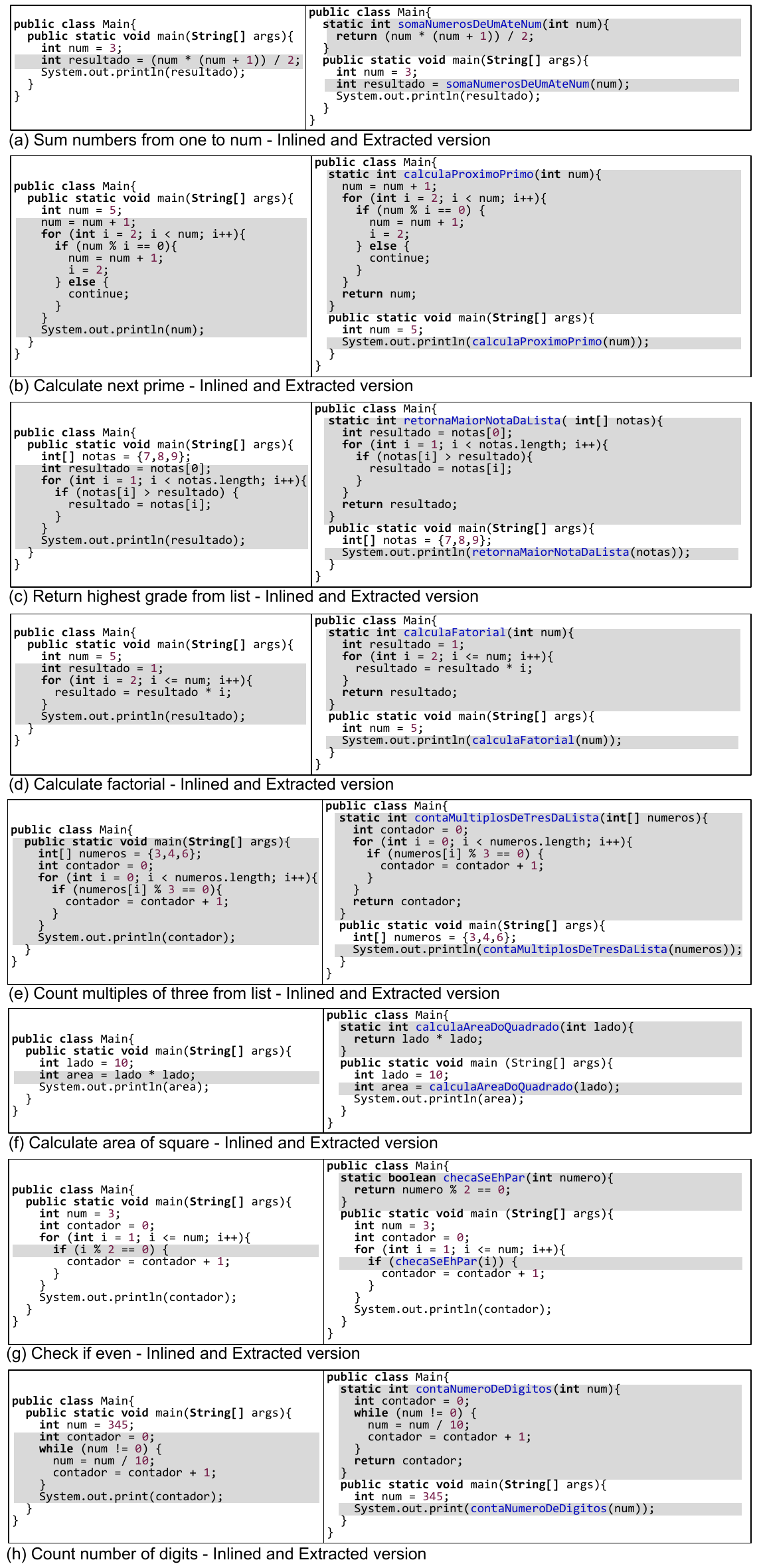}
  \caption{Programs evaluated: \textit{Sum Numbers}, \textit{Calculate Next Prime}, \textit{Return Highest Grade}, \textit{Calculate Factorial}, \textit{Count Multiples of Three}, \textit{Calculate Area of a Square}, \textit{Check If Even}, and \textit{Count Number of Digits}. Shaded areas represent our AOIs.}
  \label{fig: programs-evaluated}
\end{figure*}

% como selecionamos as tarefas
%We selected code snippets by manually analyzing code repositories of introductory programming assignments. 
{All code snippets used in the experiment were authored specifically for this study and are presented in Figure~\ref{fig: programs-evaluated}. To ground their design in realistic novice-level practices, we began by consulting common introductory programming tasks from public learning platforms such as GeeksforGeeks\footnote{\url{https://www.geeksforgeeks.org/}} and Leetcode\footnote{\url{https://leetcode.com/}}. 
These platforms served as inspiration to identify common introductory programming patterns (e.g., types of tasks, loops, conditionals, simple arithmetic). We designed the Extract and Inline Method tasks to reflect the natural variation in refactoring granularity observed in public learning platforms and open repository submissions, ranging from single-line to multi-line cases. We created the code snippets and then refined them through rounds of pilot studies with novices. These refinements helped us adjust the difficulty level, ensure that each program fit entirely on the screen for eye tracking, avoid unintended complexity or multiple refactorings, and standardize stylistic elements such as font, spacing, and indentation.}
{In addition, the tasks involve mathematical operations that are part of the standard high school curriculum in Brazil. For this reason, we assumed that these tasks were unlikely to be particularly challenging.}

{In all the programs, the subject had the task of specifying the correct output but without multiple answer options. For instance, given the input in the program, which was a number or a list of three numbers, the subject had the task to calculate the factorial, and calculate the next prime, among others. The output was also a number.} Providing information about the code, such as finding the output, is a methodology employed by 70\% of the studies in the domain of code comprehension~\cite{Oliveira2020evaluating}.

% tamanho da tela
The evaluated programs had 6--18 lines of code, with a median number of 12 lines. %We restricted the number of lines to fit completely on the screen. 
All the used programs were free of syntactic errors. 
%cuidados - estilo de codigo
The programs followed Consolas font style, font size 11, line spacing of 1.5, and eight white spaces of indentation.
% construçoes simples
Even though written in the Java language, we used simple constructions that commonly occur in many programming languages.
% apenas um refactoring por tarefa
{We made sure that each program with the inlined method contained exactly one method in its scope, the Java main method. Each program with the extracted method contained exactly two methods, the Java main method, and the extracted method}.
% evitar ter apenas dois resultados como saida
Both the inlined and extracted versions of the same task presented the same output. However, due to the use of a Latin Square design, no subject was exposed to inlined and extracted versions of the same task.
{To bring diversity to the programs, we used different styles such as assigning the methods' call to a variable and printing the variable or calling the methods in the print statement.}

In our study, we defined the AOIs as the regions of the code where the logic was affected by the inlining and extracting the method, as these areas are important for understanding the impact of the two approaches on code comprehension. In the Inlined Method version, the AOI corresponds to the specific line(s) where the logic of the method is implemented directly in the calling context, highlighting the absence of modularization. {In the Extracted Method version, the AOIs include both the new method extracted and the line of the method call. These two regions form the AOIs in the Extracted Method version, reflecting the modular structure of the extracted method version. }

\subsection{Subjects} \label{subjects}
  
% caracterizando os  novatos
We recruited 32 undergraduate students that we refer to as ``novices.''
% o que consideramos novatos
They reported having 30 months of experience with programming languages in general, which mainly included Java, Python, JavaScript, C, and C++. However, only in Java, they had on average 12 months of experience. 
% como foram recrutados
They were recruited from two distinct universities from two cities in the Northeast of Brazil, invited mainly in person or through text messages. They were Brazilian Portuguese speakers enrolled in academic universities. {Ethical approval was granted by the Research Ethics Committees of both universities involved in the study and all participants provided informed consent prior to participation.}

Although we conducted the study with 32 participants, we were able to collect other demographic data from 28 of them. The average age of the participants was 21.6 years with the standard deviation of 3.07 with ages ranging from 19 to 34 years. {Regarding their gender, all participants voluntarily opted to provide their self-identification: 21 participants selected male and 7 selected female.} %Regarding gender, 21 participants were male and 7 were female.
%Following Cohen~\cite{cohen1988spa}, we need a sample of 26 subjects for a two-sample \textit{t}-test to have a nominal power of 0.08, and a significant level of 0.05. Given that we have 32 subjects, which were exposed to both versions of the programs, our sample reached and exceeded the minimum sample size.

Regarding the sample size, we estimated the number of participants required to achieve a statistical power of 0.8 with a significance level of 0.05 using a \textit{t}-test sample size computation. Our analysis indicated that a minimum of 26 participants, divided into two samples, would be necessary to detect a large effect size. Following Cohen's guidelines for effect size interpretation~\cite{cohen1988spa} and consistent with previous work on code comprehension~\cite{gopstein2017understanding}, we assumed a large effect size (0.8).

\subsection{Treatments}\label{treatments}

\begin{figure}[]
\centering
  \includegraphics[width=0.6\columnwidth]{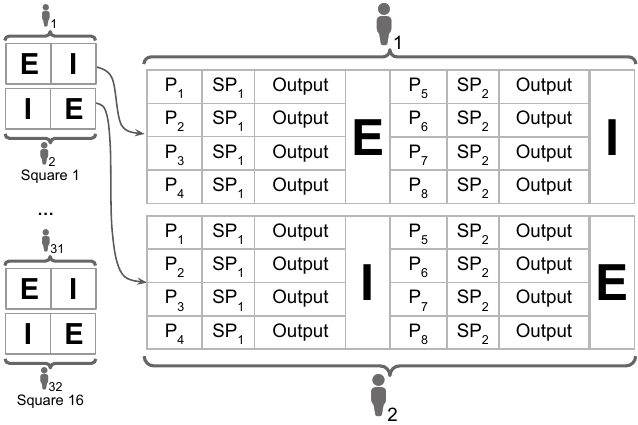}
   \caption{Structure of the experiment in terms of experimental units divided into 16 squares. Subject$_1$ takes four programs (P$_1$--P$_4$) with the Extract Method (E) of \textit{Sum Numbers} (P$_1$), \textit{Calculate Next Prime} (P$_2$), \textit{Return Highest Grade} (P$_3$), \textit{Calculate Factorial} (P$_4$). These programs are from set of programs 1 (SP$_1$). Subject$_1$ also takes four programs (P$_5$--P$_{8}$) from the set of programs 2 (SP$_2$) comprising the Inline Method (I) of \textit{Count Multiples of Three} (P$_5$), and \textit{Calculate Area of a Square} (P$_6$), \textit{Check If Even} (P$_7$), and \textit{Count Number of Digits} (P$_8$). Subject$_2$ takes the complement to that. ``Output'' describes a task in which the subject has to specify the correct output.}
   
   \label{fig: structure-treatment-extract}
\end{figure}

Each subject examined eight programs (P$_1$--P$_{8}$).
% criando dois projetos
To avoid learning effects, we designed 16 distinct programs divided into two Sets of Programs (SP$_1$ and SP$_2$). 
One subject examines four programs with the Inline Method (I) of the set SP$_1$, and four programs with the Extract Method (E) of set SP$_2$, as seen in Figure~\ref{fig: structure-treatment-extract}. Another subject examines four programs with the Extract Method version (E) of the set SP$_1$, and four programs with the Inline Method version (I) of set SP$_2$.
Being in the same set SP$_1$ but with different refactorings, both programs P$_1$ print the same output. Both programs are examined by distinct subjects as well.
% tratamentos e baseline
We designed the programs with the Inline Method version to be our baseline group, and the ones with the Extract Method version to be the treatment group. %In all the programs, the subject has the task of specifying the correct output but without multiple answer options. For instance, given the input in the program, the subject has the task to calculate the factorial, and calculate the next prime, among others.

\subsection{Eye Tracking System}\label{eye-tracking-system}
We used the Tobii Eye Tracker 4C in our experiment which has a sample rate of 90 Hz. The calibration of the eye tracker followed the standard procedure of the device driver with five points. The eye tracker was mounted on a laptop screen with a resolution of 1366 x 720 pixels, width of 30.9 cm, and height of 17.4 cm, at a distance of 50-60 cm from the subject. We displayed code as full-screen images; no IDE was used and line numbers were not shown. We computed an accuracy error of 0.7 degrees which translates to 0.6 lines of inaccuracy on the screen, considering the font size we used and the line spacing. We tested the line spacing in the pilot study. We designed it to be sufficiently large so we could overcome the eye tracker accuracy limitations. {For processing the gaze data, we implemented a script in Python, which allowed us to analyze and collect the metrics. The script includes functions that compute fixations; generate visualizations such as fixation scatterplots, scanpaths, and heatmaps; compute fixation durations; define the AOI regions; and export all computed measures into a structured spreadsheet file. We made it available in our supplementary material~\cite{our-artifacts}.}

\subsection{\proof{Fixations} and Saccades Instrumentation} \label{instrumentation}

A fixation can be understood as the stabilization of the eye on part of a visual stimulus for a period of time, and the rapid eye movements between two fixations are called saccades~\cite{salvucci2000identifying,holmqvist2011eye}. As we fixate our eyes, we trigger cognitive processes~\cite{just1980theory}. In the code comprehension scenario, source code can work as a visual stimulus over which the subject performs the task of reading to specify the output.

There is no standardized threshold of time for a fixation because it depends on the processing demands of the task. However, popular guidelines indicate a threshold between 100 and 300~ms~\cite{salvucci2000identifying,rayner1998eye}. Thus, after analyzing our programs, we used 200~ms as our threshold.
%outro parametro de espaco de dispersao
We used a Dispersion-Based algorithm to classify the fixations. Particularly, we used the Dispersion-Threshold Identification (I-DT)~\cite{salvucci2000identifying}. As its parameters, we classified gaze samples as belonging to a fixation if the samples \proof{were} located within a spatial region of approximately 0.5 degrees~\cite{nystrom2010adaptive}, which corresponded to 25 pixels on our screen.

\subsection{Analysis of the Quantitative Data} \label{analysis}

% tarefas resolvidas e nao resolvidas
From a total of 256 possible observations (32 subjects $\times$ 8 programming tasks), the subjects solved 248 (96.9\%) of them. This set includes programming tasks that were solved either in the first attempt or after many attempts. {The remaining eight tasks (3.1\%) were not solved. The novices initiated these tasks, made unsuccessful attempts, and ultimately opted to skip them. In these cases, no correct answer was provided.}
%Como imputamos dado faltante
We imputed missing data {for these} eight tasks using the Multivariate Imputation by Chained Equations (MICE) method, implemented through the Mice R package. This method employs Predictive Mean Matching (PMM), which is considered one of the most effective imputation techniques compared to others due to its ability to preserve the natural variability of the data~\cite{jadhav2019comparison}. PMM works by predicting the missing values using a regression model and imputing them with observed values that are closest to the predicted values, ensuring that imputed values come from the original data distribution. This feature is particularly advantageous, as it minimizes the risk of generating unrealistic data points.

% testes estatisticos usados
We performed statistical analysis to test the null hypotheses of our RQs using a significance level of 0.05. It means that we have a 5\% risk of finding a difference when there is no actual difference. For \textit{p}-{values smaller than} 0.05, we reject the null hypothesis that there was no difference between the median of the treatments. To minimize the risk of incorrectly rejecting a null hypothesis, resulting from multiple statistical tests, we applied a False Discovery Rate (FDR) correction~\cite{benjamini1995controlling,peitek2020drives}.

% verificação de normalidade da distribuição
We used the Shapiro-Wilk Test~\cite{shapiro65analysis} to test the distribution of the data. When \proof{the data were} normally distributed, we verified whether the variances of the two groups compared were equal~\cite{sheskin2020handbook}. Then we performed the parametric \textit{t} test for the two independent samples to verify whether there was a statistically significant difference between the two groups~\cite{sheskin2020handbook,sharafi2020practical}. When the data \proof{did} not follow a normal distribution and we could not normalize them, we used the non-parametric Mann–Whitney U test (Wilcoxon rank-sum test), which can be applied to these specific situations~\cite{sheskin2020handbook,sharafi2020practical}.
In the scenario of fixations, the mean value might not be appropriate since it can be dependent on some very high values~\cite{galley2015fixation}. In our analysis, we used the median as a measure of central tendency. We used medians for the time and visual effort metrics because they are less affected by extreme values. However, for the number of answer submissions, we used the average, as the values showed minimum variation, between 1 and 3, and the average is more sensitive to small differences. 

We also combined the metric values to analyze both perspectives. Each subject, in each square, solves four programs using the Extracted Method and four using the Inline Method. Then, we calculated the median of all values for each metric across all programs with Extracted Method and compared them to those of the Inline Method. However, for accuracy, we also used the mean.

To identify patterns in the eye gaze transitions in our data, we defined each line of code as a small region to be analyzed independently. These regions were defined in pixels on the images of the tasks according to a previous guideline~\cite{holmqvist2011eye}. 
Their positioning was precisely defined considering the camera limitations so that we could have a margin between the regions, and the regions did not overlap. In addition, we defined the white-space as a region so that we could be aware of any threat to validity given the camera limitations.
Using the chronological order of the fixations and their positions, we identified a sequence of visited regions for each participant. We then summarized these sequences by simplifying repeated transitions to go from one region to the same region. We \proof{made} the sequences and the images of the tasks with the regions identified available in our supplementary material~\cite{our-artifacts}.

\subsection{Survey Procedure} \label{survey}

We used Google Forms to conduct the survey, which took approximately 5-10 minutes to complete. Participants were informed that the data would be used for academic purposes. Recruitment was done through in-person invitations, via email to undergraduate students, and messaging apps targeting student and professor groups, who were encouraged to share the survey further. Due to the snowballing nature of the distribution, the exact number of invited participants is not available. However, we received responses from a total of 76 subjects, out of which 58 were considered novices with less than 10 years of programming experience, in general, and 18 were classified as experienced with more than 10 years of experience. Out of 58 novices, the majority (59.2\%) reported one to five years of experience. In Java specifically, the majority (80.4\%) reported one to five years of experience. All participants were from Brazil.

The survey presented side-by-side code snippets with functionally equivalent behavior, one using method extraction and the other with the method inlined, such as depicted in our motivating example, Figure~\ref{fig: motivating-example-survey-factorial}. The order of presentation was randomized to reduce bias, and no technical terms such as refactoring, extracted, or inlined were mentioned. We asked them to choose their preferred version and to justify their choice, allowing us to understand their decision-making processes.

\subsection{Analysis of the Qualitative Data} \label{analysis-quali}

To analyze the qualitative data, {we followed} the method of grounded theory proposed by Strauss and Corbin~\cite{strauss1998basics}. 
First, we conducted semi-structured interviews and broke the subjects' answers into smaller chunks of data, and assigned each chunk a code that emerges from it. {The first author performed the initial open coding in the first step}. Second, we analyzed all the codes and searched for opportunities to group them into higher-level concepts. Third, we grouped similar concepts searching for opportunities to form higher-level categories. Fourth, we derived a theory through an inductive approach. {Throughout these stages, the first, second, and third authors held rounds of discussion to refine interpretations, contributing to the consolidation of concepts, categories, and the final theoretical account while all coded data and analyses were subsequently shared with the remaining authors.}

\section{Results}
\label{sec:results}

In Sections~\ref{results:RQ1}--\ref{results:RQ6}, we address our research questions. In these sections, when we mention a statistically significant difference, it means we rejected the null hypothesis for the mentioned metric.

\subsection{\textbf{RQ$_1$: To what extent does the Extract Method refactoring affect task completion time?}}\label{results:RQ1}

\begin{table}[h!]
\setlength\tabcolsep{1.8pt}
\caption{Results for \textbf{time spent in AOI and in Code} (RQ$_{1}$). I = Inline Method; E = Extract Method; PD = Percentage Difference from Inline Method (I) to Extract Method (E); PV = \textit{p}-value after FDR correction; ES = Cliff's Delta effect size. Columns I and E are based on the median as a measure of central tendency. The All Programs row compares the median time spent across all programs between the Inline and Extracted methods.} 
\begin{center}
\resizebox{12cm}{!}{
\begin{tabular}{p{2.9cm}rrrrrrrrrr}
    \hline
    \multicolumn{1}{|c|}{\textbf{Tasks}} &\multicolumn{5}{|c|}{\textbf{In AOI}} &  \multicolumn{5}{|c|}{\textbf{In Code}}\\
    \hline
    \multicolumn{1}{|c|@{}}{} &\multicolumn{1}{|c|@{}}{\shortstack{I\\sec}} &  \multicolumn{1}{|c|@{}}{\shortstack{E\\sec}} &\multicolumn{1}{|c|@{}}{\shortstack{\\PD\\\%}} &\multicolumn{1}{|c|@{}}{PV} &\multicolumn{1}{|c|@{}}{ES}
    &\multicolumn{1}{|c|@{}}{\shortstack{I\\sec}} &  \multicolumn{1}{|c|@{}}{\shortstack{E\\sec}} &\multicolumn{1}{|c|@{}}{\shortstack{PD\\\%}}
    &\multicolumn{1}{|c|@{}}{PV} &\multicolumn{1}{|c|@{}}{ES}\\
    \hline
    
    Sum Numbers & 8.8 & 21.6 & {$\uparrow$146.2} & \textbf{0.0009} & 0.75 & 
    15.9 & 30.9 & {$\uparrow$93.9} & \textbf{0.02} & 0.57 \\
    
    Next Prime & 121.2 & 53.9 & {$\downarrow$55.5} & 0.10 & n/a & 
    132.6 & 61.6 & {$\downarrow$53.5} & 0.10 & n/a \\
    
    Highest Grade & 77.7 & 23.7 & {$\downarrow$70.0} & \textbf{0.02} & -0.50 & 
    92.6 & 32.3 & {$\downarrow$65.0} & 0.14 & n/a \\
    
    Factorial & 62.2 & 13.1 & {$\downarrow$78.8} & \textbf{0.04} & -0.47 & 
    81.3 & 22.1 & {$\downarrow$72.8}&\textbf{0.02} & -0.51 \\
    
    Multiples of Three & 24.6 & 37.5 & {$\uparrow$52.4} & 0.24 & n/a & 
    39.2 & 49.0 & {$\uparrow$24.9} & 0.22 & n/a \\
    
    Area of a Square & 2.5 & 6.9 & {$\uparrow$166.9} & \textbf{0.00000} & 0.93 & 
    7.7 & 14.9 & {$\uparrow$94.4} & \textbf{0.02} & 0.53 \\
    
    Check If Even & 4.7 & 9.8 & {$\uparrow$108.4} & \textbf{0.005} & 0.66 & 
    28.3 & 36.3 & {$\uparrow$28.3} & \textbf{0.06} & 0.42 \\
    
    Number of Digits & 34.5 & 26.0 & {$\downarrow$24.7} & 0.25 & n/a & 
    66.4 & 38.2 & {$\downarrow$42.4} & 0.17 & n/a \\
    
   \rowcolor[gray]{.8} All Programs & 127.5 & 111.2 & {$\downarrow$12.8} & 0.24 & n/a & 
    191.8 & 177.5 & {$\downarrow$7.4} & 0.19 & n/a \\
    
    \hline
\end{tabular}
}
\end{center}
\label{tab: timetable}
\end{table}

In this section, we analyze the impact of the Extract Method refactoring on task completion time by measuring the time spent examining the AOI and the complete code. The results, summarized in Table~\ref{tab: timetable}, indicate that while there were significant reductions in time for some tasks, increases were observed for others, suggesting a nuanced impact of the refactoring.

The Extract Method refactoring significantly reduced the time spent in the AOI for the tasks of determining the \textit{Highest Grade} and calculating the \textit{Factorial}, demonstrating improved efficiency.
In contrast, tasks such as \textit{Sum Numbers}, \textit{Area of a Square}, and \textit{Check If Even} exhibited increased time in AOI, indicating potential challenges or complexities introduced by refactoring in these contexts.
Overall, the effects of inlining and extracting methods appear to be task-dependent, with varying implications for novice developers' code comprehension.

\begin{tcolorbox}[colback=gray!5, colframe=black!50, title=Answering RQ$_1$, sharp corners=south,
  width=\textwidth]
The effects of Extract Method refactoring on task completion time are mixed; while some tasks showed decreased time in the AOI, others indicated increased time, highlighting that the impact of refactoring is context-sensitive and may vary based on the specific task at hand.
\end{tcolorbox}

\subsection{\textbf{RQ$_2$: To what extent does the Extract Method refactoring affect the number of attempts?}}\label{results:RQ2}

In this section, we evaluate the impact of the Extract Method refactoring on the number of attempts made by participants to solve the tasks. As the programs were relatively simple, we used the mean number of attempts as a more meaningful measure than the median. Table~\ref{tab: accuracytable} presents a summary of the results.

The Extract Method refactoring led to a notable reduction in the number of attempts for specific tasks. In particular, tasks such as determining the \textit{Highest Grade}, identifying the \textit{Multiples of Three}, and calculating the \textit{Number of Digits} required significantly fewer attempts. However, for other tasks such as \textit{Sum Numbers}, we observed an increase in the number of attempts, suggesting that the complexity introduced by the Extract Method might have varied based on the specific nature of the task. Overall, while some tasks become more straightforward to solve after applying the Extract Method, others may introduce new challenges, indicating that the refactoring benefits are not uniformly distributed across all problem types, particularly for novices.

\begin{tcolorbox}[colback=gray!5, colframe=black!50, title=Answering RQ$_2$, sharp corners=south,
  width=\textwidth]
The Extract Method refactoring generally reduces the number of attempts required to solve tasks, however, while some tasks required fewer attempts, others led to more attempts, highlighting the context-sensitive nature of the refactorings' effects on code comprehension.
\end{tcolorbox}

\begin{table}[h]
\setlength\tabcolsep{1.6pt}
\caption{Results for \textbf{number of answer attempts} (RQ$_{2}$). I = Inline Method; E = Extract Method; PD = Percentage Difference from Inline Method (I) to Extract Method (E); PV = \textit{p}-value after FDR correction; ES = Cliff's Delta effect size. Columns I and E are based on the mean as a measure of central tendency. The All Programs row compares the average number of attempts across all programs between the Inline and Extracted methods.} 
\begin{center}
\resizebox{7.0cm}{!}{\begin{tabular}{p{2.9cm}rrrrr}
    \hline
    \multicolumn{1}{|c|}{\textbf{Tasks}} &\multicolumn{5}{|c|}{\textbf{Attempts}}\\
    \hline
    \multicolumn{1}{|c|@{}}{} &\multicolumn{1}{|c|@{}}{I} &  \multicolumn{1}{|c|@{}}{E} &\multicolumn{1}{|c|@{}}{\shortstack{\\PD\\\%}} &\multicolumn{1}{|c|@{}}{PV} &\multicolumn{1}{|c|@{}}{ES}\\
    \hline
    
    Sum Numbers & 1.00 & 1.31 & {$\uparrow$31.2} & 0.98 & n/a \\
    
    Next Prime &  1.44 & 1.44 & \textcolor{black}{0.0} & n/a & n/a \\
    
    Highest Grade & 1.81 & 1.19 & {$\downarrow$34.4} & \textbf{0.05} & -0.42 \\
    
    Factorial & 1.56 & 1.31 & {$\downarrow$16.0} & 0.30 & n/a \\
    
    Multiples of Three & 1.25 & 1.00 & {$\downarrow$20.0} & \textbf{0.05} & -0.25 \\
    
    Area of a Square & 1.00 & 1.00 & 0.0 & n/a & n/a \\
    
    Check If Even & 1.06 & 1.25 & {$\uparrow$17.6} & 0.17 & n/a \\
    
    Number of Digits & 1.31 & 1.00 & {$\downarrow$23.8} & \textbf{0.05} & -0.25 \\
    
    \rowcolor[gray]{.8} All Programs & 5.2 & 4.7 & {$\downarrow$8.9} & \textbf{0.05} & -0.27\\
    \hline
\end{tabular}
}
\end{center}
\label{tab: accuracytable}
\end{table}

\subsection{\textbf{RQ$_{3}$: To what extent does the Extract Method refactoring affect fixation duration?}}\label{results:RQ3}

In this section, we evaluate how the Extract Method refactoring influences fixation duration, considering both the AOI and the complete code. As illustrated in Table~\ref{tab:fixation durationtable}, the results show varying effects depending on the nature of the task. For some tasks, the refactoring significantly reduced fixation duration, whereas in others, it led to increased time spent in the AOI.

The Extract Method refactoring led to reductions in fixation duration for tasks such as determining the \textit{Highest Grade} and calculating the \textit{Factorial}, suggesting a positive impact on visual effort for these tasks.
In contrast, tasks such as \textit{Sum Numbers}, \textit{Area of a Square}, and \textit{Check If Even} saw a notable increase in fixation duration. A similar trend was observed when analyzing the complete code, with some tasks showing improvements while others became more time-consuming.
These outcomes highlight that the influence of refactoring varies significantly across different tasks, suggesting that its effectiveness can be tied to the specific characteristics and complexity of each task. 

It is noteworthy that the tasks exhibiting the highest visual effort in the AOI were among the simplest, such as calculating the \textit{Area of a Square} and checking if a number is even. This observation suggests that, even tasks typically regarded as straightforward can incur higher visual effort in the presence of method extraction.

\begin{table}[h]
\setlength\tabcolsep{1.8pt}
\caption{Results for \textbf{fixation duration in AOI and in Code} (RQ$_{3}$). I = Inline Method; E = Extract Method; PD = Percentage Difference from Inline Method (I) to Extract Method (E); PV = \textit{p}-value after FDR correction; ES = Cliff's Delta effect size. Columns I and E are based on the median as a measure of central tendency. The All Programs row compares the median fixation duration across all programs between the Inline and Extracted methods.} 
\begin{center}
\resizebox{12cm}{!}{
\begin{tabular}{p{2.9cm}rrrrrrrrrr}
    \hline
    \multicolumn{1}{|c|}{\textbf{Tasks}} &\multicolumn{5}{|c|}{\textbf{In AOI}} &  \multicolumn{5}{|c|}{\textbf{In Code}}\\
    \hline
    \multicolumn{1}{|c|@{}}{} &\multicolumn{1}{|c|@{}}{\shortstack{I\\sec}} &  \multicolumn{1}{|c|@{}}{\shortstack{E\\sec}} &\multicolumn{1}{|c|@{}}{\shortstack{\\PD\\\%}} &\multicolumn{1}{|c|}{PV} &\multicolumn{1}{|c|@{}}{ES}
    &\multicolumn{1}{|c|@{}}{\shortstack{I\\sec}} &  \multicolumn{1}{|c|@{}}{\shortstack{E\\sec}} &\multicolumn{1}{|c|@{}}{\shortstack{PD\\\%}}
    &\multicolumn{1}{|c|}{PV} &\multicolumn{1}{|c|}{ES}\\
    \hline
    
    Sum Numbers & 6.2 & 14.2 & {$\uparrow$130.1} & \textbf{0.01} & 0.60 & 
    8.7 & 18.4 & {$\uparrow$110.5} & \textbf{0.03} & 0.53 \\
    
    Next Prime & 63.7 & 35.3 & {$\downarrow$44.5} & 0.14 & n/a & 
    65.6 & 38.8 & {$\downarrow$40.8} & \textbf{0.07} & -0.39 \\
    
    Highest Grade & 45.1 & 11.8 & {$\downarrow$73.6} & \textbf{0.03} & -0.42 & 
    47.1 & 16.3 & {$\downarrow$65.8} & \textbf{0.07} & -0.53 \\
    
    Factorial & 40.2 & 8.4 & {$\downarrow$78.9} & \textbf{0.01} & -0.64 & 
    46.0 & 11.7 & {$\downarrow$74.5} & \textbf{0.07} & -0.51 \\
    
    Multiples of Three & 11.7 & 19.9 & {$\uparrow$69.1} & 0.38 & n/a & 
    18.1 & 28.9 & {$\uparrow$60.0} & 0.30 & n/a \\
    
    Area of a Square & 1.4 & 3.0 & {$\uparrow$121.1} & \textbf{0.01} & 0.65 & 
    4.1 & 6.5 & {$\uparrow$59.8} & 0.12 & n/a \\
    
    Check If Even & 2.9 & 5.0 & {$\uparrow$73.1} & \textbf{0.06} & 0.46 & 
    14.5 & 19.6 & {$\uparrow$35.1} & 0.09 & n/a \\
    
    Number of Digits & 21.6 & 14.5 & {$\downarrow$32.7} & 0.23 & n/a & 
    40.5 & 17.8 & {$\downarrow$55.9} & \textbf{0.07} & -0.38 \\
    
    \rowcolor[gray]{.8} All Programs & 70.9 & 59.2 & {$\downarrow$16.4} & 0.21 & n/a & 
    102.4 & 84.5 & {$\downarrow$17.4} & 0.12 & n/a \\
    
    \hline
\end{tabular}
}
\end{center}
\label{tab:fixation durationtable}
\end{table}

\begin{tcolorbox}[colback=gray!5, colframe=black!50, title=Answering RQ$_3$, sharp corners=south,
  width=\textwidth]
The impact of Extract Method refactoring on fixation duration differs across tasks; while some tasks benefited from reduced fixation duration, others experienced an increase. Notably, tasks that are generally considered straightforward exhibited increased visual effort in processing. This underscores that the effects of refactoring can be linked to the specific task characteristics, including their complexity and simplicity.
\end{tcolorbox}

\subsection{\textbf{RQ$_{4}$: To what extent does the Extract Method refactoring affect fixations count?}}\label{results:RQ4}

In this section, we investigate the impact of the Extract Method refactoring on the count of fixations by analyzing the results in the AOI and in the complete code. The findings, summarized in Table~\ref{tab: fixations counttable}, reveal a complex interplay between refactoring and fixation count, with significant reductions observed in some tasks and increases in others.

The Extract Method refactoring led to reductions in fixation count within the AOI for tasks such as calculating the \textit{Factorial} and determining the \textit{Highest Grade}.
Conversely, tasks including \textit{Sum Numbers}, \textit{Area of a Square}, and \textit{Check If Even}, similar to the previous eye tracking metric, exhibited an increase in fixation count in the AOI, indicating additional visual effort.
Additionally, the analysis of fixation count in the complete code mirrored this pattern, with significant reductions for the \textit{Factorial} and \textit{Highest Grade} tasks, while increases were observed for \textit{Sum Numbers} and \textit{Check If Even}.

\begin{table}[h]
\setlength\tabcolsep{1.8pt}
\caption{Results for \textbf{fixations count in AOI and in Code} (RQ$_{4}$). I = Inline Method; E = Extract Method; PD = Percentage Difference from Inline Method (I) to Extract Method (E); PV = \textit{p}-value after FDR correction; ES = Cliff's Delta effect size. Columns I and E are based on the median as a measure of central tendency. The All Programs row compares the median fixations count across all programs between the Inline and Extracted methods.} 
\begin{center}
\resizebox{12cm}{!}{
\begin{tabular}{p{2.9cm}rrrrrrrrrr}
    \hline
    \multicolumn{1}{|c|}{\textbf{Tasks}} &\multicolumn{5}{|c|}{\textbf{In AOI}} &  \multicolumn{5}{|c|}{\textbf{In Code}}\\
    \hline
    \multicolumn{1}{|c|@{}}{} &\multicolumn{1}{|c|@{}}{I} &  \multicolumn{1}{|c|@{}}{E} &\multicolumn{1}{|c|@{}}{\shortstack{\\PD\\\%}} &\multicolumn{1}{|c|@{}}{PV} &\multicolumn{1}{|c|@{}}{ES}
    &\multicolumn{1}{|c|@{}}{I} &  \multicolumn{1}{|c|@{}}{E} &\multicolumn{1}{|c|@{}}{\shortstack{PD\\\%}}
    &\multicolumn{1}{|c|@{}}{PV} &\multicolumn{1}{|c|@{}}{ES}\\
    \hline
    
    Sum Numbers & 17.5 & 44.0 & {$\uparrow$151.4} & \textbf{0.005} & 0.62 & 
    23.5 & 47.5 & {$\uparrow$102.1} & \textbf{0.003} & 0.56 \\
    
    Next Prime & 186.0 & 108.0 & {$\downarrow$41.9} & 0.19 & n/a & 
    191.5 & 111.0 & {$\downarrow$42.0} & 0.17 & n/a \\
    
    Highest Grade & 141.5 & 45.6 & {$\downarrow$67.7} & \textbf{0.02} & -0.53 & 
    166.0 & 50.0 & {$\downarrow$69.8} & \textbf{0.010} & -0.51 \\
    
    Factorial & 141.0 & 34.0 & {$\downarrow$75.8} & \textbf{0.04} & -0.44 & 
    156.0 & 40.0 & {$\downarrow$74.3} & \textbf{0.02} & -0.40 \\
    
    Multiples of Three & 38.5 & 62.5 & {$\uparrow$62.3} & 0.33 & n/a & 
    47.5 & 80.5 & {$\uparrow$69.4} & 0.26 & n/a \\
    
    Area of a Square & 4.6 & 11.0 & {$\uparrow$138.8} & \textbf{0.005} & 0.59 & 
    12.0 & 20.0 & {$\uparrow$66.6} & 0.07 & n/a \\
    
    Check If Even & 8.5 & 20.1 & {$\uparrow$137.1} & \textbf{0.005} & 0.58 & 
    39.0 & 56.5 & {$\uparrow$44.8} & \textbf{0.04} & 0.42 \\
    
    Number of Digits & 96.4 & 49.0 & {$\downarrow$49.2} & 0.19 & n/a & 
    99.0 & 49.0 & {$\downarrow$50.5} & 0.06 & n/a \\
    
    \rowcolor[gray]{.8} All Programs & 252.0 & 189.5 & {$\downarrow$24.8} & 0.26 & n/a & 
    288.0 & 282.5 & {$\downarrow$1.9} & 0.24 & n/a \\
    
    \hline
\end{tabular}
}
\end{center}
\label{tab: fixations counttable}
\end{table}

\begin{tcolorbox}[colback=gray!5, colframe=black!50, title=Answering RQ$_4$, sharp corners=south,
  width=\textwidth]
In line with previous observations regarding visual effort, some tasks demonstrated a reduction in fixation count, while others saw an increase. Notably, tasks that are generally considered straightforward showed a significant increase in number of times the novices fixated on the AOI.
\end{tcolorbox}

\subsection{\textbf{RQ$_{5}$: To what extent does the Extract Method refactoring affect regressions count?}}\label{results:RQ5}

In this section, we assess the influence of the Extract Method refactoring on the count of regressions by examining both the AOI and the complete code. The results, presented in Table~\ref{tab: regressions counttable}, highlight a significant variation in regression behavior across different tasks.

The Extract Method refactoring led to a substantial reduction in regression counts in the AOI for tasks such as calculating the \textit{Factorial} and determining the \textit{Highest Grade}.
In contrast, an increase in regressions count was observed for the tasks of \textit{Sum Numbers} and \textit{Check If Even} within the AOI, suggesting a hindered task performance.
This pattern is echoed in the complete code analysis, where significant reductions were noted for the \textit{Factorial} and \textit{Highest Grade} tasks, while an increase was recorded for \textit{Sum Numbers}.

\begin{table}[h]
\setlength\tabcolsep{1.8pt}
\caption{Results for \textbf{regressions count in AOI and in Code} (RQ$_{5}$). I = Inline Method; E = Extract Method; PD = Percentage Difference from Inline Method (I) to Extract Method (E); PV = \textit{p}-value after FDR correction; ES = Cliff's Delta effect size. Columns I and E are based on the median as a measure of central tendency. The All Programs row compares the median regressions count across all programs between the Inline and Extracted methods.} 
\begin{center}
\resizebox{12cm}{!}{
\begin{tabular}{p{2.9cm}rrrrrrrrrr}
    \hline
    \multicolumn{1}{|c|}{\textbf{Tasks}} &\multicolumn{5}{|c|}{\textbf{In AOI}} &  \multicolumn{5}{|c|}{\textbf{In Code}}\\
    \hline
    \multicolumn{1}{|c|@{}}{} &\multicolumn{1}{|c|@{}}{I} &  \multicolumn{1}{|c|@{}}{E} &\multicolumn{1}{|c|@{}}{\shortstack{\\PD\\\%}} &\multicolumn{1}{|c|@{}}{PV} &\multicolumn{1}{|c|@{}}{ES}
    &\multicolumn{1}{|c|@{}}{I} &  \multicolumn{1}{|c|@{}}{E} &\multicolumn{1}{|c|@{}}{\shortstack{PD\\\%}}
    &\multicolumn{1}{|c|@{}}{PV} &\multicolumn{1}{|c|@{}}{ES}\\
    \hline
    
    Sum Numbers & 6.0 & 12.5 & {$\uparrow$108.3} & \textbf{0.03} & 0.57 & 
    9.5 & 18.0 & {$\uparrow$89.4} & \textbf{0.03} & 0.52 \\
    
    Next Prime & 80.5 & 41.0 & {$\downarrow$49.0} & 0.12 & n/a & 
    84.5 & 44.0 & {$\downarrow$47.9} & 0.12 & n/a \\
    
    Highest Grade & 43.0 & 11.0 & {$\downarrow$74.4} & \textbf{0.03} & -0.50 & 
    75.0 & 19.0 & {$\downarrow$74.6} & \textbf{0.03} & -0.53 \\
    
    Factorial & 52.0 & 8.0 & {$\downarrow$84.6} & \textbf{0.03} & -0.67 & 
    70.5 & 15.5 & {$\downarrow$78.0} & \textbf{0.03} & -0.49 \\
    
    Multiples of Three & 13.5 & 19.0 & {$\uparrow$40.7} & 0.41 & n/a & 
    21.5 & 28.5 & {$\uparrow$32.5} & 0.53 & n/a \\
    
    Area of a Square & 1.0 & 2.0 & {$\uparrow$100.0} & 0.11 & n/a & 
    5.0 & 7.0 & {$\uparrow$40.0} & 0.12 & n/a \\
    
    Check If Even & 1.0 & 3.0 & {$\uparrow$200.0} & 0.07 & 0.43 & 
    18.5 & 24.5 & {$\uparrow$32.4} & 0.15 & n/a \\
    
    Number of Digits & 21.0 & 13.0 & {$\downarrow$38.0} & 0.15 & n/a & 
    46.0 & 21.5 & {$\downarrow$53.2} & 0.12 & n/a \\
    
    \rowcolor[gray]{.8} All Programs & 75.0 & 54.0 & {$\downarrow$28.0} & 0.12 & n/a & 
    125.5 & 114.0 & {$\downarrow$9.16} & 0.12 & n/a \\
    
    \hline
\end{tabular}
}
\end{center}
\label{tab: regressions counttable}
\end{table}

\begin{tcolorbox}[colback=gray!5, colframe=black!50, title=Answering RQ$_5$, sharp corners=south,
  width=\textwidth]
The Extract Method refactoring reduced the count of regressions for more complex tasks, aligning with the previous eye-tracking metrics. Conversely, for simpler tasks, it resulted in a higher frequency of regressions as novices go back in the code more often.
\end{tcolorbox}

\subsection{\textbf{RQ$_{6}$: How does the Extract Method refactoring affect the perceptions {and motivations}
of the subjects?}}\label{results:RQ6}

\begin{figure}[h!]
\centering
  \includegraphics[width=0.7\textwidth]{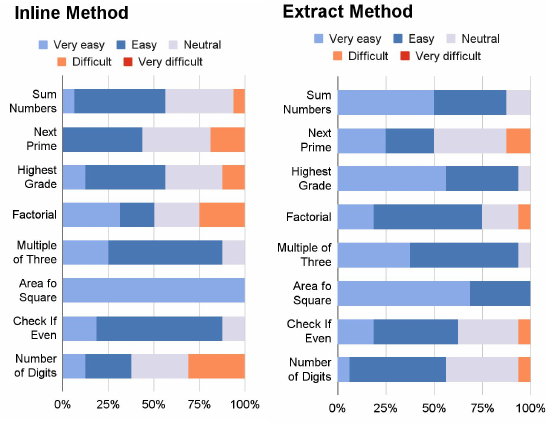}
  \caption{Perception of difficulties of the subjects with the Inline Method and Extract Method of the tasks.}
  \label{fig: perception}
\end{figure}

Through a combination of interviews and an online survey with novices in Java, we aimed to understand the difficulties they encountered with Extracted and Inlined Method versions and their underlying motivations for selecting either approach. The 32 novices in Java who solved the code programs rated each program individually using a five-point scale \proof{ranging} from very easy, easy, neutral, difficult, or very difficult to solve. We \proof{compared} their perceived difficulties between Inline and Extract Method in Figure~\ref{fig: perception}. We found that subjects perceived tasks with the Extract Method \proof{as easier} to solve than those with the Inline Method, especially for tasks such as determining the \textit{Highest Grade}, \proof{calculating} the \textit{Factorial}, and \proof{determining} the \textit{Number of Digits}. 

{To strengthen the interpretation of the stacked bar charts in Figure~\ref{fig: perception}, we complemented the visual analysis with statistical testing. First, a global Wilcoxon test comparing difficulty ratings between the Extracted and Inlined Methods indicated no overall difference (\textit{p-value} = 0.1698). We then conducted paired Wilcoxon tests for each task, applying a Bonferroni correction. Only the task to compute Highest Grade remained significant after correction (p = 0.0244), indicating that, except for this case, difficulty ratings did not differ across conditions.}

\begin{figure}[h!]
    \centering
    \includegraphics[width=0.8\textwidth]{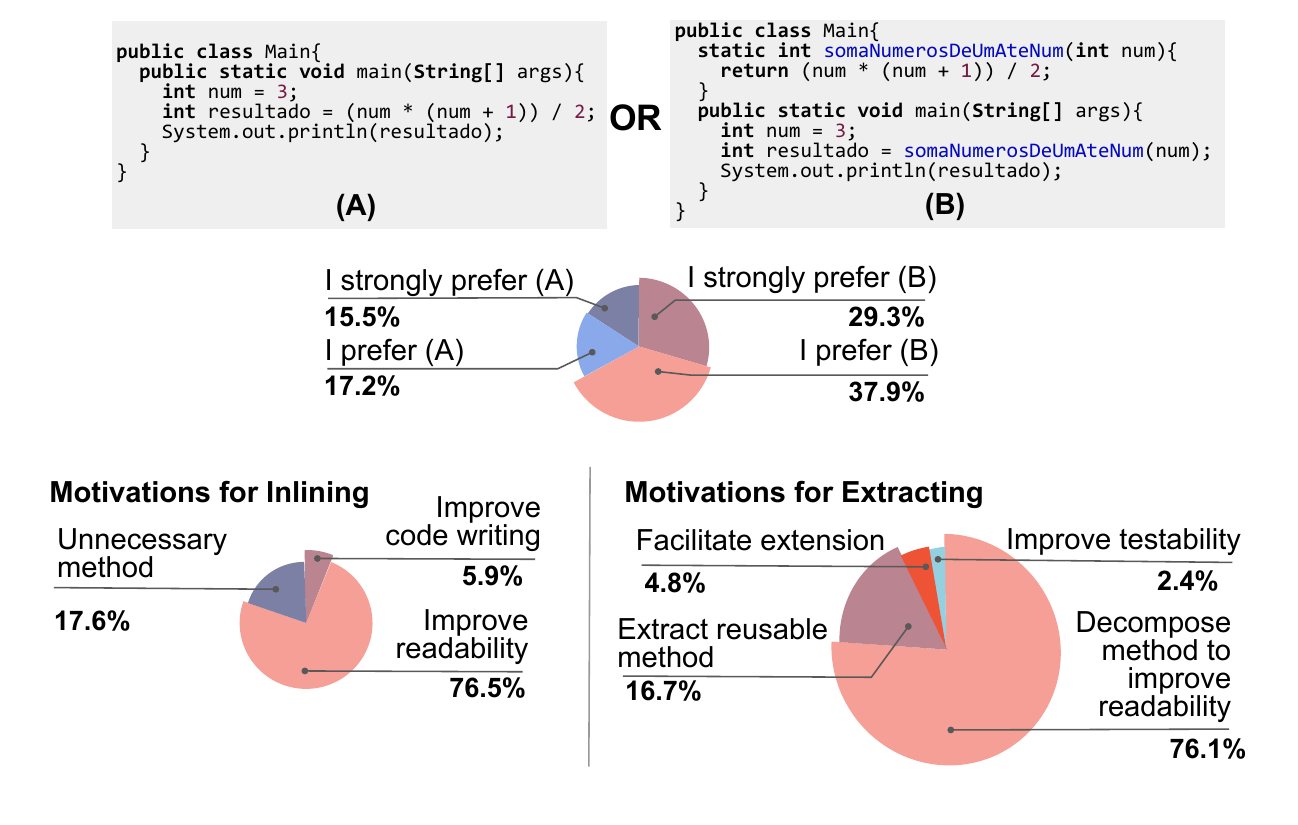}
    \caption{Analysis of the results of the survey with motivations for extracting and inlining a method in the task to sum numbers from one to num.}
    \label{fig: survey-sum-numbers}
\end{figure}

We conducted an online survey with \proof{another} 58 novices in Java comparing code snippets with the method extracted and inlined. In Figure~\ref{fig: survey-sum-numbers}, we depict the results for the comparison of the task to sum numbers from one to num and the motivations for their preferences. For 76.1\% of the subjects, the extracted method version was preferred mainly because it decomposes the method to improve the readability of the code, followed by extracting a reusable method (16.7\%), facilitating extension (4.8\%) and improving testability (2.4\%). With respect to improving readability, we found terms such as ``\textit{clean},'' ``\textit{organized},'' and ``\textit{simple}.'' On the other hand, 32.7\% of the subjects preferred the inlined one mainly because they find the version simpler, easier to understand, or more readable, or because the method becomes unnecessary.  One participant mentioned that ``\textit{It allows a continuous reading of the code, without needing to go back to the function call}.'' Indeed, this aligned with an increase in the number of regressions in the code with the extracted method version. In general, the underlying reasons align well with those obtained by Silva et al.~\cite{silva2016why} who investigated the motivations of professionals for applying the Extract Method or the Inline Method. Interestingly, novices seemed to prioritize code readability over reuse.

\begin{figure}[h!]
    \centering
    \includegraphics[width=0.8\textwidth]{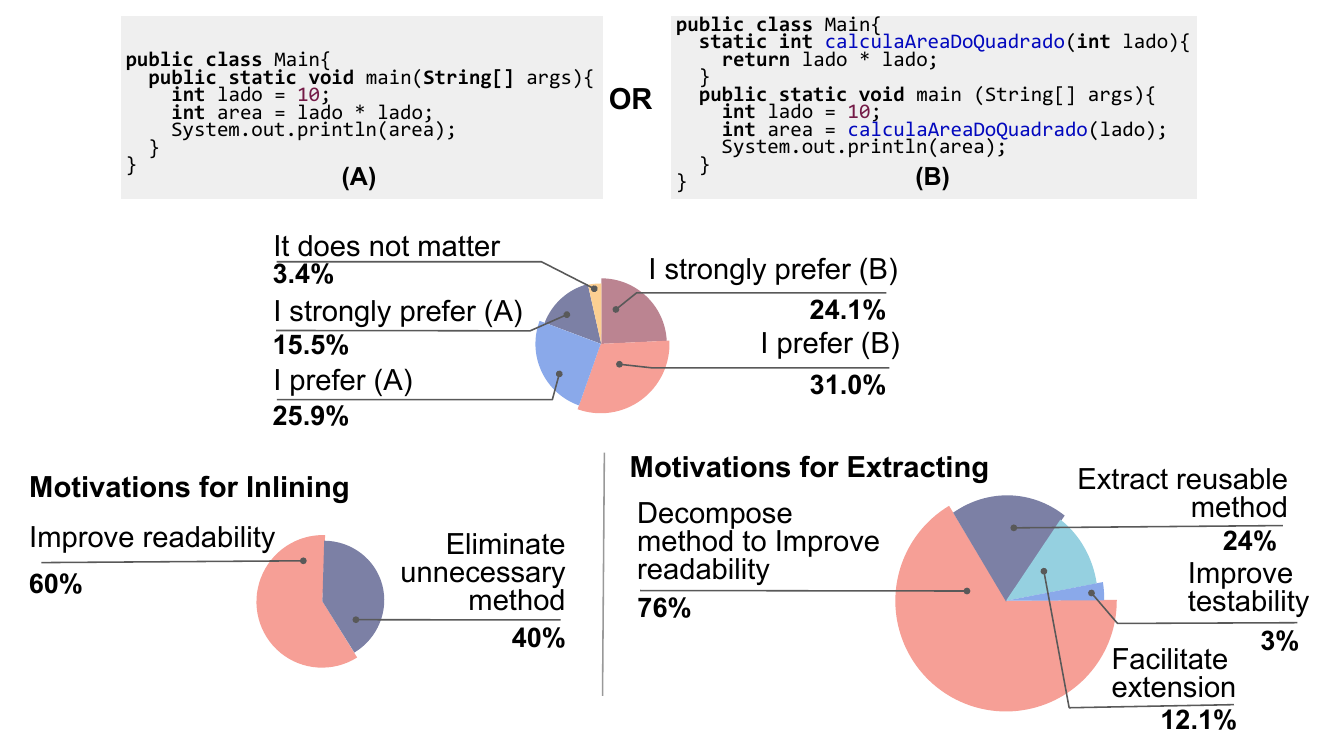}
    \caption{Analysis of the results of the survey with motivations for extracting and inlining a method in the task to calculate the area of the square.}
    \label{fig: survey-square}
\end{figure}

In Figure~\ref{fig: survey-square}, we compare the preferences for both versions of the task to calculate the area of a square. 
The novices showed a split preference between the inlined and extracted versions. The Extracted Method version was preferred by 31.0\%, with an additional 24.1\% strongly preferring it, whereas the Inlined Method version was preferred by 25.9\% and strongly preferred by 15.5\%, with only 3.4\% reporting no preference. These findings reveal a nuanced trade-off between the two styles. Although readability was a key factor motivating both inlining and \proof{extraction}, the extracted version was additionally valued for supporting reuse and future extensions, which aligns with Fowler's argument that shorter, well-named methods can improve code maintainability~\cite{fowler2018refactoring}. On the other hand, a considerable portion of the subjects favored the inlined version, reflecting a preference for directness and avoiding what they perceived as unnecessary indirection. We found that despite the structural and long-term benefits of shorter methods, novice developers tend to perceive code as more readable when the logic is concise and immediately visible in one place.

\begin{figure}[h!]
    \centering
    \includegraphics[width=0.8\textwidth]{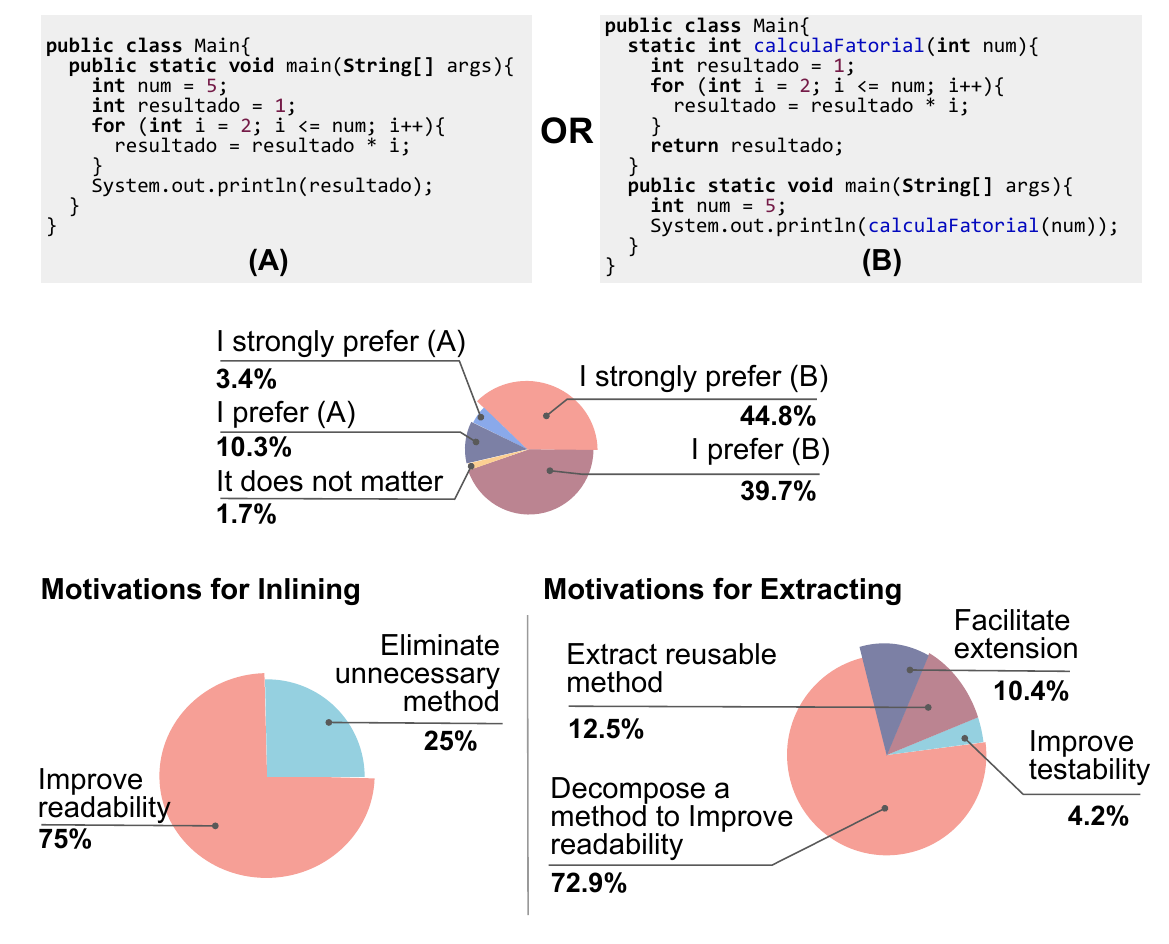}
    \caption{Survey results with motivations for extracting and inlining a method in the task to calculate the factorial of a number.}
    \label{fig: motivating-example-survey-factorial}
\end{figure}

In a task to calculate the factorial of a number, the novices expressed a preference for the extracted method version, as depicted in Figure~\ref{fig: motivating-example-survey-factorial}. While 39.7\% preferred the extracted version, 44.8\% strongly preferred it. In contrast, only 10.3\% preferred the inlined version and 3.4\% strongly preferred it, with 1.7\% reporting no preference. The main motivation for extracting the method was decomposing the method to improve readability (72.9\%), followed by enabling reuse (12.5\%), facilitating extension (10.4\%), and improving testability (4.2\%). The novices who preferred the extracted method often emphasized its role in reducing cognitive load. One subject even pointed out that the iterative nature of the factorial calculation justified extraction, as ``\textit{a larger operation replaced by a method with a clear name becomes easier to understand.}'' Another subject mentioned that this ``\textit{extracting would make more sense given that this algorithm was more complex.}'' These statements indicate that method extraction helps the novices manage the mental effort required to process non-trivial logic.

Other tasks such as returning highest grade from the list, computing next prime, counting the number of digits, checking if a number is even, and counting multiples of three presented similar perceptions always with higher preferences for the Extracted Method version. The main motivations were decomposing a method to improve readability, with rates ranging from 65.4\% to 77.1\% followed by extracting a reusable method with rates ranging from 12.1\% to 17.1\%. Across these five tasks, the novices consistently favored extracted methods as the algorithms were perceived as more complex. For instance, one subject mentioned ``\textit{I believe Option A [extracted method version] is better because the logic in the function is not so trivial.}'' They often mentioned that extraction facilitated understanding, that the method name improved readability, and that the code became more organized, clear, and reusable. In contrast, inlining was preferred for trivial operations such as checking an even number. {One subject mentioned }``\textit{I don't see a reason why to make a whole method just to check if the variable is even or odd.}'' This suggests that method extraction is perceived as essential for managing non-trivial logic.

Fowler~\cite{fowler2018refactoring} identifies several motivations for the Extract Method refactoring, such as improving readability, supporting reuse, and clarifying code responsibilities. These motivations also appeared in studies with professionals. For instance, Silva et al.~\cite{silva2016why} found that reuse and decomposing a method to improve readability were the main drivers for extraction, while eliminating unnecessary methods and improving readability were the main drivers for inlining. Interestingly, readability was not the primary motivation among professionals. In contrast, in our study with Java novices, readability emerged as the central reason for both extraction and inlining. Novices frequently described extracted methods as making the code simpler, easier to understand, more organized, and cleaner. 

While Fowler's original motivations~\cite{fowler2018refactoring} and the professionals' accounts~\cite{silva2016why} emphasize maintenance and comprehension, the novices' perspective reveals a complementary dimension: {the Inline Method} version is preferred because it improved their code writing, as depicted in Figure~\ref{fig: survey-sum-numbers}. The subject mentioned that ``\textit{Although the second one [method extracted] is more organized, I prefer the first one [method inlined] for small codes because it's easier to code.}'' We observed a contrast between novices and professionals: while novices tend to view refactoring as a way to simplify their immediate task, professionals frame it as a long-term structural investment in the system's overall health.

% etapa de codigos
{We coded the novices' answers in the interviews. In the code step, we found a set of 35 codes across 256 answers. The issues included ``Didn't pay attention to the function'', ``Difficulties with the repetition'', and ``The name helped to infer.'' We added more details in Table~\ref{tab: grounded-theory-process} and the complete process in our supplementary material~\cite{our-artifacts}.
In the concept step, we connected and grouped multiple codes. We found 16 concepts that emerged from the codes. For instance, we grouped ``Names were suggestive'' and ``The name helped infer'' in the same underlying concept, which is ``Meaningfulness.''
% etapa de categorias
In the category step, seven main categories emerged from the concepts, namely, 
``Attention'', ``Knowledge'', ``Control Flow'', ``Names'', ``Mathematics'', ``Memory load'', and ``No difficulties.'' 
% etapa de teoria
In the theory step, we abstracted the observations and difficulties identified in the earlier steps into generalized theoretical insights. }

\begin{table}[h]
\centering
{\renewcommand{\arraystretch}{1.3}
\begin{tabular}{@{}p{3.1cm}p{2.4cm}p{1.8cm}p{3cm}}
\toprule 
\thead{\textbf{Code}\\ \textbf{Step}} & \thead{\textbf{Concept}\\ \textbf{Step}} & \thead{\textbf{Category}\\ \textbf{Step}} & \thead{\textbf{Theory}\\ \textbf{Step}}\\
\midrule 
Didn't pay attention to the function &  Lack of attention & \multirow{4}{2cm}{Attention}  & \multirow{5}{3cm}{{Limited cognitive load impairs cognitive processing}} \\ 
\cline{1-2}
Didn't see the equals operator & Missed details \\
\cline{1-3}
Difficulties in remembering the values & short-term memory & \multirow{2}{2cm}{Memory load} \\
\cline{1-4}
Familiarity with factorial & Knowledge of Domain & \multirow{5}{2cm}{Knowledge} & \multirow{6}{3cm}{{Prior knowledge supports the development of cognitive schemas}}\\
\cline{1-2}
Not used to this iteration & Knowledge of Iterator \\
\cline{1-2}
Confused the variable type & Knowledge of Language \\
\cline{1-4}
Difficulties with modulo & Decision structure & \multirow{8}{2cm}{Control Flow} & \multirow{7}{3cm}{{Control flow structures demand more cognitive resources}} \\
\cline{1-2}
Difficulties with the repetition & Loop structure \\
\cline{1-2}
Loop with decision is difficult & Nesting \\
\cline{1-2}
Going back and forth takes time & Reading flow \\
\cline{1-2}
Had to read line by line & Linear reading \\
\cline{1-4}
The name helped to infer & Meaningfulness & \multirow{3}{2cm}{Names} & \multirow{4}{3cm}{{Meaningful names help to construct mental structures}} \\
\cline{1-2}
Didn't understand the name & Lack of clearness \\
\cline{1-4}
Confused in the multiplication & Math operation & \multirow{4}{2cm}{Mathematics} & \multirow{4}{3.5cm}{{Interpreting mathematical operations in code increases mental effort}} \\
\cline{1-2}
Confused the order of the operations & Precedence order \\

\cline{1-4}
No difficulties & No difficulties & \multirow{1}{4cm}{No difficulties} \\

\bottomrule
\end{tabular}
}
\caption{{Coding process of the subjects' answers, illustrating the connections between observed difficulties, abstract concepts, and derived theories.}}
\label{tab: grounded-theory-process}
\end{table}

{Method names emerged as a key factor in code comprehension, with 44 responses referencing either the presence or absence of suggestive names. The subjects consistently reported that descriptive method names helped them infer code behavior and form hypotheses about functionality. One subject mentioned \textit{''To solve the task, I used only the name of the method and the input, otherwise it would take longer.''} Method names also increased subjects' confidence, as reported: ``\textit{I answered correctly because of the name}.''
In the absence of method names, the variable names acted as cues for understanding the code. For instance, a subject mentioned ``\textit{Since there were no method names, the variable names provided hints}.'' }

{We also learned that subjects check whether the method does what it says. While in three tasks, the subjects mentioned not needing to check the method, in 22 tasks, the subjects checked the method. Some of the reasons were \textit{``Checked the code to see if the body matches the name''}, \textit{''You cannot always trust the code others wrote''}, and \textit{``To check if there is some kind of trick.''}}

% codigo sem metodo
{Inlining the method favored a bottom-up reading. In the Control Flow category, covering decision and repetition structures, nesting, and linear reading, subjects reported difficulties in 31 tasks, 25 of them with method inlined. One subject mentioned ``\textit{Since I had no method to help me, I had to read the code line by line}.'' This careful examination increases both time and visual effort.}

\begin{tcolorbox}[colback=gray!5, colframe=black!50, title=Answering RQ$_6$, sharp corners=south,
  width=\textwidth]
The preference for the extracted method version is predominant across the tasks. The novices were mainly motivated by improved readability and reuse. In contrast, the preferences for the inlined version were driven by perceptions of readability and elimination of unnecessary methods. In addition, despite the structural and long-term benefits of shorter methods, novices perceive code as more readable when the logic is concise and immediately visible in one place. {Clear method names emerged as a relevant factor influencing novices' comprehension, guiding top-down reasoning and reducing cognitive effort; when names were unclear or absent, students shifted to slower, bottom-up processing.}
\end{tcolorbox}

\section{Discussion}
\label{sec:discussion}

In this section, we examine and discuss each task individually: Sum numbers from one to num (Section~\ref{disc:sum-numbers}), calculate factorial (Section~\ref{disc:calc-factorial}), calculate the area of the square (Section~\ref{disc:area-square}), calculate the next prime (Section~\ref{disc:next-prime}), return the highest grade from the list (Section~\ref{disc:highest-grade}), count multiples of three (Section~\ref{disc:mult-three}), check if even (Section~\ref{disc:check-even}), and count number of digits (Section~\ref{disc:num-digits}). We also present the coding of the subjects' answers (Section~\ref{disc:coding}) and propose an integrative theory of how programmers comprehend code when comparing inline and extract-method representations (Section~\ref{sec:integrative-theory}).

\subsection{{Sum numbers from one to num}} \label{disc:sum-numbers} 

Participants' perceptions align with the online-survey responses and with the findings of Silva et al.~\cite{silva2016why}, in that developers extract a piece of code into a separate method to make the original method easier to understand. {Although the participants perceived the extracted version as easier, this perception did not fully align with their actual performance. While the refactored version felt simpler, their behavior indicated increased cognitive effort, reflected in longer processing time, more attempts, and greater visual activity.} To shed light on this discrepancy, we examine the qualitative data collected in the interviews.

Three subjects mentioned ``difficulties in the order of arithmetic operations'' while one mentioned difficulty interpreting the return expression. A participant who took four attempts to solve it mentioned the return as difficult. When the task involves an unfamiliar formula, the inlined version may be more effective, as it presents the computation directly rather than abstracting it into another region of the code via a return. Not everyone is familiar with the formula for the sum of the first \textit{n} positive integers, which we intentionally used to contrast with a more familiar formula, such as calculating the area of a square. In both cases, the impact was negative for the time and visual effort, except for the number of attempts, which increased more substantially for this task.

\begin{figure}[h!]
    \centering
    \includegraphics[width=\textwidth]{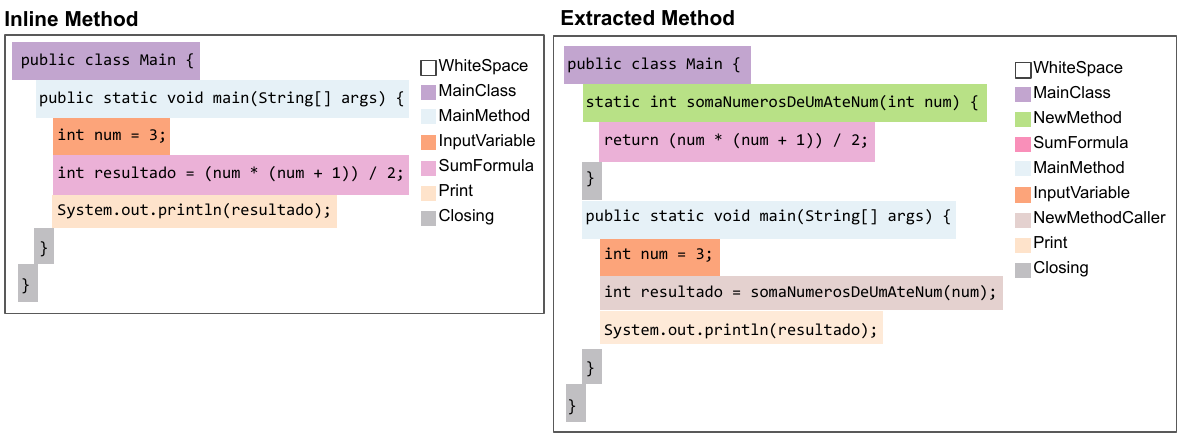}
    \caption{Set of regions inside the inlined and extracted versions for the task to sum numbers from one to num.}
    \label{fig: sum-number-regioes}
\end{figure}

\begin{figure}[ht]
  \subfloat[Transitions of Subject 5]{
	\begin{minipage}[c][0.2\width]{
	   0.3\textwidth}
	   \centering
	   \includegraphics[width=1\textwidth]{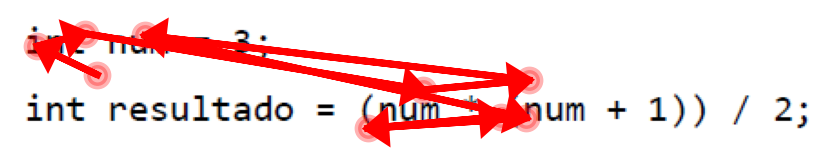}
	\end{minipage}}
 \hfill 	
  \subfloat[Transitions of Subject 7]{
	\begin{minipage}[c][0.2\width]{
	   0.3\textwidth}
	   \centering
	   \includegraphics[width=1\textwidth]{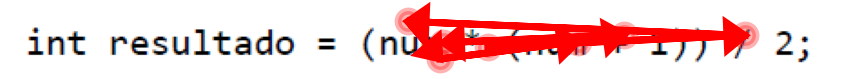}
	\end{minipage}}
 \hfill	
  \subfloat[Transitions of Subject 9]{
	\begin{minipage}[c][0.2\width]{
	   0.3\textwidth}
	   \centering
	   \includegraphics[width=1\textwidth]{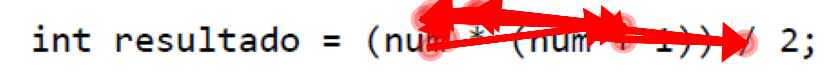}
	\end{minipage}}
\caption{Sequence of transitions of subjects on the inlined version for the task to sum numbers from one to num.}
\label{fig: gaze-sum-numbers-inlined}
\end{figure}

To deepen our analysis and search for reading patterns, we identified a set of regions in the code as depicted in Figure~\ref{fig: sum-number-regioes}. This strategy allowed us to find gaze transitions, which are visual transitions between these regions. For instance, in the inlined method, after reading the line of the formula, \texttt{SumFormula}, 87\% of the subjects go back to the previous line where the main variable involved was assigned, \texttt{SumFormula} $\rightarrow$ \texttt{InputVariable}, possibly to refresh the memory with the input value. An example is depicted in Figure~\ref{fig: gaze-sum-numbers-inlined}(a). The transition \texttt{SumFormula} $\rightarrow$ \texttt{Print} was made by 93.7\% reaching 15 times in total, indicating a linear reading to deliver the solution. The region \texttt{SumFormula} plays the most important role to solve the task which is confirmed by the number of fixations in it. Each subject fixates their eyes on this area 19.6 times on average. Half of these fixations are characterized by visual movements going back and forth concentrated inside the region, such as depicted in Figure~\ref{fig: gaze-sum-numbers-inlined}(b) and Figure~\ref{fig: gaze-sum-numbers-inlined}(c).

\begin{figure}[ht]
  \subfloat[Transitions of Subject 8]{
	\begin{minipage}[c][0.3\width]{
	   0.3\textwidth}
	   \centering
	   \includegraphics[width=1\textwidth]{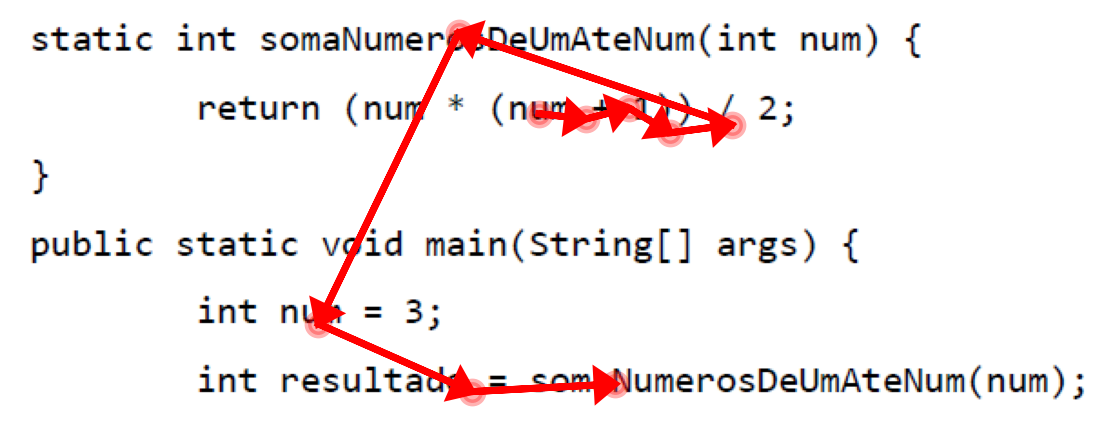}
	\end{minipage}}
 \hfill	
  \subfloat[Transitions of Subject 10]{
	\begin{minipage}[c][0.3\width]{
	   0.3\textwidth}
	   \centering
	   \includegraphics[width=1\textwidth]{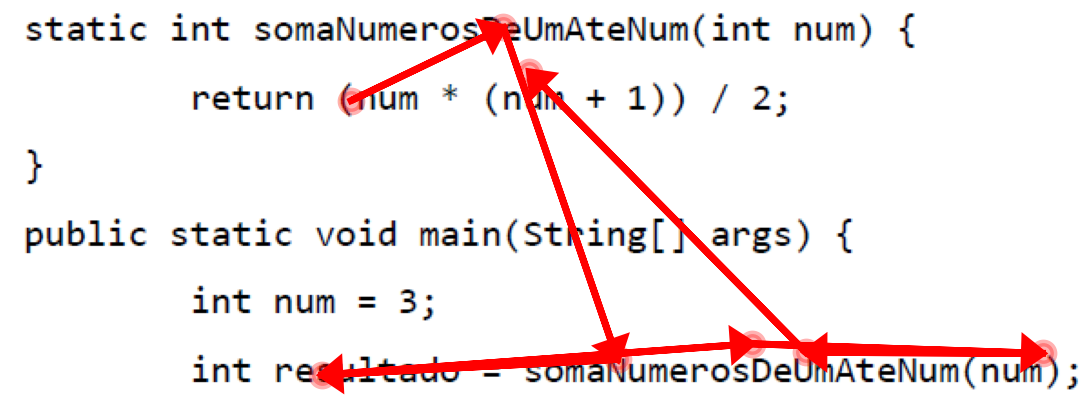}
    \end{minipage}}
 \hfill
\subfloat[Transitions of Subject 26]{
	\begin{minipage}[c][0.3\width]{
	   0.3\textwidth}
	   \centering
	   \includegraphics[width=1\textwidth]{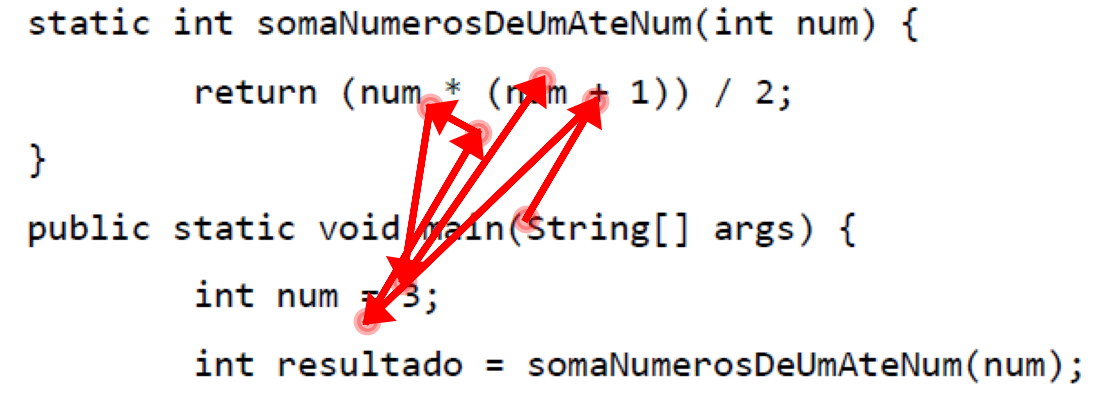} 
	\end{minipage}}
 
\caption{Sequence of transitions of subjects on the extracted version for the task to sum numbers from one to num.}
\label{fig: gaze-sum-numbers-extracted}
\end{figure}

Concerning the extracted method, we had more visual regressions horizontally and vertically in the AOI. The horizontal regressions might relate to longer lines because of the methods' names. The vertical regressions in the tasks might relate to going back and forth between caller and method. This behavior can be confirmed by the patterns in the following gaze transitions. For instance, 100\% of the subjects fixated on the regions \texttt{NewMethod} and {\texttt{NewMethodCaller}}, an average of 9.7 and 4.5 times, respectively. Some of these fixations are characterized by visual movements going back and forth inside the region, as depicted in Figure~\ref{fig: gaze-sum-numbers-extracted}(b). Similar to the inlined method, the subjects tend to go back to \texttt{InputVariable}, where the main variable is assigned a value. We observed this area being fixated 4.4 times on average, with the transition {\texttt{NewMethodCaller}} $\rightarrow$ \texttt{InputVariable} appearing 19 times and \texttt{SumFormula} $\rightarrow$ \texttt{InputVariable} appearing 34 times. In Figure~\ref{fig: gaze-sum-numbers-extracted}(c), we have examples of subjects going back and forth between these two areas and performing longer transitions. Similar behavior was observed for other subjects. 

\begin{figure}[h!]
    \centering
    \includegraphics[width=\textwidth]{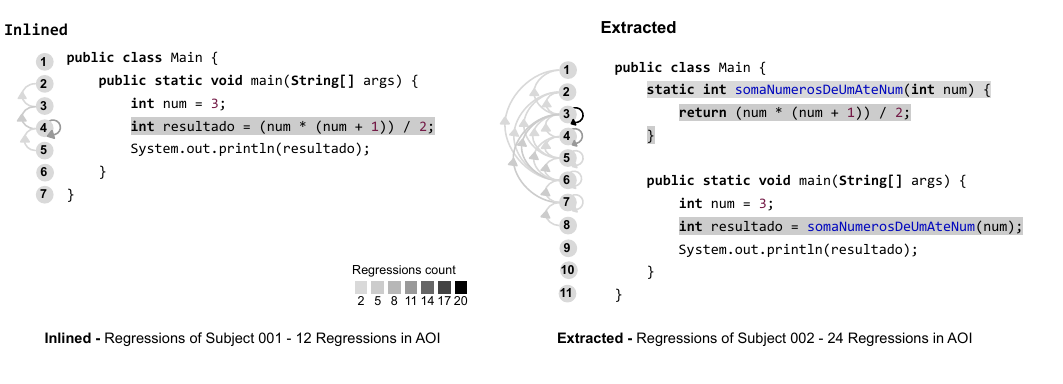}
    \caption{Visual regressions for the inlined and extracted versions for the task to sum numbers from one to num.}
    \label{fig: 11N_vs_11R}
\end{figure}

We developed transition graphs in our study to depict the distribution of the regressions for two subjects while examining a program. We selected two subjects whose individual results for the time, attempts, and visual metrics are coherent with the median results for the other subjects.  
One subject examined the program with the inlined version and the other examined the extracted version. In the graph, each edge represents a regression with a direction to a previous line of code or to the same line (self-loop edge). Each node represents a line of code. The grayscale intensity of the edge represents the number of times such regression was repeated. In Figure~\ref{fig: 11N_vs_11R}, the Extracted version exhibits several long regressions, whereas the inlined version shows fewer and shorter regressions. This reinforces the idea of a more linear and localized navigation pattern in the Inlined version, suggesting that participants were able to follow the control flow with fewer disruptions and less need to backtrack, which may indicate lower visual effort during comprehension.

As a takeaway, while inlining keeps all the logic in one place, reducing navigation overhead, method extraction may increase the need to visually search and switch between different regions of code. In the context of novices, these choices can impact the effort involved in understanding the code, highlighting the importance of evaluating the specific context of each task when deciding which approach to adopt.

\subsection{{Calculate Factorial}} \label{disc:calc-factorial}

\begin{figure}[h!]
    \centering
    \includegraphics[width=\textwidth]{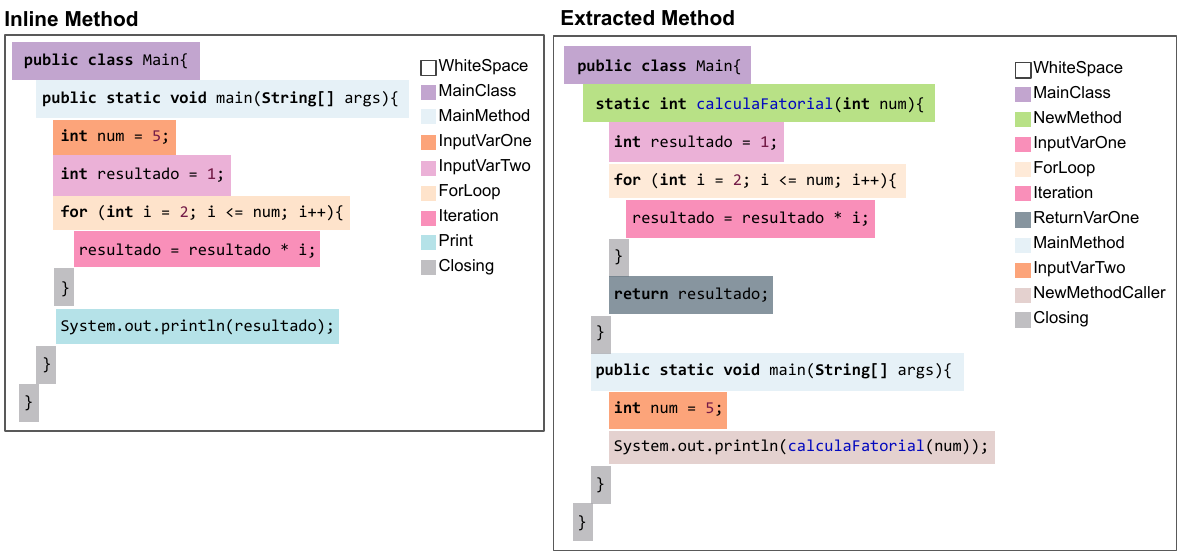}
    \caption{Set of regions inside the inlined and extracted versions for the task to calculate the factorial.}
    \label{fig: calc-factorial-regioes}
\end{figure}

{We noted a general alignment between participants’ subjective perceptions and their overall performance patterns. Rather than focusing solely on quantitative differences, we need to look more closely at how method extraction affected the way novices visually navigated the code. To deepen this interpretation and better understand which regions of the code were most affected, we now turn to Figure}~\ref{fig: calc-factorial-regioes}.

\begin{figure}[ht]
  \subfloat[Transitions of Subject 1]{
	\begin{minipage}[c][0.3\width]{
	   0.3\textwidth}
	   \centering
	   \includegraphics[width=1\textwidth]{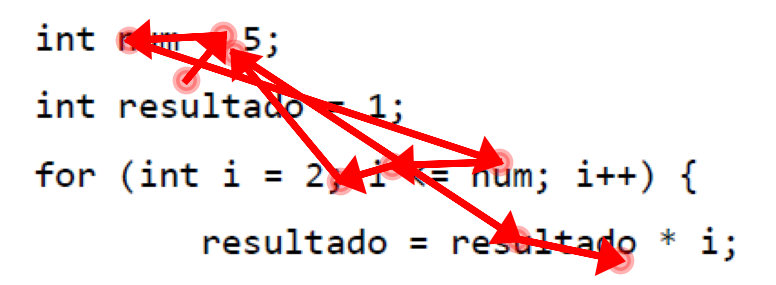}
	\end{minipage}}
 \hfill 	
  \subfloat[Transitions of Subject 5]{
	\begin{minipage}[c][0.3\width]{
	   0.3\textwidth}
	   \centering
	   \includegraphics[width=1\textwidth]{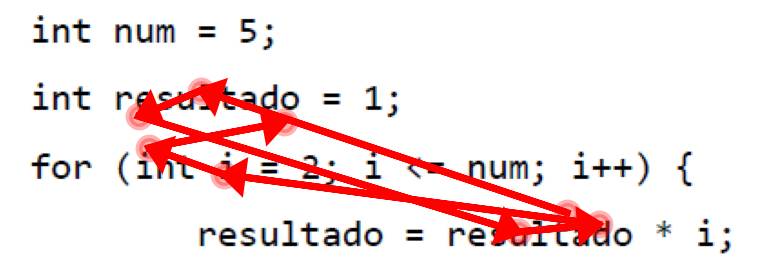}
	\end{minipage}}
 \hfill	
  \subfloat[Transitions of Subject 9]{
	\begin{minipage}[c][0.3\width]{
	   0.3\textwidth}
	   \centering
	   \includegraphics[width=1\textwidth]{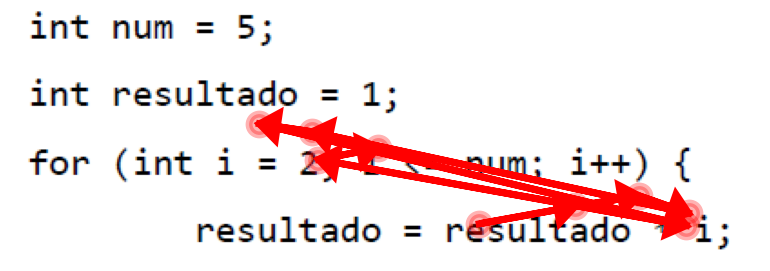}
	\end{minipage}}
\caption{Sequence of transitions of subjects on the inlined version for the task to calculate the factorial.}
\label{fig: gaze-calc-factorial-inlined}
\end{figure}

In the inlined version, the subjects made more visual regressions mainly in the \texttt{for} loop. The loop variable iterates four times, which means that the gaze transitions  \texttt{ForLoop} $\rightarrow$ \texttt{Iteration} can occur up to four times. {However, the novices make back-and-forth transitions between these two regions almost three times more often than what would be expected from simply tracing its execution, which is beyond what the structure of the code would demand. This might indicate that novices were not merely following the loop but repeatedly checking and rechecking its logic.}
%the forward transition \texttt{ForLoop} $\rightarrow$ \texttt{Iteration} on average 11.2 times and the backward transition \texttt{Iteration} $\rightarrow$ \texttt{ForLoop} 12.5 times. 
In Figure~\ref{fig: gaze-calc-factorial-inlined}(b) and Figure~\ref{fig: gaze-calc-factorial-inlined}(c), we observe two examples of this transition. Regressive movements from the loop reaching the variable involved were also frequent. For instance, they made the backward transition \texttt{ForLoop} $\rightarrow$ \texttt{InputVarOne} %3.7 times 
and \texttt{ForLoop} $\rightarrow$ \texttt{InputVarTwo}
{frequently, reinforcing the idea that the inlined version might be associated with constant variable confirmation.} %11.3 times, on average. 
An example can be seen in Figure~\ref{fig: gaze-calc-factorial-inlined}(a).

\begin{figure}[ht]
  \subfloat[Transitions of Subject 6]{
	\begin{minipage}[c][0.6\width]{
	   0.3\textwidth}
	   \centering
	   \includegraphics[width=1\textwidth]{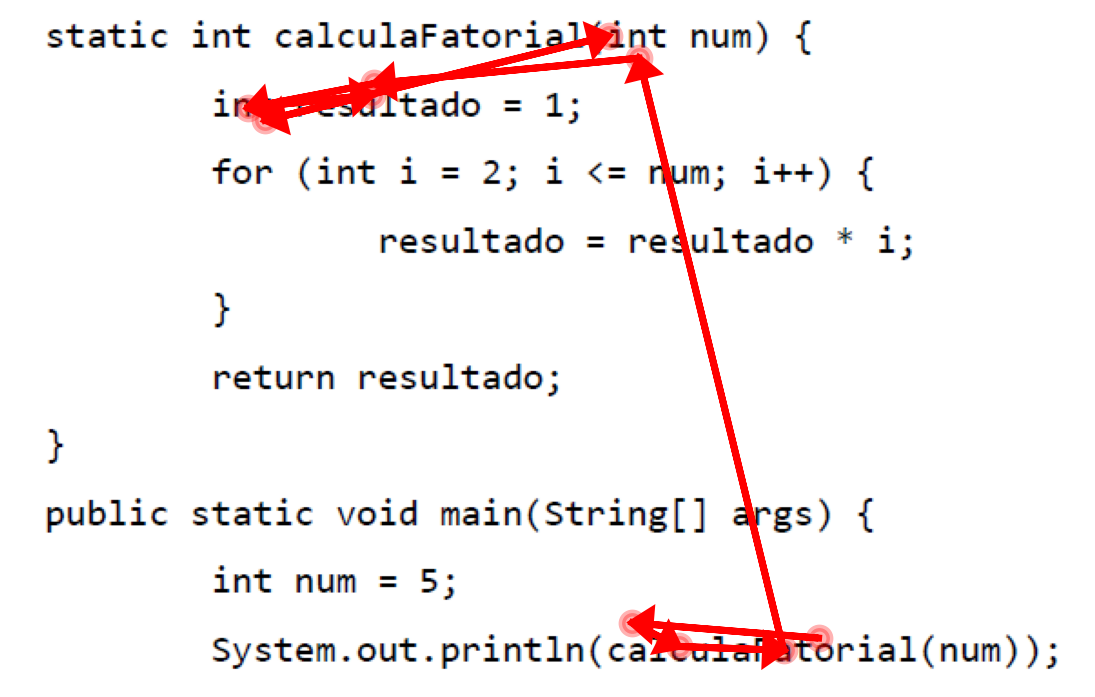}
	\end{minipage}}
 \hfill 	
  \subfloat[Transitions of Subject 8]{
	\begin{minipage}[c][0.6\width]{
	   0.3\textwidth}
	   \centering
	   \includegraphics[width=1\textwidth]{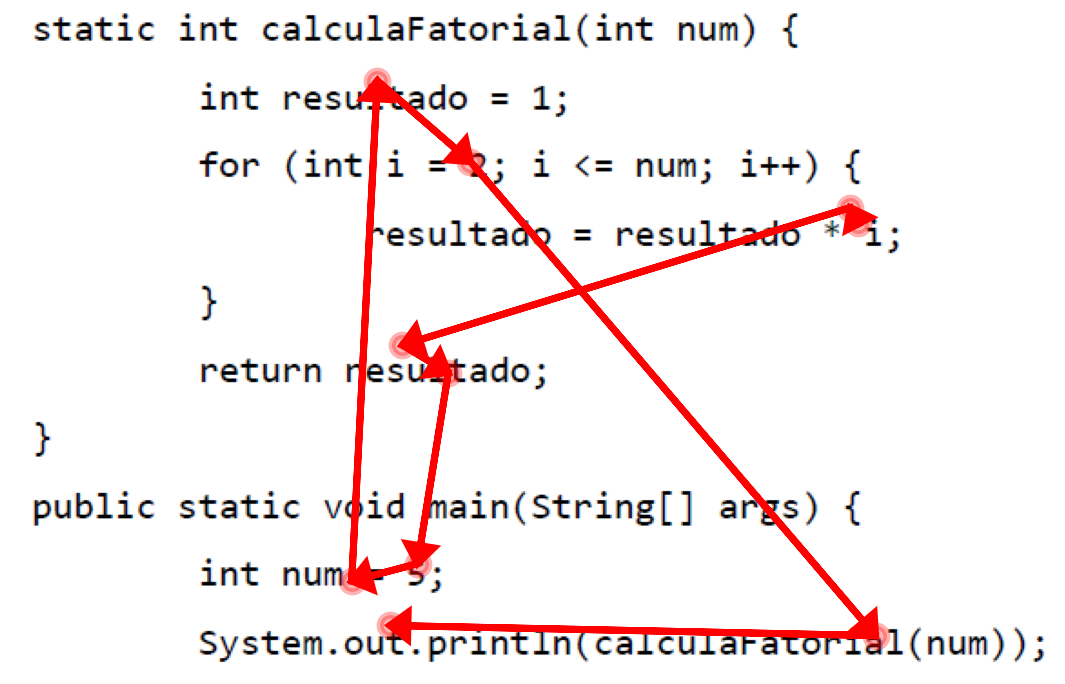}
	\end{minipage}}
 \hfill	
  \subfloat[Transitions of Subject 22]{
	\begin{minipage}[c][0.6\width]{
	   0.3\textwidth}
	   \centering
	   \includegraphics[width=1\textwidth]{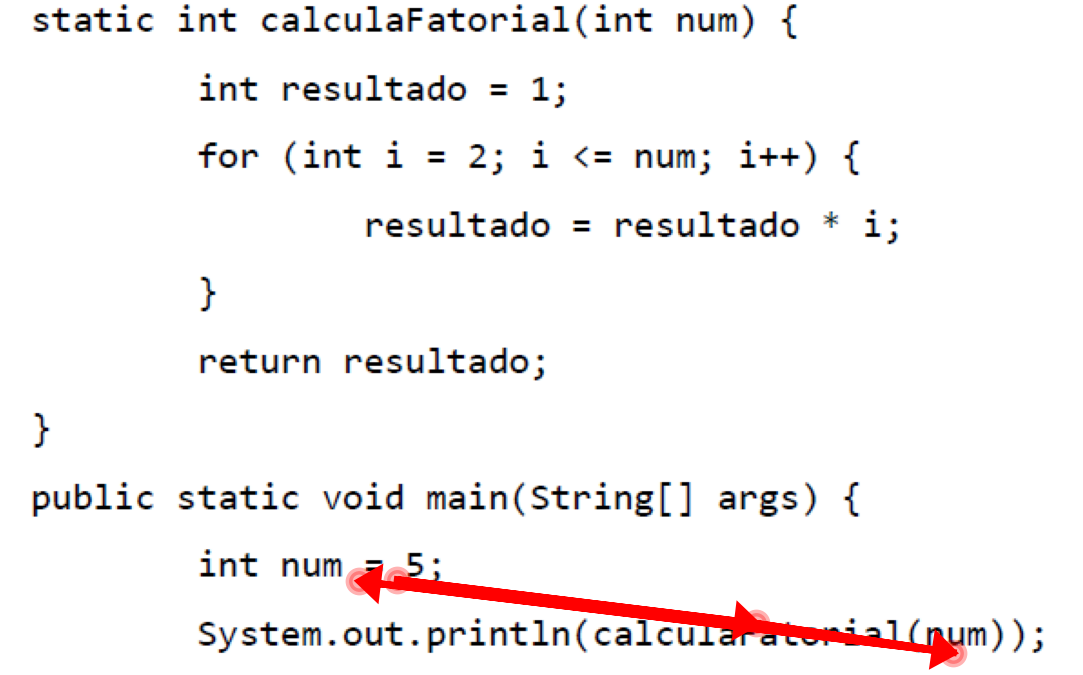}
	\end{minipage}}
\caption{Sequence of transitions of subjects on the extracted version for the task to calculate the factorial.}
\label{fig: gaze-calc-factorial-extracted}
\end{figure}

{In the extracted version, the visual behavior of novices suggests fewer checking and rechecking in the loop structure. Although they still moved between \texttt{ForLoop} and \texttt{Iteration}, these transitions occurred approximately half as often as in the inlined version, indicating that the refactoring eased the cognitive demand associated with tracking the loop's execution.} The name of the method plays a relevant role. The subjects fixated on the \texttt{NewMethod} region (containing the method name) about three times on average, as in Figure~\ref{fig: gaze-calc-factorial-extracted}(a). {A recurring pattern appeared around how novices connected the input variable to the extracted method, with transitions from \texttt{InputVarTwo} to the lines inside the body of the method},%, which reached 1.6 times, 
such as depicted in Figure~\ref{fig: gaze-calc-factorial-extracted}(b). In addition, %37\% of the 
subjects transited between \texttt{InputVarTwo} and \texttt{NewMethodCaller} passing the variable to the new method, as in Figure~\ref{fig: gaze-calc-factorial-extracted}(c), {indicating an effort to reconcile the flow of data across method boundaries}. In general, the transitions on the inlined method version seemed more concentrated on the loop and variables while in the extracted, we observed more scattered transitions reaching the top and bottom of the code. 

As a takeaway, the linearity of the inlined code led subjects to navigate back and forth between the loop and variables, with higher visual effort and a more concentrated gaze pattern. In contrast, the navigation in the extracted method seemed more scattered, with subjects engaging with both the top and bottom of the code. Novices both perceived the extracted version as easier to follow and performed better with it.

\subsection{Calculate the Area of the Square} \label{disc:area-square}

{Although novices reported a similar perceived ease when working with both code versions, their interaction behavior revealed a different pattern. In tasks involving simple logic, the extracted-method version demanded greater visual effort, as reflected in more extensive reading and navigation behavior. This mismatch between perceived difficulty and observed effort suggests that self-reports may not fully capture the cognitive demands imposed by structural transformations. In such cases, introducing an additional method can increase the need for context switching between the method call and the method body, even when the underlying computation is trivial. Analyzing gaze transitions helps explain how separating the logic into another method affects how novices move through the code, shedding light on the discrepancy between perceived and actual effort.}

\begin{figure}[h!]
    \centering
    \includegraphics[width=\textwidth]{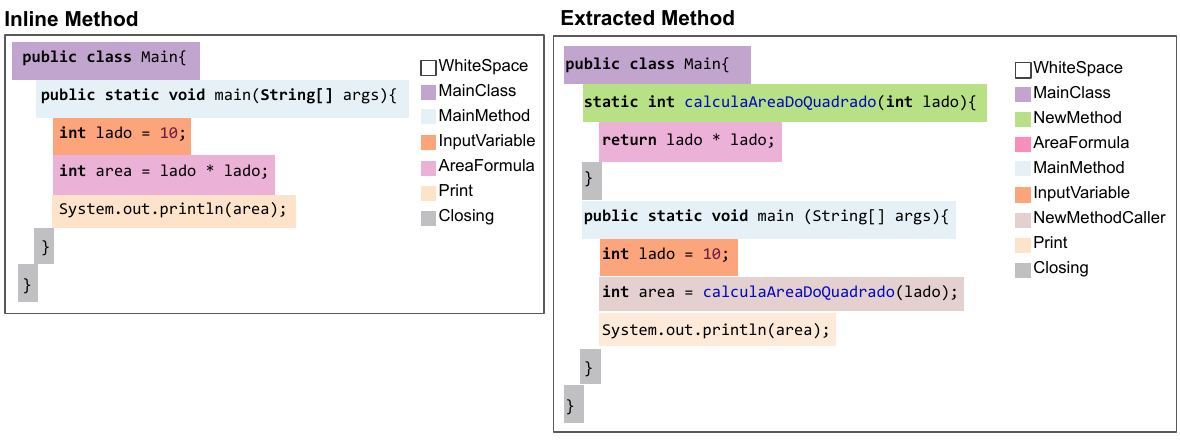}
    \caption{Set of regions inside the inlined and extracted versions for the task to calculate the area of the square.}
    \label{fig: area-square-regioes}
\end{figure}

\begin{figure}[ht]
  \subfloat[Transitions of Subject 4]{
	\begin{minipage}[c][0.3\width]{
	   0.3\textwidth}
	   \centering
	   \includegraphics[width=0.7\textwidth]{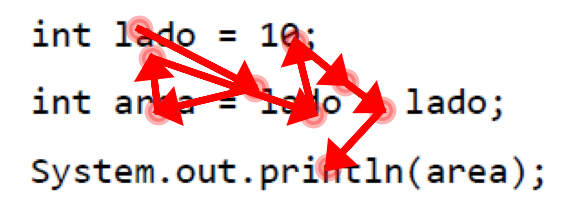}
	\end{minipage}}
 \hfill 	
  \subfloat[Transitions of Subject 6]{
	\begin{minipage}[c][0.3\width]{
	   0.3\textwidth}
	   \centering
	   \includegraphics[width=0.7\textwidth]{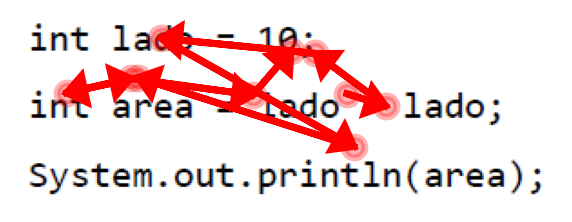}
	\end{minipage}}
 \hfill	
  \subfloat[Transitions of Subject 26]{
	\begin{minipage}[c][0.3\width]{
	   0.3\textwidth}
	   \centering
	   \includegraphics[width=0.7\textwidth]{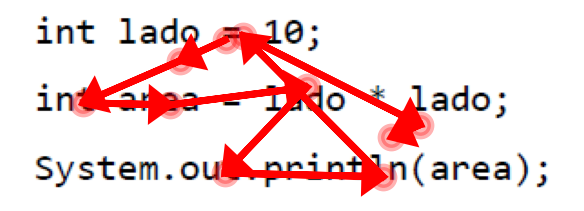}
	\end{minipage}}
\caption{Sequence of transitions of subjects on the inlined version for the task to calculate the area of the square.}
\label{fig: gaze-area-square-inlined}
\end{figure}

\begin{figure}[ht]
  \subfloat[Transitions of Subject 1]{
	\begin{minipage}[c][0.4\width]{
	   0.3\textwidth}
	   \centering
	   \includegraphics[width=1\textwidth]{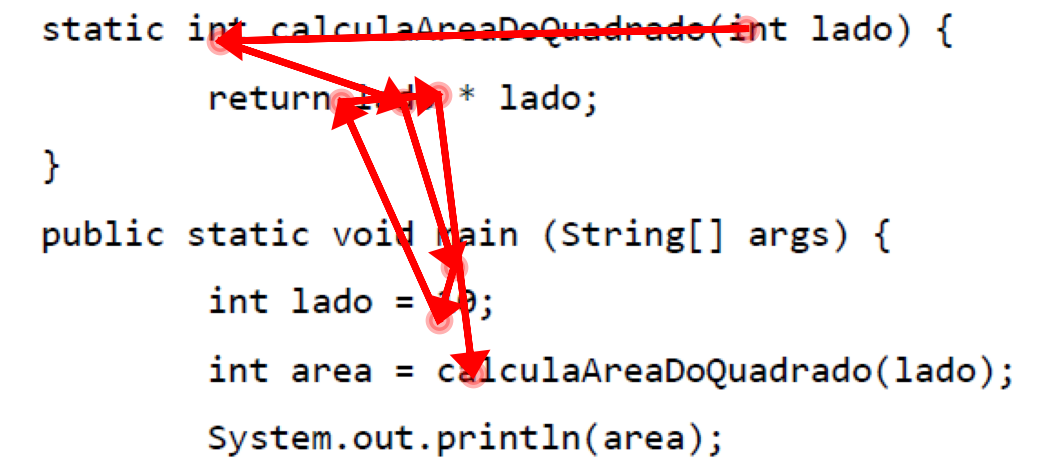}
	\end{minipage}}
 \hfill 	
  \subfloat[Transitions of Subject 11]{
	\begin{minipage}[c][0.4\width]{
	   0.3\textwidth}
	   \centering
	   \includegraphics[width=1\textwidth]{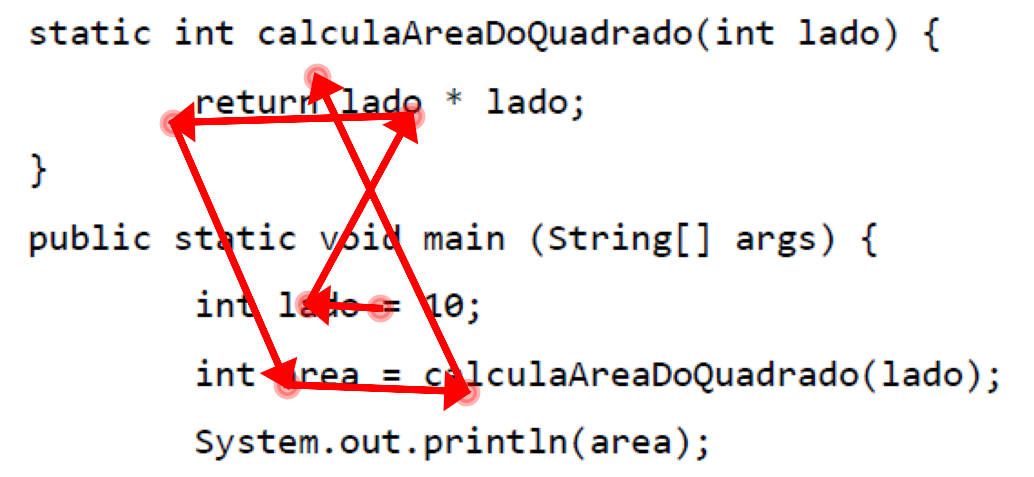}
	\end{minipage}}
 \hfill	
  \subfloat[Transitions of Subject 13]{
	\begin{minipage}[c][0.4\width]{
	   0.3\textwidth}
	   \centering
	   \includegraphics[width=1\textwidth]{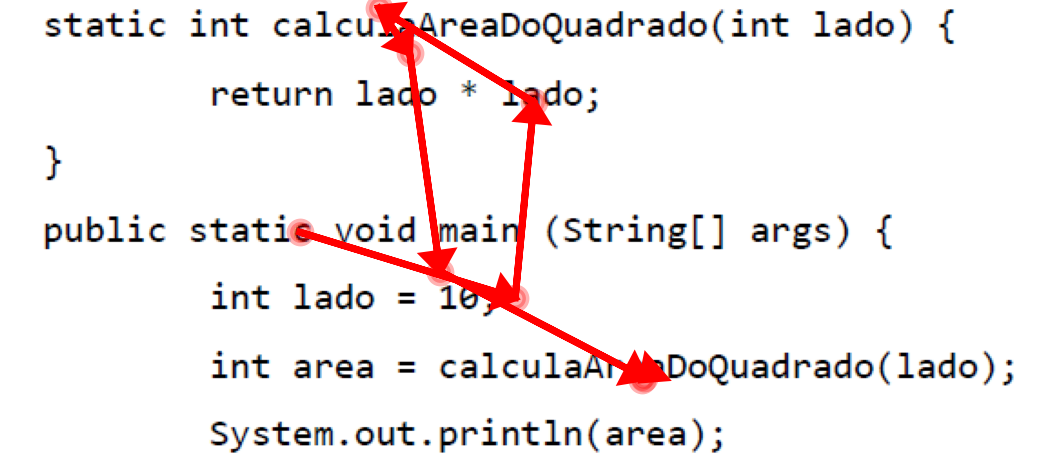}
	\end{minipage}}
\caption{Sequence of transitions of subjects on the extracted version for the task to calculate the area of the square.}
\label{fig: gaze-area-square-extracted}
\end{figure}

\begin{figure}[h!]
\centering
  \includegraphics[width=\textwidth]{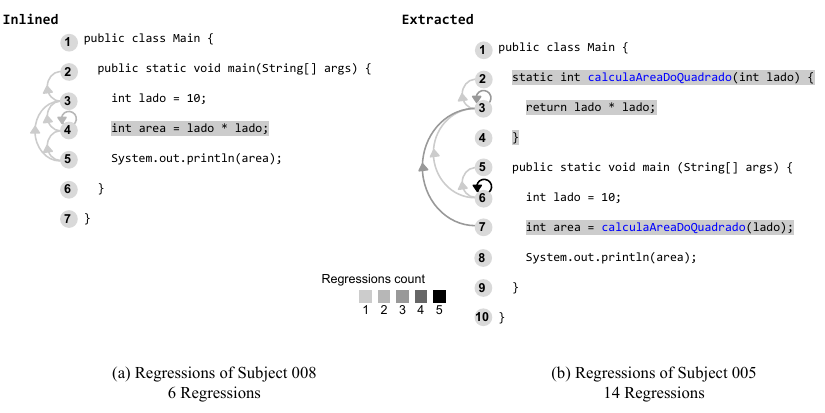}
  \caption{Eye movement regressions for the inlined and extracted versions for the task to calculate the area of the square.}
  \label{fig: 61N_vs_61R_005_008}
\end{figure}

The most visited regions in the inlined method were the \texttt{InputVariable} and \texttt{AreaFormula}, which contain the input value for the computation and the computation of the formula for the area. They fixated on the \texttt{InputVariable}, on average, 3.5 times and the \texttt{AreaFormula} 4.0 times. In addition, the average number of gaze transitions between these two regions going forward and backward was 1.1 and 1.5, respectively. As seen in Figure~\ref{fig: gaze-area-square-inlined}(a)--(c), most fixations concentrate on these areas, which makes sense given the relevance of the information on them to solve the task. This is also illustrated in Figure~\ref{fig: 61N_vs_61R_005_008}, where the inlined method exhibits regressions, especially in these regions, Lines 3 and 4. On the other hand, in the extracted method, the most fixated region was the \texttt{NewMethodCaller} with 4.5 times, on average, followed by \texttt{InputVariable}, with 4.2 times. The number of gaze transitions between these two regions going forward and backward was 1.0 and 0.8, respectively. The novices moved between \texttt{NewMethodCaller} and the body of the method as well, as depicted in Figure~\ref{fig: gaze-area-square-extracted}(a)--(c), and also in Figure~\ref{fig: 61N_vs_61R_005_008}, where the extracted method presents long regressions back to the method body.

While method extraction is generally believed to improve modularity and facilitate understanding, our results for this task suggest that this benefit is not guaranteed for novices. In this task, participants spent more visual effort on the extracted version compared to the inlined one, despite reporting similar perceived difficulty. This discrepancy between subjective perception and objective effort indicates that the extracted method introduced cognitive overhead, particularly through increased gaze transitions between the method call and its definition. These findings align with Crosby et al.~\cite{crosby2002theroles} conclusion that novices discriminate very little between areas of the program and thus do not seem to use beacons. Although the extracted method included a meaningful name, a semantic cue or beacon~\cite{siegmund2017measuring}, the novices in Java failed to use it effectively to guide comprehension. Instead, their eye movements revealed regressions suggesting inability to rely on the method name alone for semantic chunking. This underscores that for novices, beacons such as descriptive method names may not reduce cognitive load unless they are supported by sufficient experience to interpret them correctly.

\subsection{Calculate Next Prime} \label{disc:next-prime} 

{Although the number of attempts was similar for both versions, participants’ interaction patterns indicate that the extracted-method version reduced the cognitive and visual effort required to solve the task. This reduction was particularly evident in how novices engaged with the loop structure.} While one subject reported that the ``\textit{for loop takes effort}'', another mentioned trouble with attention because she ``\textit{confused the iterator in the \texttt{for} loop}'' and made three attempts until solving the problem. On the other hand, with the extracted method, a subject mentioned ``\textit{to solve the problem, I used the name and the input, otherwise, it would take longer}.'' Even adding more lines of code, the reduction in the time and visual effort seemed to be attributed to the clarity provided by the method's name, which helped participants better understand its behavior. 

\begin{figure}[h!]
\centering
  \includegraphics[width=0.9\textwidth]{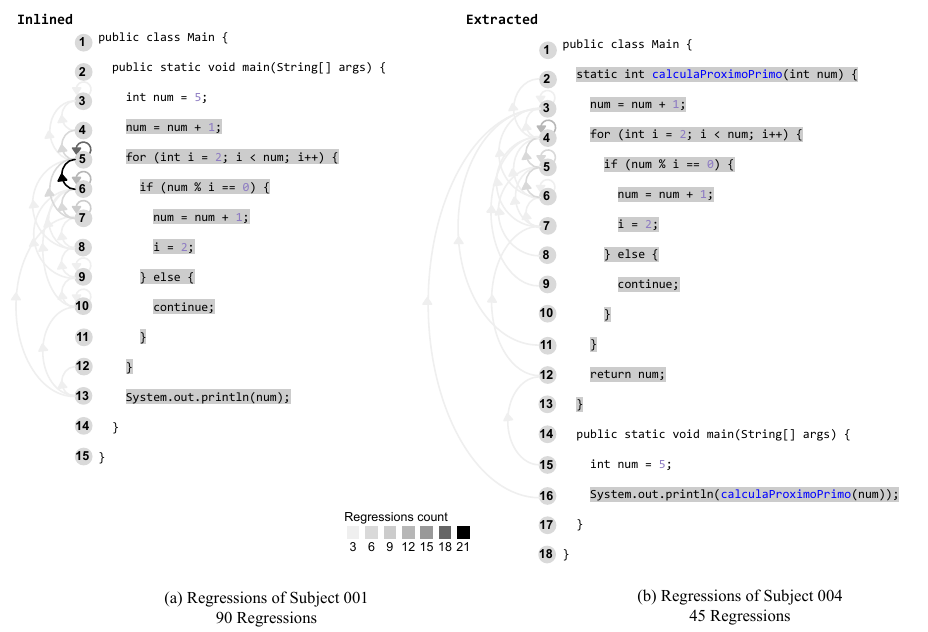}
  \caption{Visual regressions for the inlined and extracted versions for the task to calculate
the next prime of a number.}
  \label{fig: Regressions-Subjects-001-004}
\end{figure}

{This} aligns with Fowler~\cite{fowler2018refactoring}'s recommendation that a well-chosen method name can ease comprehension because it clarifies the intent of the code. This is illustrated in Figure~\ref{fig: Regressions-Subjects-001-004}, where in the Extracted version, with fewer regressions, subjects relied on reading the function name as their main strategy. In contrast, in the inlined version, the subject performs several regressions, particularly between lines 4 and 8, focusing on the \texttt{if} and \texttt{for} statements and their related variables. This difficulty may be associated with the high concentration of statements.

\subsection{Return Highest Grade from the List} \label{disc:highest-grade} 

\begin{figure}[h!]
\centering
  \includegraphics[width=0.9\textwidth]{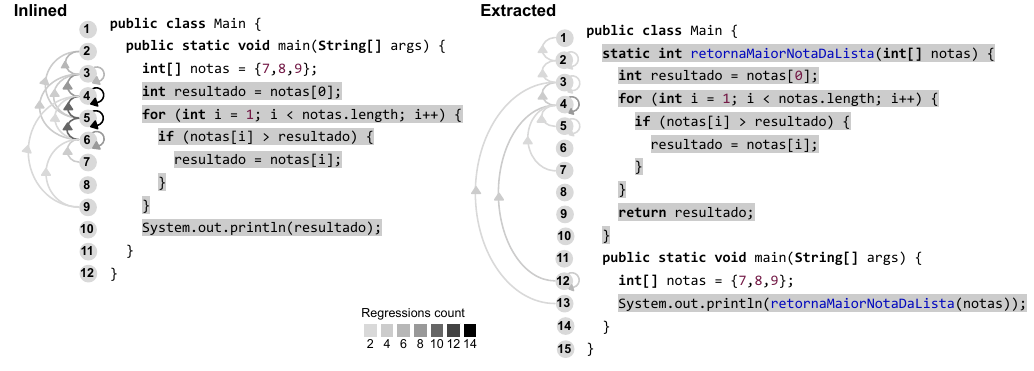}
  \caption{Visual regressions for the inlined and extracted versions to determine the highest grade from the list.}
  \label{fig: compare-regs}
\end{figure}

The extracted method has three more lines of code. We observed that this change led participants to revisit earlier lines more often, especially in the lines where the variables were assigned values, and in the loop followed by the decision control, as seen in Figure~\ref{fig: compare-regs}.
Four subjects mentioned that they needed to pay more attention to the loop, especially in ``\textit{finding the iterator variable in the loop}'' and in the condition inside the loop. One of them gave up on solving and the other needed two attempts. In the extracted version, we observed regressions between the call of the method and the method. However, the subject \proof{made} fewer regressions while examining the lines inside the method when we compare them with the same lines on the inlined method. Five subjects mentioned that the name was helpful to them.

As a takeaway, the method extraction reduced the visual effort especially in areas involving the loops and conditionals. This reduction seemed to be impacted by the combination of both structure changing and clear name, with novices making fewer regressions. The inlined version led to more back-and-forth navigation particularly around the loop and decision control.

\subsection{Count Multiples of Three} \label{disc:mult-three} 

\begin{figure}[ht]
  \subfloat[Transitions of Subject 6]{
	\begin{minipage}[c][0.3\width]{
	   0.3\textwidth}
	   \centering
	   \includegraphics[width=0.95\textwidth]{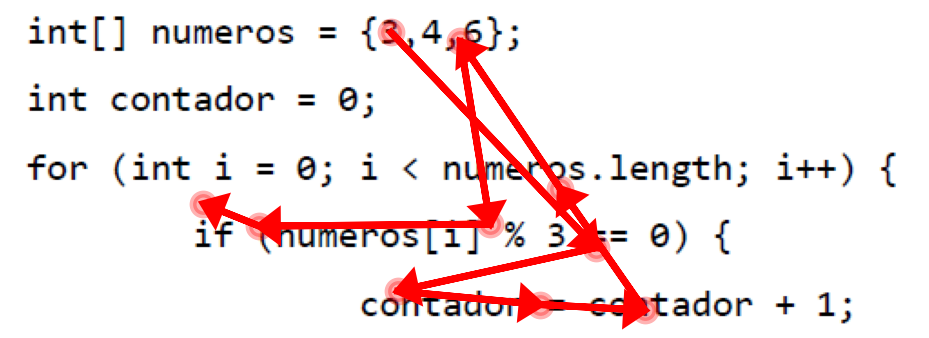}
	\end{minipage}}
 \hfill 	
  \subfloat[Transitions of Subject 10]{
	\begin{minipage}[c][0.3\width]{
	   0.3\textwidth}
	   \centering
	   \includegraphics[width=0.95\textwidth]{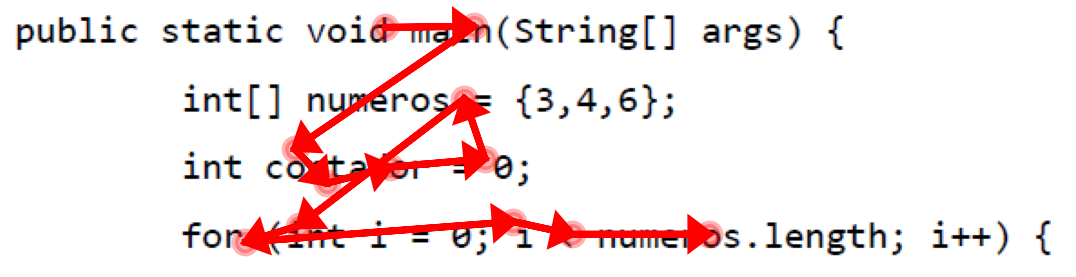}
	\end{minipage}}
 \hfill	
  \subfloat[Transitions of Subject 22]{
	\begin{minipage}[c][0.3\width]{
	   0.3\textwidth}
	   \centering
	   \includegraphics[width=0.95\textwidth]{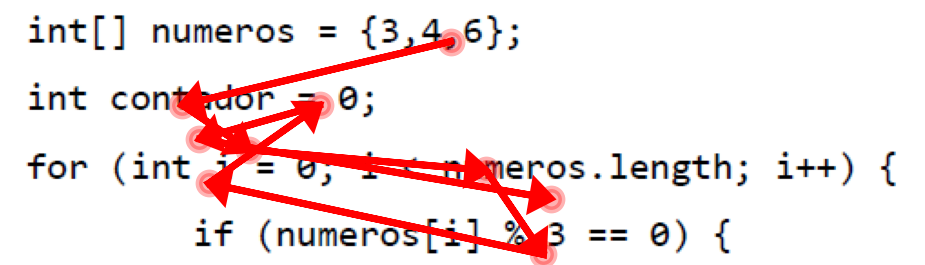}
	\end{minipage}}
\caption{Sequence of transitions of subjects on the inlined method for the task to count the multiples of three from a list.}
\label{fig: gaze-mult-three-inlined}
\end{figure}

In the inlined method, two subjects mentioned ``\textit{difficulties with examining line by line because there is no function}.'' An example of such a linear reading can be seen in Figure~\ref{fig: gaze-mult-three-inlined}(b). We also noticed that they tend to go back and forth between the iteration and the list, as in Figure~\ref{fig: gaze-mult-three-inlined}(a) and (c) to check whether the number iterated is \proof{a multiple} of three. In the extracted version, 12 subjects mentioned that the name helped them to infer what the answer was and 9 of them mentioned checking the function to see if it was doing what they inferred by the name. One subject mentioned:``\textit{The name gave me a hint but I checked to see if the method's body corresponded to the name}.'' We observed a reduction in the number of attempts with the extracted version by 20\%. 

\subsection{Check if Even} \label{disc:check-even} 

\begin{figure}[ht]
  \subfloat[Transitions of Subject 1]{
	\begin{minipage}[c][0.5\width]{
	   0.3\textwidth}
	   \centering
	   \includegraphics[width=0.95\textwidth]{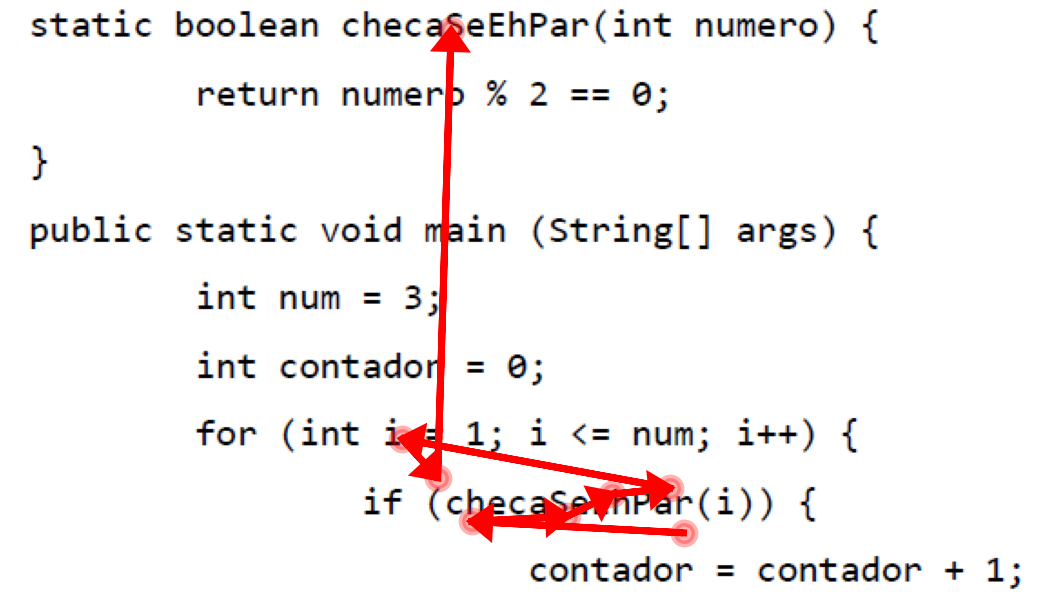}
	\end{minipage}}
 \hfill 	
  \subfloat[Transitions of Subject 7]{
	\begin{minipage}[c][0.5\width]{
	   0.3\textwidth}
	   \centering
	   \includegraphics[width=0.95\textwidth]{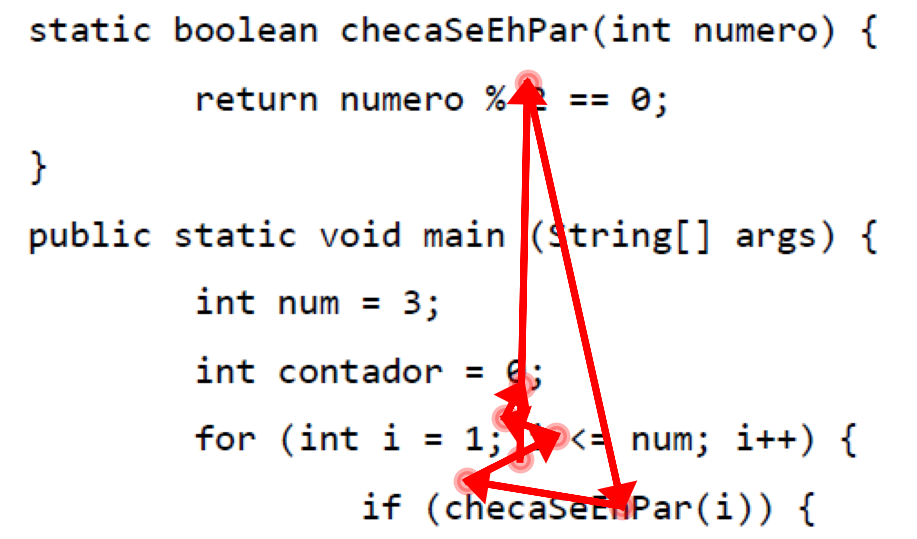}
	\end{minipage}}
 \hfill	
  \subfloat[Transitions of Subject 11]{
	\begin{minipage}[c][0.5\width]{
	   0.3\textwidth}
	   \centering
	   \includegraphics[width=0.95\textwidth]{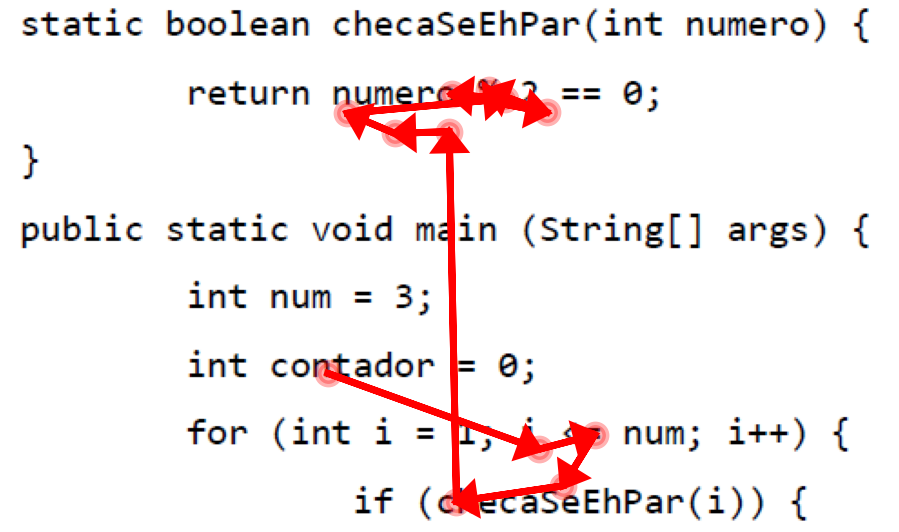}
	\end{minipage}}
\caption{Sequence of transitions of subjects on the extracted version for the task to check if a number is even.}
\label{fig: gaze-check-even-inlined}
\end{figure}

\begin{figure}[h!]
\centering
  \includegraphics[width=0.8\textwidth]{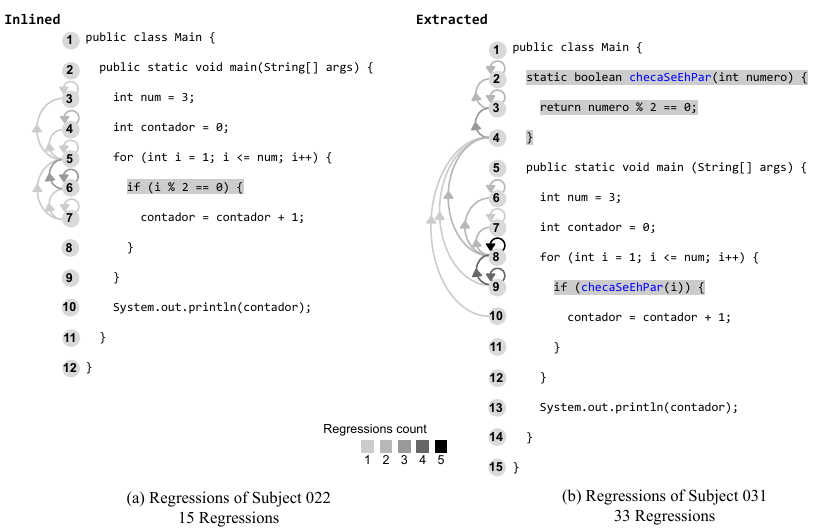}
  \caption{Eye movement regressions for the inlined and extracted versions for the task to check if a number is even.}
  \label{fig: 71N_vs_71R_031_022}
\end{figure}

{The extracted-method version prompted more frequent revisits to earlier code regions in the AOI, suggesting increased navigation effort.}
A subject mentioned that she ``\textit{started with the function from top to bottom, then went back to the function. The back-and-forth movement would be faster and simpler if it was in one line. It would also save space}.'' Examples of subjects going back in the code to the method can be seen in Figures~\ref{fig: gaze-check-even-inlined}(a)--(c). These reading patterns are observed for other subjects as well, which can indicate an impact on the reading flow, hindering it. This is also illustrated in Figure~\ref{fig: 71N_vs_71R_031_022}, where the extracted method exhibits long regressions back to the method body, whereas the inlined method keeps instructions closer and shows shorter regressions. Indeed, one participant reported needing to pay more attention in the extracted version; overall, participants required 17.6\% more attempts. In addition, even though five subjects mentioned that the name was helpful, they often checked whether the function \proof{did} what it \proof{said}, which \proof{impacted} the time. 

As a takeaway, for simpler tasks such as checking if a number is even, the extra modularization of the extracted method may slow down the problem-solving process due to the increased navigation effort. While clear names aim to help understanding of the code, it may pose extra effort in the visual navigation process. Thus, there might be a balance between modularity and ease of reading. While names can aid understanding, they can also hinder the performance of novices.

{Complementing this observation, we noticed an additional pattern related to the nature of the tasks themselves. Interestingly, similar to the check if even task, the tasks that showed the smallest benefit from the Extract Method refactoring were also the tasks whose logic consisted of a single line of code (e.g., sum numbers from one to num in Section~\ref{disc:sum-numbers} and calculating the area of a square in Section~\ref{disc:area-square}). These tasks offer no internal structure that could be clarified or abstracted by extracting a method; their logic is already compact enough to be fully apprehended inline. In such cases, introducing an auxiliary method may actually increase the cognitive load by requiring an additional navigation step. This suggests that the benefits of modularization depend on a minimum threshold of complexity: while multi-line tasks (e.g., factorial, finding the highest grade) benefit from extraction, trivial computations do not. This nuance helps refine our understanding of when Extract Method improves novice comprehension and when inline implementations are more appropriate.
}

\subsection{Count Number of Digits} \label{disc:num-digits} 
\begin{figure}[ht]
  \subfloat[Transitions of Subject 6]{
	\begin{minipage}[c][0.5\width]{
	   0.3\textwidth}
	   \centering
	   \includegraphics[width=0.95\textwidth]{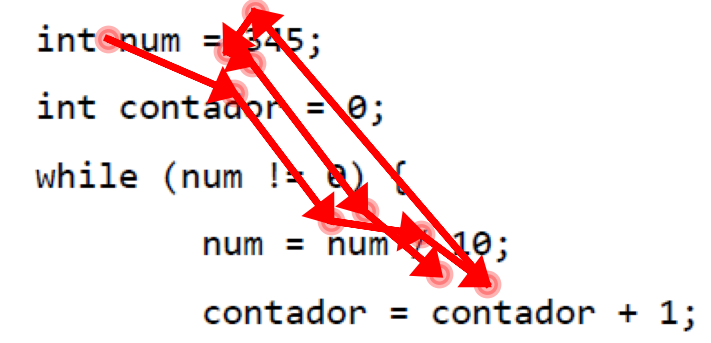}
	\end{minipage}}
 \hfill 	
  \subfloat[Transitions of Subject 8]{
	\begin{minipage}[c][0.5\width]{
	   0.3\textwidth}
	   \centering
	   \includegraphics[width=0.95\textwidth]{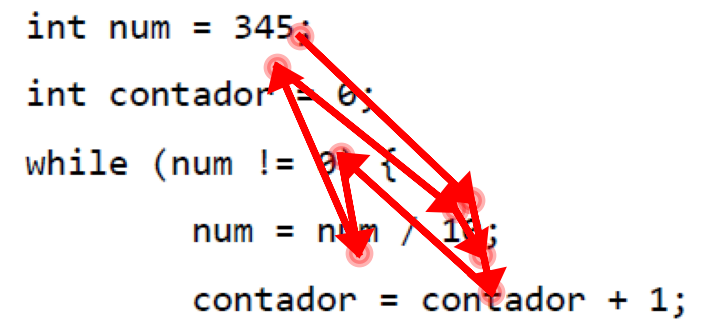}
	\end{minipage}}
 \hfill	
  \subfloat[Transitions of Subject 16]{
	\begin{minipage}[c][0.5\width]{
	   0.3\textwidth}
	   \centering
	   \includegraphics[width=0.95\textwidth]{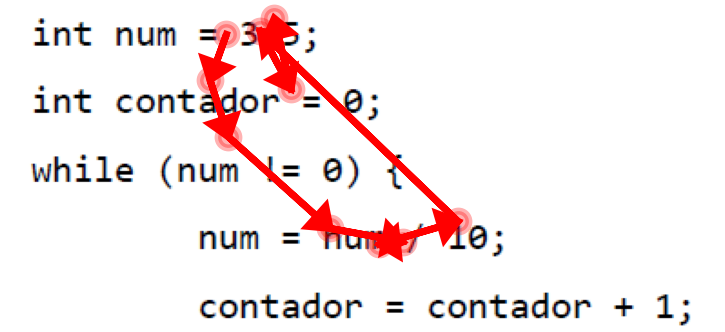}
	\end{minipage}}
\caption{Sequence of transitions of subjects on the inlined version for the task to count the number of digits.}
\label{fig: number-digits-inlined}
\end{figure}

In the inlined version, we observed that the subjects \proof{went} back and forth between the loop and the variables defined before the loop, which were updated within the loop, as depicted in Figure~\ref{fig: number-digits-inlined}(a)--(c). To solve the task, the loop iterates three times until the \texttt{while} condition is satisfied. However, six subjects mentioned difficulties with variable type, which \proof{confused them} about when the condition could be satisfied. One subject mentioned that the inlined version was ``\textit{not suggestive without a name or a function. I did not see that it was an integer number. I had difficulties with the mathematical division.}'' Another one mentioned that she ``\textit{thought this was going to be an infinite loop}.'' 

\begin{figure}[h!]
\centering
  \includegraphics[width=0.9\textwidth]{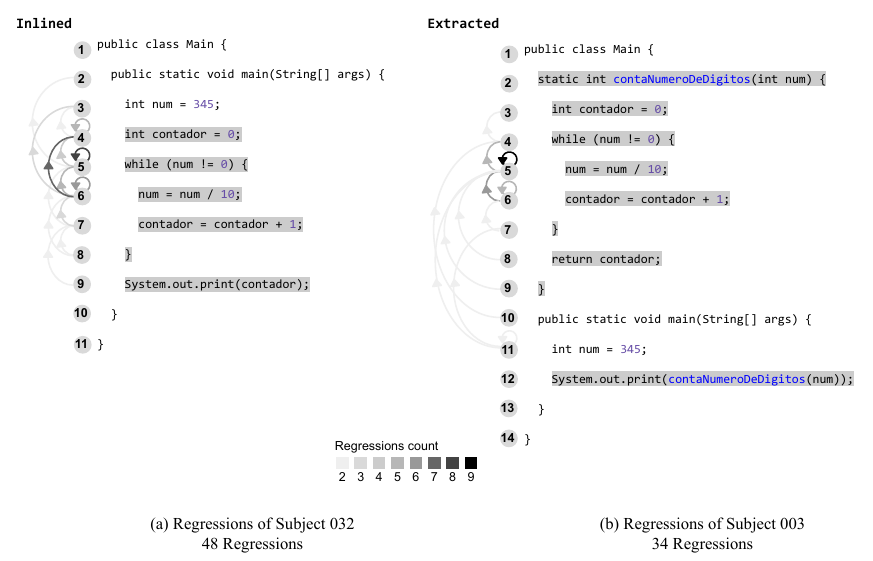}
  \caption{Eye movement regressions for the inlined and extracted versions for the task to count the number of digits.}
  \label{fig: 003 vs 032 EMSE}
\end{figure}

In Figure~\ref{fig: 003 vs 032 EMSE}(a), the inlined version shows a dense pattern of regressions, particularly in the loop region. In contrast, the extracted version displays more evenly distributed regressions. This suggests that moving the digit-counting instructions to a separate method allows the subject to focus more on what the code does rather than on the details of how each step is executed.

In this task, the method name could help confirm hypotheses about the loop's behavior. The novices could more easily infer the logic behind the loop and understand when it should stop. In the inlined method, novices were left to navigate through the code repeatedly to understand the variables and logic, which could explain the slowdown in task completion time to solve the task and increased visual effort.

\subsection{Coding Subjects' Answers}\label{disc:coding}

Overall, method names seemed to facilitate a top-down approach to code reading, while the absence of names led to a bottom-up approach, requiring deduction. This aligns with Brooks'~\cite{brooks1977towards} top-down model of problem-solving in programming, which emphasizes that developers often start by forming high-level hypotheses based on the overall structure of the program and refine their understanding through detailed analysis. When method names are present, programmers can anticipate the general purpose of the code before delving into the specifics.

The behavior of checking whether the method does what it says indicates a persistent level of distrust toward unfamiliar code, even when method names are clear. Thus, meaningful naming can ease comprehension, but the implementation must also be carefully aligned with the names to foster trust and reduce the need for extensive verification.

Behavioral theories have already been applied in software engineering research~\cite{shneiderman1979syntactic,brooks1978using}. The bottom-up approach is consistent with Shneiderman and Mayer's~\cite{shneiderman1979syntactic} theory of bottom-up processing, which suggests that understanding is constructed incrementally, starting with the details and building up to a higher-level understanding. In the absence of descriptive method names, programmers rely on identifiers and must assemble meaning incrementally through detailed, line-by-line analysis.

The subjects reported difficulties in the category of Mathematics, namely math operations and precedence order. Among 24 tasks, 16 used the inlined version. Since no hypotheses could be inferred from a name, the subjects had to formulate it from a careful examination, which seemed to be hampered by the operations. The most common difficulties were \textit{``confusion in the math operation''}, particularly in the task to count the number of digits, in both extracted and inlined versions. This might explain the lack of statistical differences in the time and visual effort.

\subsection{{Integrative Theory of Method Extraction Comprehension}}
\label{sec:integrative-theory}

\begin{figure}[h!]
\centering
  \includegraphics[width=1\textwidth]{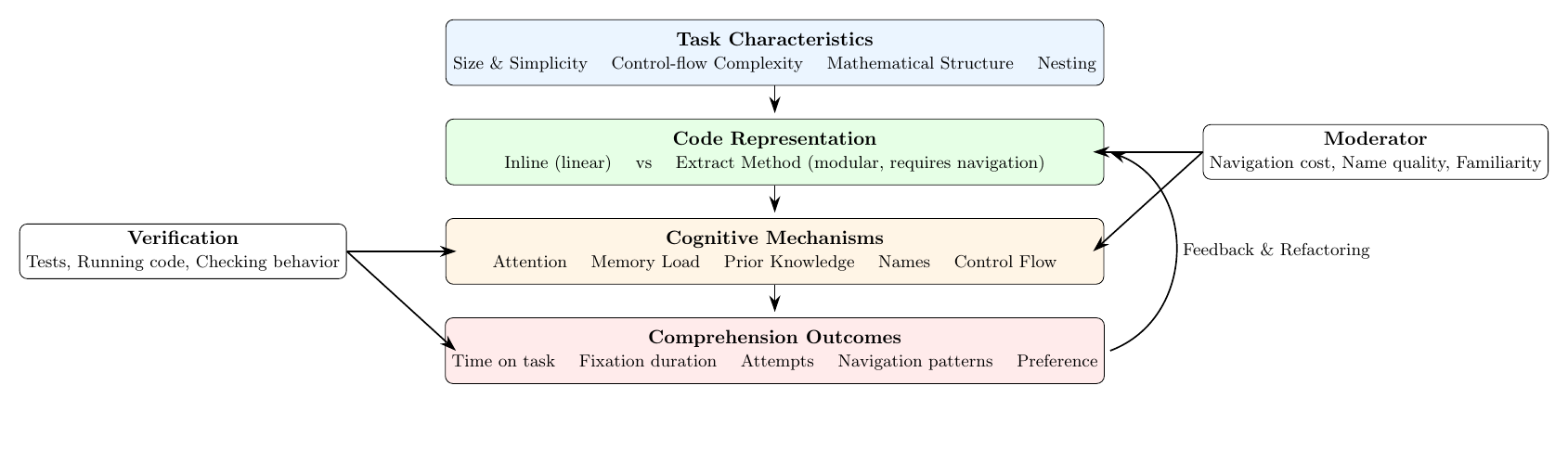}
  \caption{{Integrative visual model structured into four interacting conceptual layers: Task Characteristics, Code Representation, Cognitive Mechanisms, and Comprehension Outcomes.}}
  \label{fig:integrative-model}
\end{figure}

{Based on the categories, concepts, and relationships identified through our coding analysis, we propose an integrative theory explaining how programmers comprehend code when comparing inline and extract method representations. Figure~\ref{fig:integrative-model} visually summarizes this theory, which is structured into four interacting conceptual layers: \textit{Task Characteristics}, \textit{Code Representation}, \textit{Cognitive Mechanisms}, and \textit{Comprehension Outcomes}, moderated by factors such as navigation cost, name quality, and snippet familiarity.}

{\textit{Task Characteristics.} This foundational layer captures the intrinsic properties of each snippet that define its baseline cognitive difficulty. Elements such as control-flow complexity, mathematical structure, size and simplicity, and the presence of nested or iterative logic determine the initial cognitive demands placed on the programmer. These characteristics set the stage upon which all subsequent influences operate, as depicted in the top layer of Figure~\ref{fig:integrative-model}.}

{\textit{Code Representation.} The second layer reflects how the snippet is structurally presented to the reader. Inline versions provide a linear flow of information with minimal navigation, while extract method versions encapsulate logic into named units that enable top-down comprehension. This layer functions as a structural modulator: depending on the complexity of the underlying task, modularization may reduce cognitive load by organizing logic into meaningful chunks, or it may introduce navigation overhead that outweighs potential benefits. The diagram illustrates this modulation through the arrows connecting this layer to both Task Characteristics and Cognitive Mechanisms.}

{\textit{Cognitive Mechanisms.} This intermediate layer represents the cognitive processes activated during comprehension, including attention allocation, working-memory load, prior knowledge, control-flow reasoning, name meaningfulness, and mathematical interpretation. These mechanisms translate the structural properties of the task and its representation into actual comprehension behavior. As Figure~\ref{fig:integrative-model} shows, moderators such as navigation cost, name quality, and snippet familiarity influence this layer directly. Importantly, our theory posits that Extract Method refactoring reduces cognitive load primarily when task complexity is sufficiently high to benefit from abstraction; for simple tasks, this advantage diminishes or reverses due to navigation costs.}

{\textit{Comprehension Outcomes.} The final layer captures the observable results of the comprehension process: time on task, number of attempts, fixation duration, navigation patterns, and subjective preference. These outcomes emerge from the interplay between task characteristics, code representation, and cognitive mechanisms rather than from any single factor alone. Extract method versions generally lead to reduced time, fewer attempts, and lower visual effort for complex tasks, consistent with lowered cognitive load. For simple tasks, however, increased navigation introduces additional effort, which can neutralize or even hinder performance.}

{We distinguish between two types of influencing factors: moderators and verifiers. Moderators are variables that shape the strength or direction of the relationships between layers; for example, navigation cost and name quality do not act independently but modify how code representation affects cognitive mechanisms. Verifiers, in contrast, are elements that do not alter the relationships themselves but help confirm or falsify the expected behavioral patterns predicted by the theory. Measures such as fixation duration, number of visits, or validation steps performed by participants serve as verifiers by providing empirical evidence that aligns with or challenges the theorized cognitive processes.}

{In general, the integrative theory highlights that comprehension does not depend solely on whether Extract Method refactoring is applied, but rather on how well the representation aligns with the underlying complexity of the task and the cognitive aspects provided to the programmer. Extract \proof{Method} is most beneficial when it meaningfully reduces cognitive burden relative to task demands, and may be neutral or detrimental when the task is already simple enough to be processed inline with minimal effort.}

\section{Threats to Validity}
\label{sec:threats}

In this section, we describe the internal validity (Section~\ref{internal-validity}), external validity (Section~\ref{external-validity}), and construct validity (Section~\ref{construct-validity}).

\subsection{Internal validity} \label{internal-validity}

We understand that the use of descriptive method names in the extract method version may influence the participants' comprehension process by facilitating top-down comprehension. In our study, the extracted method version provided an additional clue through the method name, or a ``beacon,''~\cite{siegmund2017measuring}. This beacon may lead participants to recognize the functionality of the method more quickly, relying on top-down comprehension processes, where the overall purpose is known and only the details need to be clarified. In contrast, the inline version may require more bottom-up comprehension, as participants need to analyze the code line by line to infer its purpose. Therefore, the differences observed in our experiment could also reflect a combination of top-down and bottom-up comprehension processes triggered by the presence or absence of descriptive method names.

However, it is important to highlight that the Extract Method refactoring inherently has an entangled effect, as it performs two tasks simultaneously: 1) it restructures the code by creating a new method, and 2) it provides a clear name (a beacon) to the newly created method. While we acknowledge that simple coding tasks performed by students may not necessarily involve assigning highly descriptive names to code fragments, the use of clear method names is a fundamental part of the Extract Method refactoring mechanics. This practice is widely recommended by authors such as Fowler~\cite{fowler2018refactoring} and is considered a best practice in refactoring.

%ambiente
We conducted the controlled experiment in two different locations to increase diversity, which may have an influence on the visual attention of the subjects. To mitigate it, we carefully arranged the rooms to have similar conditions of light, temperature, and quietness.
%author presence
The presence of a researcher in the room may have unintentionally influenced the visual attention or performance of the subjects. To mitigate it, we let the subjects feel comfortable and avoided any interaction while they were examining the programs.

%o problema da camera
The eye tracker equipment has limitations. Despite careful calibration and recalibration, we still needed to adjust gaze points. 
% exemplo do problema 
For specific subjects, the heatmap and plot of fixations showed fixations concentrated over a blank area not touching the code.  
% o que fizemos para corrigir o problema
For these specific subjects, a small adjustment was sufficient to correct the gaze points. All the fixations for a particular program received the same adjustment.
% quanto foi o ajuste
The adjustment in the \textit{y}-coordinate was a median \proof{of} 12.5 pixels. We did not adjust the \textit{x}-coordinate.
% metodo
The adjustment in the points may influence their interpretation. The authors discussed these adjustments for each subject. However, we decided that the threat of adjusting the points would be preferable to a threat of analyzing the data with points not touching the code. The fixations and adjustment strategy are available in our replication package~\cite{our-artifacts}. 

{A potential threat to the internal validity of our study concerns the data analysis process of the eye-tracking metrics. As highlighted by Dörzapf et al.~\cite{dorzapf2024data}, the choice of analysis tools and scripts can significantly influence the outcomes of eye-tracking studies. To mitigate this threat and enhance transparency and replicability, we have made the analysis script available in our replication package~\cite{our-artifacts}.}

% cadeira
During pilot studies, a swivel chair impaired gaze data collection. To mitigate it, we used chairs without swiveling capability in the controlled experiment. 
% tempo
We allocated one hour for each subject and assigned them 10 programs, which may have influenced the visual effort. To minimize it, we designed simple and short programs.

% quadrado latino
With the Latin Square design, we blocked the set of programs to control for noise. We analyzed the programs combined in the squares and individually. The analysis of individual programs may violate the design. However, analyzing them combined and individually can give a more nuanced and complete understanding of the effects of the refactorings.

% Externas
\subsection{External validity} \label{external-validity}

% tamanho do programa
We used small programs to ensure the code fit on the screen, which may restrict generalization to larger programs. In particular, our static setup does not capture IDE-based navigation behaviors (e.g., scrolling, jump-to-definition), which may influence how developers interact with extracted methods. Additionally, the extraction sizes evaluated in this study cover only a limited range, which may restrict the applicability of our findings to larger extractions. However, if we observe differences in small code snippets, we expect greater differences in larger ones. Nevertheless, we need studies with larger code snippets. 

% novatos
We focused on novices in Java, which may restrict generalization to more experienced developers in Java. Other eye tracking studies have also focused on novices to understand code comprehension~\cite{busjahn2015eye}. In the future, we intend to focus on experienced developers. The participants' prior exposure to different programming paradigms may have influenced the way the code was understood.

% linguagem de programação
We focused on Java, which may restrict generalization to other programming languages. To mitigate it, we used constructions commonly employed in other languages. Most of our subjects reported some experience with other languages. Our programs were designed to contain identifiers in Portuguese given that the subjects were Brazilians. 

% tarefas
We assigned the subjects a task to specify the correct output of the program, which was a numeric value. They answered the output aloud after reading the code with no syntax highlighting. This task may not generalize to other tasks, such as finding a bug or adding a feature. 

% codigos
The number of methods in the program may influence the visual effort of the subjects. To minimize this threat, we consistently used only one method extracted in the program except for the main method. The methods' names may influence the comprehension and visual effort of the subjects. Confusing names can hamper comprehension and require more effort. To minimize this threat, we used lessons from previous studies and guidelines, refined the names through a pilot study, and discussed the names among the authors.

% Construct
\subsection{Construct validity} \label{construct-validity}

% tempo e acuracia
Code comprehension has been often measured through time and answer correctness~\cite{Oliveira2020evaluating}.
% esforço visual
These metrics have been combined with the visual effort to investigate code comprehension~\cite{sharif2012aneye,oliveira2020atoms,costa2021evaluating,melo2017vari}.
% tempo e numero de fixacoe, e regressoes
In particular, visual effort has been measured before with fixation duration and fixation count~\cite{sharif2012aneye,binkley2013impact}. In addition, eye movement regressions have been associated with visual effort~\cite{sharafi2015eye}.

We used an imputation method for 3.1\% of the data. 
To assess its impact, we compared key metrics, such as fixation counts and regressions, between imputed and non-imputed datasets. The differences were minimal, suggesting that the imputation introduced no significant bias and did not affect the study's conclusions.

% olhos rastreados
When inviting participants, we had to inform them that their eye movements would be tracked. This may influence where or how much they look at some regions of the code. To minimize this threat, we did not inform them about the precise goals of the study to avoid hypothesis guessing. Variation in individual learning styles and cognitive strategies may have influenced how the code was understood.

\section{Related Work}  
\label{sec:related-work}
% overview
In this section, we discuss related work on code refactoring (Section~\ref{code-refactoring}), code comprehension (Section~\ref{code-comprehension}), and eye tracking in code activities (Section~\ref{eye-tracking-in-code activities}).

\subsection{Code Refactoring}
\label{code-refactoring}

% o que fizeram
Aiming to investigate why developers refactor their code, 
Silva et al.~\cite{silva2016why}  
% como eles fizeram
monitored Java projects to detect applied refactorings and asked the developers about their motivations.
% o que acharam
They found that the Extract Method was the most common refactoring, mainly used for code reuse, introducing alternative method signatures, and improving readability.
% relacionamento com o nosso
In our study, we conducted a controlled experiment to assess the impact of the Extract Method and Inline Method refactorings on the subjects' code comprehension and perceptions. We found that novices predominantly preferred the extracted version motivated by readability, clearer organization, and meaningful names. However, they favored inlined versions in simpler tasks motivated by simplicity, ease of writing, and elimination of unnecessary methods, which is consistent with the quantitative results.

% o que fizeram
Cedrim et al.~\cite{cedrim2017understanding} studied the impact of refactorings, such as the Extract Method and Inline Method, on code smells through static code metrics.
% como eles fizeram
They conducted a longitudinal study observing how refactorings impacted 13 types of code smells along the version histories of 23 projects.
% o que acharam
They found that 57\% of the refactorings, including the Inline Method, were neutral in the sense that they did not reduce the occurrences of smells, and 33.3\%, including the Extract Method, were negative, meaning that they induced the introduction of new smells instead.  
% relacionamento com o nosso
We studied the impact of the Extract Method and Inline Method refactorings from a dynamic perspective, focusing on the human visual effort of novices through a controlled experiment. Specifically, our study examined tasks involving small code snippets to understand how these refactorings influence novices' code comprehension in a practical and behavioral context. 

% o que fizeram e como
Hora and Robbes~\cite{hora2020characteristics} analyzed 70K instances of the Extract Method in 124 Java systems to characterize its use in terms of magnitude, content, transformation, size, and degree. They found that (i) the Extract Method is the third most frequent refactoring; (ii) the Extract Method refactorings concentrate on operations related to creation, validation, and setup; (iii) methods that are targets of the extractions are 2.2x longer than the average ones, and they are reduced by one statement after the extraction; and (iv) single method extraction represents most of the cases.
% relacionamento com o nosso
In contrast, we examined the impact of the Extract Method and Inline Method refactorings on code comprehension of novices, focusing on small, simple tasks. Our findings indicate that, although the Extract Method may offer better organization and readability in general, the Inline version tends to be more effective in simpler and more straightforward scenarios.

Golubev et al.~\cite{golubev-fse-2021} conducted a large-scale survey with 1,183 users of IntelliJ-based IDEs aiming to investigate how developers refactor code, which refactorings are most popular, and why some developers avoid IDE refactoring tools. They found that nearly two-thirds spend more than an hour per refactoring session, refactoring types vary in popularity, and many developers are interested in IDE refactoring features, but lack guidance on how to use them. Our study reinforces the importance of code refactoring for novices, who are not always familiar with refactoring techniques and the tools that support them. Therefore, we investigated the impact of specific refactorings, namely Extract and Inline, on novices' code comprehension.

\subsection{Code comprehension}
\label{code-comprehension}

{Source code comprehension is a fundamental cognitive activity in software engineering, supporting tasks such as program understanding, maintenance, and evolution. Despite its central role, the concept itself has historically been defined and operationalized in diverse and often inconsistent ways.}
{To address this conceptual fragmentation, Wyrich~\cite{Wyrich2023source} proposes a contemporary definition of source code comprehension together with a conceptual model aimed at supporting empirical investigation. The model distinguishes between the code and contextual artifacts involved, the cognitive processes performed by developers, and the observable outcomes of comprehension. By making these dimensions explicit, this work provides a theoretical foundation for reasoning about what is being measured when comprehension is studied empirically.}

{Building on this conceptual perspective, Wyrich et al.~\cite{Wyrich2024forty} present a systematic mapping study covering 40 years of code comprehension experiments. Their analysis reveals a wide variety of experimental tasks, measurements, and contexts, as well as substantial variation in how comprehension is implicitly defined. The study highlights recurring challenges such as limited comparability across experiments and the lack of shared assumptions about what constitutes successful comprehension.}

{These challenges are examined in more depth by Wyrich et al.~\cite{Wyrich2024apples}, who argue that source code comprehension should be treated as a latent variable. From this perspective, differences in operationalization and measurement can lead to problematic study selection and aggregation in secondary research. The authors show how insufficient conceptual alignment can result in ``apples-to-oranges'' comparisons, even when studies are nominally labeled as addressing code comprehension.}

{Related human-centric constructs are also discussed in the work of Oliveira et al.~\cite{Oliveira2020evaluating}, who investigate how code readability and legibility are evaluated in empirical studies. Although not focused exclusively on comprehension, their analysis demonstrates how closely related constructs are often defined and measured inconsistently, further complicating the interpretation and comparison of findings across the literature. This reinforces the need for clear conceptual boundaries and explicit definitions when studying source code comprehension.}

{Concrete examples of how code comprehension has been operationalized at the cognitive level can be found in primary studies such as Siegmund et al.~\cite{siegmund2017measuring}.} They investigated the role of semantic cues in code comprehension using functional Magnetic Resonance Imaging (fMRI). Their study involved 11 participants including computer science students, mathematics students, and professional programmers. They designed controlled experimental conditions to isolate cognitive processes related to bottom-up comprehension and comprehension based on semantic cues, or beacons. Their findings showed that semantic cues facilitate code comprehension by activating brain regions associated with semantic processing.
In our study, we conducted a controlled experiment with 32 Java novices using eye tracking to measure task completion time, number of attempts, and visual effort. By integrating these quantitative metrics with participant feedback, we found cognitive patterns that show how novices approached and understood the code. Our results indicate that, in tasks such as calculating the area of a square, modularization introduced by method extraction and beacons actually hindered the code comprehension of novices due to increased visual navigation effort, disrupting their linear code reading.

% o que fizeram
Sharafi et al.~\cite{sharif2010aneye} studied the impact of camel case and underscore on code comprehension by 
% como eles fizeram
measuring time, accuracy, and visual effort using eye tracking.
% o que acharam
They found significantly less time and lower visual effort with the underscore style. 
% o que fizeram
Sharafi et al.~\cite{sharafi2012women} explored whether gender influenced these effects. However, 
% como eles fizeram
%however, measuring the impact of the subjects' gender on the time, accuracy, and visual effort. 
% o que acharam
no differences were observed.
% relacionamento com o nosso
In our study, we used similar metrics---time, number of attempts, and visual effort---however in a different context. We investigated the behavior of novices when working with the extracted and inlined versions, aiming to understand how these refactorings affect their ability to comprehend and navigate code through simple tasks.

Hofmeister et al.~\cite{hofmeister2019shorter} 
studied the impact of identifier naming styles, namely, letters, abbreviations, and full words, on program comprehension using a within-subjects design with 72 professional C\# developers. 
They found that full-word identifiers were associated with 19\% faster defect detection compared to letters or abbreviations, with no significant difference between the latter two. Our study used time as one of the metrics to assess code comprehension, but in the context of refactoring styles. We also incorporated additional measures such as number of attempts and eye tracking data, namely, fixation count, duration, and regressions, to gain a deeper understanding of how novices engage with code during task completion. 

Yeh et al.~\cite{yeh2022identifying} conducted an experiment using an electroencephalogram (EEG) device to measure developers' cognitive load while they reviewed C code snippets. They found that although cognitive load can influence code comprehension performance, other human factors also play an important role, especially for novice developers, such as the tendency to forget certain programming rules or to misinterpret code instructions. In our study, we collected behavioral data, such as time spent on tasks and the number of attempts, along with participants' subjective perceptions, including preferences and comments about the strategies and difficulties they encountered while examining the code. 

Hermans and Aivaloglou~\cite{hermans2016code} 
investigated the impact of code smells on novice Scratch programmers in a controlled experiment with 61 participants. They were divided into three groups: one received a clean version of a program, while the others received versions exhibiting either the Duplication or Long Method smell. The subjects that worked with code containing Duplication or Long Method smells performed worse on comprehension tasks, although task completion time remained unaffected. Long Method impaired understanding, while Duplication made modification harder. The findings relate to ours by showing how code quality influences novices' comprehension. Our findings suggest that the Inline version can be more effective than the Extract version in simpler tasks for novices, as it may reduce cognitive overhead.

Martins et al.~\cite{martins2024eyes} explored the use of eye-tracking to analyze how the presence of code smells affects code comprehension of Java developers. The authors focused on three specific types of code smells: long method, feature envy, and data class. They observed that certain smells, such as long method and feature envy, impose a higher cognitive effort on developers, while others, such as data class, result in lower cognitive demands. The study also presents complementary eye-tracking indicators that reveal additional aspects of how code smells hinder comprehension. Although their focus is on code smells more broadly, the insights highlight the importance of investigating the cognitive impact of specific refactoring techniques. Building on this, our work aims to address this gap by comparing inlined and extracted method versions using eye-tracking data to assess their effects on code comprehension of Java novices.

\subsection{Eye tracking in code activities}
\label{eye-tracking-in-code activities}

Costa et al.~\cite{da2023seeing} conducted controlled experiments using eye tracking to assess how small confusing code patterns, known as atoms of confusion, affect code comprehension. Using eye tracking, they evaluated the visual effort of 32 Python novices. Oliveira et al.~\cite{oliveira2020atoms} conducted a similar study focusing on the visual attention of the 30 subjects solving problems in the C language. In a controlled experiment, we evaluate how code changes impact the comprehension of 32 novices in Java. We used similar metrics, including time, number of answer attempts, and measures of visual effort. In addition, we combined these metrics with participant feedback on perceived difficulty, strategies used, and preferences. The triangulation of these data helped us identify patterns in novices' code comprehension.

Crosby et al.~\cite{crosby2002theroles} presented the concept of beacons as features in code that act as cognitive cues to facilitate program comprehension. They investigated how programmers with different experience levels use beacons to comprehend code. They found that expert programmers tend to rely on beacons to focus their attention on semantically meaningful parts of the code, whereas novices discriminate little between areas of the code and do not seem to use beacons. Our study builds on this perspective by using eye tracking to show how Extract Method, despite introducing potentially meaningful identifiers, can disrupt novice comprehension and increase visual effort when compared to inlined versions of the same code.

Abid et al.~\cite{abid2019developer} conducted an eye tracking study with 18 developers who read and summarized 63 Java methods using the Eclipse IDE. The data collected included eye gazes, written summaries, and the time to complete each summary. The results indicate that developers tend to focus more on the body of a method than on its signature. These findings align with ours, as we found that in simpler tasks, the modularization introduced by an extracted method can slow down problem-solving for novices due to the increased navigation effort it requires. Our study employs eye tracking equipment to investigate Java code and involves 32 novices. By focusing on short methods presented outside of an IDE, this controlled setup helps us minimize potential confounding factors related to tool familiarity.

Jessup et al.~\cite{jessup2021using} compared code comprehension and perceptions between expert and novice programmers using eye tracking data. 
They found that experts exhibited more \proof{fixations;} however, no evidence was found for differences in average fixation duration, perceptions of reliability and performance, or willingness to reuse the code. Similarly, Kather et al.~\cite{kather2021through} conducted an eye-tracking study examining how code composition and familiarity affect comprehension at a more abstract level. 
Using quantitative data alongside retrospective interviews, they analyzed students' reading patterns. 
They showed that familiar code structures help students form mental schemas that aid understanding. In our study, we focused on Java novices. By combining eye tracking, interviews, and subjective feedback, we analyzed reading behavior and perceptions to better understand how novices interpret and respond to different code structures.

Alakmeh et al.~\cite{alakmeh2024predicting} developed a deep neural network that aligns a developer's gaze with the code tokens they observe to predict code comprehension and perceived difficulty. To train and evaluate their model, they conducted an experiment using an eye tracker with 27 participants and 16 short code comprehension tasks. Similarly, we conducted our study with 32 participants and 16 short code comprehension tasks. However, instead of focusing on training a predictive model, we employed established metrics to evaluate participants' comprehension and visual effort. Our study design enabled a controlled comparison of comprehension performance across different code variants, specifically contrasting the extracted and inlined versions.

% o que fizeram
Melo et al.~\cite{melo2017variability} studied the impact of compile-time variability implemented through preprocessor directives (e.g., \#ifdef) on the debugging process.
% como eles fizeram
They conducted an experiment in which developers were asked to debug programs with and without variability, while their eye movements were recorded using an eye tracker.
% o que acharam
They found that debugging time increases for code fragments containing variability, and also for nearby fragments without variability. Variability was associated with more frequent gaze transitions between definitions and usages, and a longer initial scanning phase.
% relacionamento com o nosso
While their study investigated how variability affects the cognitive process of debugging using eye tracking, we focused on the Extract Method and Inline Method with eye tracking but examined the visual effort and accuracy in tasks to be solved rather than debugging. 

% o que fizeram
Turner et al.~\cite{turner2014eye} compared how the programming languages C++ and Python affect students' comprehension of source code, using two task types: overview and bug-finding. 
% como eles fizeram
Thirty-eight students read short C++ and Python code to complete overview and bug finding tasks while their eye movements were tracked. 
% o que acharam
Although no significant differences were found in accuracy or time, students showed a significantly different gaze behavior, particularly in how often they looked at buggy lines. These findings suggest that programming language may influence attention patterns in early programming education. We, on the other hand, focused on exploring tasks only in the Java language. The 32 Java novices performed short tasks and reported the corresponding code output. Afterwards, they answered questions about their approach and reasoning while completing the tasks.

% o que fizeram
Busjahn et al.~\cite{busjahn2015eye} explored how programmers read code by introducing gaze-based measures of linearity drawing inspiration from natural language reading. 
% como eles fizeram
By comparing novices and experts reading both natural language text and Java code, the authors found that novices read code less linearly than text, and experts even less linearly than novices, indicating that programming expertise is associated with increasingly non-linear reading. These findings highlight fundamental differences between reading code and natural language. Our work also investigated code reading patterns, especially in terms of visual regressions in the same line or to previous lines of code in the context of Java novices.

\section{Conclusions} 
\label{sec:conclusion}

We conducted a controlled experiment using eye tracking to evaluate the impact of the Extract Method and Inline Method refactorings on code comprehension. We compared these techniques by measuring their impact on task completion time, number of attempts, visual effort, and by collecting qualitative feedback from 32 novice Java programmers. We also surveyed \proof{an} additional 58 Java novices to investigate their preferences and motivations. Our study adopted a more dynamic perspective by tracking participants' eye movements as they interacted with the code, allowing us to examine how {Extract Method refactoring} influences comprehension. Through eye tracking, we analyzed key code areas to compare visual effort, reading behaviors, fixation durations, and regression patterns.

The Extract Method refactoring reduced the time, number of attempts, and visual effort for two tasks, aligning with the preferences of the novices in the survey. Most novices preferred the Extract Method version, motivated by improved code understanding, reuse, and extension. For instance, to compute the factorial, in the inlined version, the subjects made more visual regressions inside the \texttt{for} loop, with more concentrated transitions, while in the Extract Method version, the modularization allowed the novices to make less concentrated transitions with fixations on the method's name and caller. This may indicate that, for more complicated scenarios where modularization is possible, the Extract Method version is a better technique. 

On the other hand, we found a negative impact of the Extract Method refactoring leading to longer completion times and more visual regressions for another three tasks. Even though most of the novices perceived method extraction as beneficial for overall understanding, we observed that it might depend on the applied scenario. For instance, in a task that calculates the \textit{Area of a Square}, the code version with the method extracted exhibited a 166.9\% higher median time spent in the AOI and 138.8\% more median visual regressions, both calculated from participants who performed the task, compared to the version with the method inlined. When the calculation was moved to a separate location in the code, the subjects tended to keep coming back to those lines interrupting the reading flow. Indeed, in the survey, novices found the method unnecessary in addition to the code being small and simpler and already contained meaningful names. This indicates that, for simpler scenarios, the inlined version consists of a better technique. 

Our findings suggest that meaningful method names, or beacons, do not always improve novices' code comprehension in Java in the context of refactorings. Even when meaningful names were present, participants still navigated back and forth between the method call and its body, reflecting increased visual effort and extended cognitive processing time. Previous studies with fMRI have found that semantic cues, such as beacons, ease bottom-up comprehension by enabling semantic segmentation in Java~\cite{siegmund2017measuring}. However, they did not evaluate the refactorings or employ eye tracking. Our eye-tracking findings in the context of refactorings indicate that meaningful method names are not always effective in improving code understanding among novices in Java. Our results are consistent with Crosby et al.~\cite{crosby2002theroles}. They observed that novices tend not to use beacons effectively in the context of tasks such as binary search written in an algorithmic language.

Regarding the practical implications, we observed that even small changes in code organization can impact novices' code comprehension and visual regression behavior. While extracted methods offer modularity, they may hinder performance in simpler tasks due to increased navigation effort. This effect could be even more pronounced in larger programs with greater distance between calls and declarations. Thus, our findings suggest that novices should consider the specific context of each task and weigh the trade-offs between code understanding and visual effort.
Our study contributes to raising awareness among educators about the potential of these techniques to ease or hinder code comprehension for novices in Java. Introductory courses should be more selective in choosing programs that do not impact the visual effort negatively. For researchers, our results show the potential of visual metrics to reveal effects of refactorings that cannot be captured by static code metrics. Other approaches such as fMRI-based experiments could possibly reveal other nuances~\cite{peitek2021program}. 

As future work, the impact revealed through eye tracking may indicate the need for tools to assist novices when applying the Extract Method or the Inline Method. Such tools could, for example, provide context-sensitive recommendations, highlight potential comprehension costs, or visualize the navigation overhead introduced by certain transformations. Moreover, we aim to evaluate larger code artifacts using IDE-based eye tracking toolchains, such as iTrace~\cite{itrace2018}, and systematically vary the number of lines extracted or inlined to assess how refactoring granularity affects gaze behavior. We also plan to evaluate additional refactorings from Fowler's catalog~\cite{fowler2018refactoring} focusing on scenarios where these transformations may offer greater cognitive or structural benefits. Furthermore, we aim to conduct more controlled experiments involving experienced developers to examine whether their familiarity with abstraction mitigates the comprehension costs observed in novices. We also plan to explore other programming languages to evaluate whether their specific characteristics influence the effectiveness of the Extract Method and Inline Method refactorings. In addition, we can design diverse task types that go beyond simply producing the correct program output, incorporating activities such as bug fixing, feature extension, and code review, which may involve different comprehension strategies. {We also aim to investigate larger and more realistic codebases}, where navigation distances could impact visual effort. Finally, integrating eye tracking with complementary techniques, such as think-aloud protocols, biometric sensing, and sentiment analysis, may provide a more nuanced understanding of how different refactorings affect developers' code comprehension.

\section*{Conflict of Interest Statement}
The authors declared that they have no conflict of interest.

\section*{Declaration of generative AI and AI-assisted technologies in the writing process}

During the preparation of this work the authors used ChatGPT in order to improve language, readability, and revising grammar, spelling, and punctuation. After using this tool, the authors reviewed and edited the content as needed and take full responsibility for the content of the publication.

\section*{Acknowledgements}
We thank the anonymous reviewers for their valuable feedback and suggestions, which helped improve the quality of this article.
This work was partially supported by CNPq (403719/2024-0, 310313/2022-8, 408040/2025-4, 404825/2023-0, 443393/2023-0, 312195/2021-4) and FAPESQ-PB (268/2025). It was also partially supported by INES.IA (National Institute of Science and Technology for Software Engineering Based on and for Artificial Intelligence) and by CNPq grant 408817/2024-0.

\appendix
\newpage
\begin{figure*}[ht]
  \subfloat[Data distribution of metrics for task to sum numbers.]{
	\begin{minipage}[c][0.4\width]{
	   1\textwidth}
	   \centering
	   \includegraphics[width=1\textwidth]{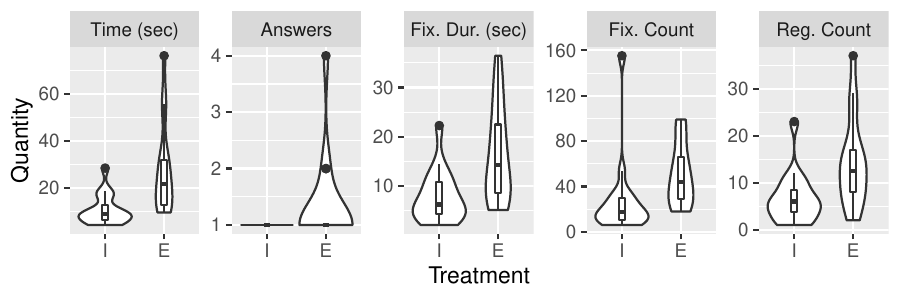}
	\end{minipage}}
 \hfill 	
  \subfloat[Data distribution of metrics for task to calculate next prime.]{
	\begin{minipage}[c][0.4\width]{
	   1\textwidth}
	   \centering
	   \includegraphics[width=1\textwidth]{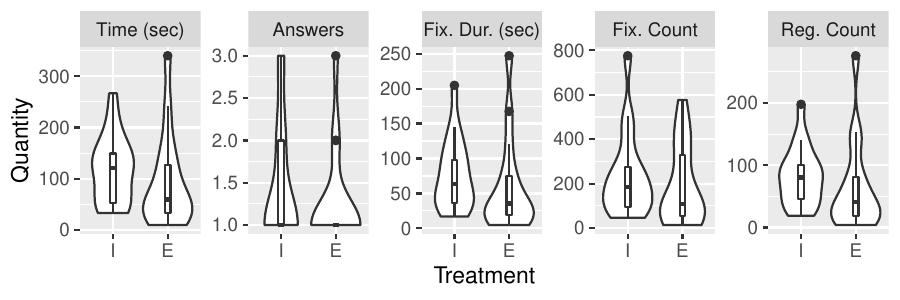}
	\end{minipage}}
 \hfill	
  \subfloat[Data distribution of metrics for task to return the highest grade.]{
	\begin{minipage}[c][0.4\width]{
	   1\textwidth}
	   \centering
	   \includegraphics[width=1\textwidth]{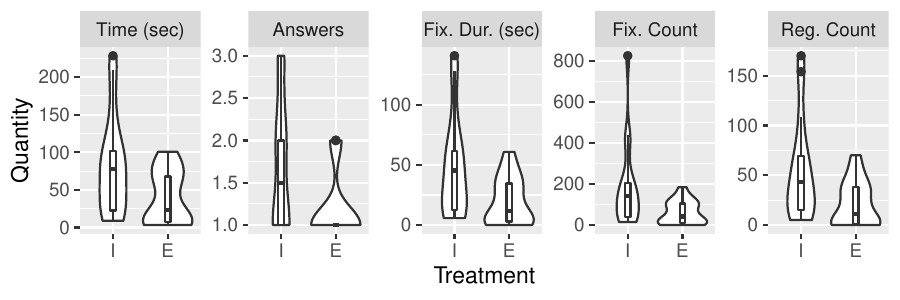}
	\end{minipage}}

 \phantomcaption
\label{fig: distribution}
\end{figure*}

\begin{figure*}[ht]
\ContinuedFloat
  \subfloat[Data distribution of metrics for task to calculate the factorial.]{
	\begin{minipage}[c][0.4\width]{
    %[0.4\width]{
	   1\textwidth}
	   \centering
	   \includegraphics[width=1\textwidth]{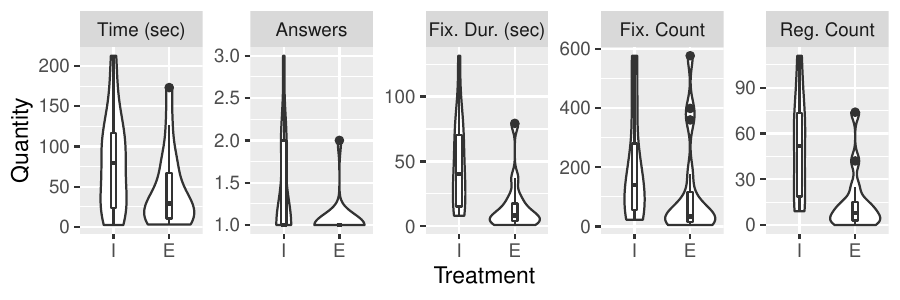}
	\end{minipage}}
\hfill
  \subfloat[Data distribution of metrics for task to count multiples of three.]{
	\begin{minipage}[c][0.4\width]{
	   1\textwidth}
	   \centering
	   \includegraphics[width=1\textwidth]{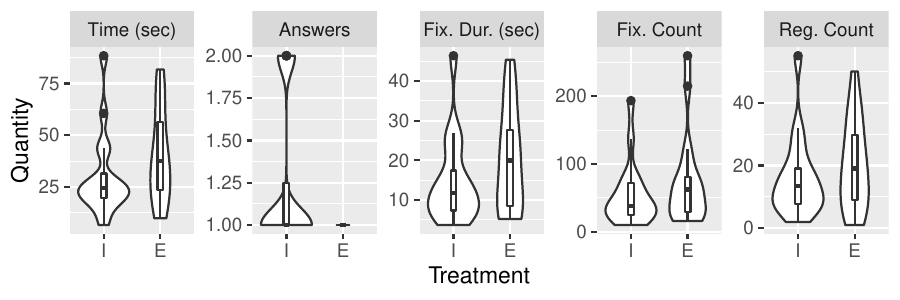}
	\end{minipage}}
 \hfill 	
  \subfloat[Data distribution of metrics for task to calculate the area of the square.]{
	\begin{minipage}[c][0.4\width]{
	   1\textwidth}
	   \centering
	   \includegraphics[width=1\textwidth]{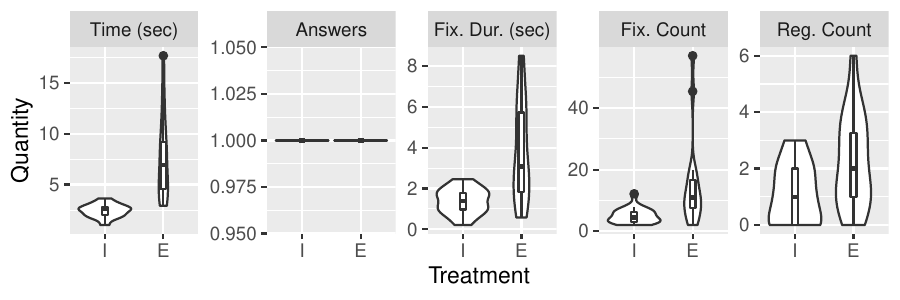}
	\end{minipage}}
  
\phantomcaption
\label{fig: distribution1}
\end{figure*}

\begin{figure*}[ht]
\ContinuedFloat
  \subfloat[Data distribution of metrics for task to check if a number is even.]{
	\begin{minipage}[c][0.4\width]{
	   1\textwidth}
	   \centering
	   \includegraphics[width=1\textwidth]{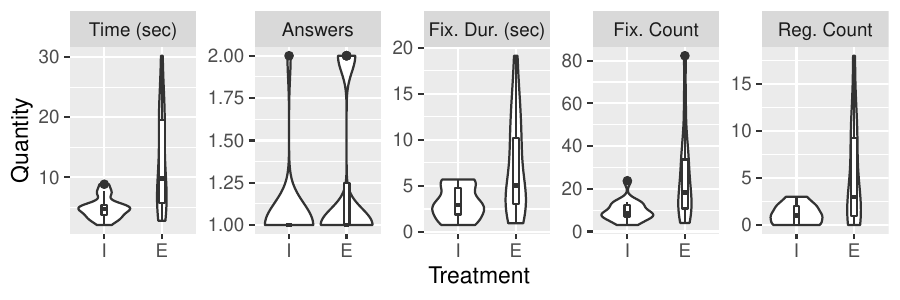}
	\end{minipage}}
\hfill	
  \subfloat[Data distribution of metrics for task to compute the number of digits.]{
	\begin{minipage}[c][0.4\width]{
	   1\textwidth}
	   \centering
	   \includegraphics[width=1\textwidth]{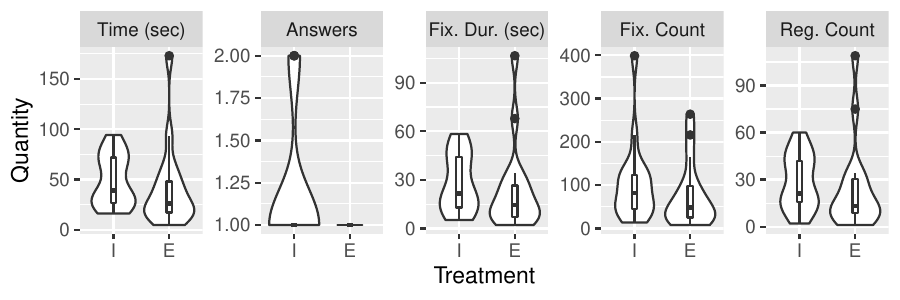}
	\end{minipage}}
 \hfill	
  \subfloat[Data distribution of metrics comparing Inline and extracted method of all tasks together.]{
	\begin{minipage}[c][0.4\width]{
	   1\textwidth}
	   \centering
	   \includegraphics[width=1\textwidth]{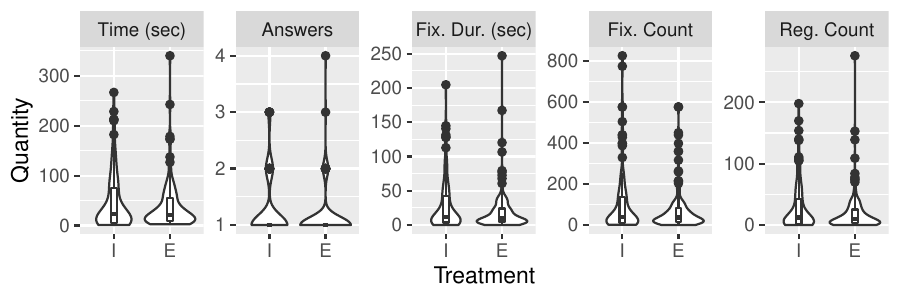}
	\end{minipage}}
\caption{Summary of the data distribution by tasks. I = Inlined method version; E = Extracted method version.}
\end{figure*}

% antes \begin{minipage}{2\textwidth}

\begin{table}[h]
\setlength\tabcolsep{2pt}
\rotatebox{90}{% Rotate the minipage by the desired angle
\begin{minipage}{1\textwidth}
\caption{\newline Results for \textbf{time spent in AOI and in Code} (RQ$_{1}$). I = Inline Method; E = Extract Method; PD = percentage difference; PV = \textit{p}-value {after FRD correction}; ES = Cliff's Delta effect size; SW = Shapiro-Wilk; SP = Shapiro \textit{p}-value; IQR = Interquartile Range; SD = Standard Deviation. Columns I and E are based on the median as a measure of central tendency. AOI/Code presents the proportion of time spent in AOI divided by the time in the Code. {The All Programs row provides a comparison of all values between the Inline and Extract methods.} }

\begin{center}
\resizebox{23cm}{!}{
\begin{tabular}{p{2.5cm}rrrrrrrrrrrrrrrrrrrrrrrr}
    \hline
    \multicolumn{1}{|c|}{\textbf{Tasks}} &\multicolumn{11}{|c|}{\textbf{In AOI}} &  \multicolumn{11}{|c|}{\textbf{In Code}}&\multicolumn{2}{|c|}{\textbf{AOI/Code}}\\
    \hline
    \multicolumn{1}{|c|@{}}{} &\multicolumn{1}{|c|@{}}{\shortstack{I\\sec}} &  \multicolumn{1}{|c|@{}}{\shortstack{E\\sec}} &\multicolumn{1}{|c|@{}}{\shortstack{\\PD\\\%}} &\multicolumn{1}{|c|@{}}{\shortstack{I\\IQR}} &\multicolumn{1}{|c|@{}}{\shortstack{I\\SD}}
    &\multicolumn{1}{|c|@{}}{\shortstack{E\\IQR}}&\multicolumn{1}{|c|@{}}{\shortstack{E\\SD}}
    
    &\multicolumn{1}{|c|@{}}{PV} &\multicolumn{1}{|c|@{}}{ES}&\multicolumn{1}{|c|@{}}{SW}&\multicolumn{1}{|c|@{}}{SP}
    
    &\multicolumn{1}{|c|@{}}{\shortstack{I\\sec}} &  \multicolumn{1}{|c|@{}}{\shortstack{E\\sec}} &\multicolumn{1}{|c|@{}}{\shortstack{PD\\\%}}
    
    &\multicolumn{1}{|c|@{}}{\shortstack{I\\IQR}} &\multicolumn{1}{|c|@{}}{\shortstack{I\\SD}}
    &\multicolumn{1}{|c|@{}}{\shortstack{E\\IQR}}&\multicolumn{1}{|c|@{}}{\shortstack{E\\SD}}
    
    &\multicolumn{1}{|c|@{}}{PV} &\multicolumn{1}{|c|@{}}{ES}&\multicolumn{1}{|c|@{}}{SW}&\multicolumn{1}{|c|@{}}{SP}
    
     &\multicolumn{1}{|c|@{}}{\shortstack{I\\\%}} &\multicolumn{1}{|c|@{}}{\shortstack{E\\\%}} \\
    \hline
    
    Sum Numbers & 8.8 & 21.6 & {$\uparrow$146.2} &6.2 & 6.5 &19.1&19.3& \textbf{0.0009}  & 0.75 & 0.77 & 1.34e-05 & 
    15.9 & 30.9 & {$\uparrow$93.9} &10.0&12.4&23.8&31.1  & \textbf{0.02}  & 0.57 & 0.71 & 1.29e-06 &0.55&0.70\\
    
    Next Prime & 121.2 & 53.9 & {$\downarrow$55.5} &96.1&66.7&93.7&92.2 & 0.10 & n/a & 0.90 & 6.00e-03 & 
    132.6 & 61.6 & {$\downarrow$53.5} &113.8&78.0&105.8&99.7 & 0.10 & n/a & 0.90 & 6.35e-03 & 0.91 & 0.87\\
    
    Highest Grade & 77.7 & 23.7 & {$\downarrow$70.0} &79.0&66.7&61.1&36.0 & \textbf{0.02} & -0.50 & 0.86 & 6.68e-04  & 
    92.6 & 32.3 & {$\downarrow$65.0} &93.3&77.5&68.3&45.4 & 0.14 & n/a & 0.86 & 7.83e-04 & 0.84 & 0.73\\
    
    Factorial & 62.2 & 13.1 & {$\downarrow$78.8} &93.6&60.6&56.1&49.0 & \textbf{0.04} &-0.47 & 0.83 & 1.86e-04 & 
    81.3 & 22.1 & {$\downarrow$72.8} &103.0&60.6&61.4&43.1 &\textbf{0.0PV2}  & -0.51 & 0.85 & 4.34e-04 & 0.76 & 0.59\\
    
    Multiples of Three & 24.6 & 37.5 & {$\uparrow$52.4} &11.8&20.2&32.5&22.6 & 0.24 & n/a & 0.89 & 3.23e-03 & 
    39.2 & 49.0 & {$\uparrow$24.9} &17.9&27.0&39.9&33.7 & 0.22 & n/a & 0.88 & 2.02e-03 & 0.63 & 0.76\\
    
    Area of a Square & 2.5 & 6.9 & {$\uparrow$166.9} &0.75&0.66&4.6&4.2 & \textbf{0.00000} & 0.93 & 0.80 & 3.49e-05 & 
    7.7 & 14.9 & {$\uparrow$94.4} &3.3&2.6&11.3&6.8 & \textbf{0.02} & 0.53 & 0.86 & 8.23e-04 & 0.32 & 0.46\\
    
    Check If Even & 4.7 & 9.8 & {$\uparrow$108.4} &1.7&1.8&13.8&8.2 & \textbf{0.005} & 0.66 & 0.78 & 1.60e-05 & 
    28.3 & 36.3 & {$\uparrow$28.3} &12.2&8.4&20.8&18.4 & {0.06} & 0.42 & 0.94 & 6.78e-02 & 0.17 & 0.27\\
    
    Number of Digits & 34.5 & 26.0 & {$\downarrow$24.7} &45.2&25.4&30.7&42.0 & 0.25 & n/a & 0.84 & 3.19e-04 & 
    66.4 & 38.2 & {$\downarrow$42.4} &48.6&37.2&42.1&56.8 & 0.17 & n/a & 0.87 & 9.92e-04 & 0.52 & 0.68\\
    
    All Programs & 127.5 & 111.2 & {$\downarrow$12.8} &69.6&56.8&46.0&48.8 & 0.24 & n/a & 0.97 & 0.27 & 
    191.8 & 177.5 & {$\downarrow$7.4} &75.6&62.1&50.2&54.1 & 0.19 & n/a & 0.98 & 0.40 & 0.66 & 0.63\\
    
    \hline
\end{tabular}
}
\end{center}
\label{tab: time}
 \end{minipage}}
\end{table}

\begin{table}[h]
\setlength\tabcolsep{2pt}
\rotatebox{90}{% Rotate the minipage by the desired angle
\begin{minipage}{1\textwidth}
\caption{\newline Results for \textbf{number of attempts of the answers} (RQ$_{2}$). I = Inline Method; E = Extract Method; PD = percentage difference; PV = \textit{p}-value {after FDR correction}; ES = Cliff's Delta effect size; SW = Shapiro-Wilk; SP = Shapiro \textit{p}-value; IQR = Interquartile Range; SD = Standard Deviation. Columns I and E are based on the mean as a measure of central tendency. {The All Programs row compares the average number of attempts across all programs between the Inline and Extract methods.} }
\begin{center}
\resizebox{13.0cm}{!}{\begin{tabular}{p{2.5cm}rrrrrrrrrrr}
    \hline
    \multicolumn{1}{|c|}{\textbf{Tasks}} &\multicolumn{11}{|c|}{\textbf{Attempts}}\\
    \hline
    \multicolumn{1}{|c|@{}}{} &\multicolumn{1}{|c|@{}}{I} &  \multicolumn{1}{|c|@{}}{E} &\multicolumn{1}{|c|@{}}{\shortstack{\\PD\\\%}} 
    
    &\multicolumn{1}{|c|@{}}{\shortstack{I\\IQR}} &\multicolumn{1}{|c|@{}}{\shortstack{I\\SD}}
    &\multicolumn{1}{|c|@{}}{\shortstack{E\\IQR}}&\multicolumn{1}{|c|@{}}{\shortstack{E\\SD}}
    
    &\multicolumn{1}{|c|@{}}{PV}  &\multicolumn{1}{|c|@{}}{ES}&\multicolumn{1}{|c|@{}}{SW}&\multicolumn{1}{|c|@{}}{SP}\\
    \hline
    
    Sum Numbers & 1.00 & 1.31 & {$\uparrow$31.2} 
    &0&0&0&0.79 
    & 0.98 & n/a & 0.31 & 3.47e{-11}\\
    
    Next Prime &  1.44 & 1.44 & \textcolor{black}{0.0} 
    &1.0&0.72&0&0.57 
    & n/a & n/a & n/a & n/a\\
    
    Highest Grade & 1.81 & 1.19 & {$\downarrow$34.4} 
    &1.0&0.79&0&0.40 
    & {0.05} & -0.42 & 0.69 & 5.69e{-7}\\
    
    Factorial & 1.56 & 1.31 & {$\downarrow$16.0} 
    &1.0&0.61&0&0.25 
    & 0.30 & n/a & 0.61 & 5.34e{-8}\\
    
    Multiples of Three & 1.25 & 1.00 & {$\downarrow$20.0} 
    &0.25&0.44&0&0 
    & {0.05} & -0.25 & 0.39 & 2.01e{-10}\\
    
    Area of a Square & 1.00 & 1.00 & 0.0 
    &0&0&0&0 
    & n/a & n/a & n/a & n/a\\
    
    Check If Even & 1.06 & 1.25 & {$\uparrow$17.6} 
    &0&0.25&0.25&0.44 
    & 0.17 & n/a & 0.44 & 5.80e{-10}\\
    
    Number of Digits & 1.31 & 1.00 & {$\downarrow$23.8} 
    &0&0.40&0&0 
    & {0.05} & -0.25 & 0.40 & 2.48e{-10}\\
    
    All Programs & 5.2 & 4.7 & {$\downarrow$8.9} 
    &0&0.53&0&0.42 
    & {0.05} & -0.27 & 0.63 & 2.60e-11 \\
    \hline
\end{tabular}
}
\end{center}
\label{tab: accuracy}
 \end{minipage}}
\end{table}

\begin{table}[h]
\setlength\tabcolsep{2pt}
\rotatebox{90}{% Rotate the minipage by the desired angle
\begin{minipage}{1\textwidth}
\caption{\newline Results for \textbf{fixations count in AOI and in Code} (RQ$_{4}$). I = Inline Method; E = Extract Method; PD = percentage difference; PV = \textit{p}-value {after FDR correction}; ES = Cliff's Delta effect size; SW = Shapiro-Wilk; SP = Shapiro \textit{p}-value; IQR = Interquartile Range; SD = Standard Deviation. Columns I and E are based on the median as a measure of central tendency. AOI/Code presents the proportion of fixations count in AOI divided by the fixations count in the Code. {The All Programs row compares the median fixations count across all programs between the Inline and Extract methods.} }
\begin{center}
\resizebox{23cm}{!}{
\begin{tabular}{p{2.5cm}rrrrrrrrrrrrrrrrrrrrrrrr}
    \hline
    \multicolumn{1}{|c|}{\textbf{Tasks}} &\multicolumn{11}{|c|}{\textbf{In AOI}} &  \multicolumn{11}{|c|}{\textbf{In Code}}&\multicolumn{2}{|c|}{\textbf{AOI/Code}}\\
    \hline
    \multicolumn{1}{|c|@{}}{} &\multicolumn{1}{|c|@{}}{I} &  \multicolumn{1}{|c|@{}}{E} &\multicolumn{1}{|c|@{}}{\shortstack{\\PD\\\%}} 
    
    &\multicolumn{1}{|c|@{}}{\shortstack{I\\IQR}} &\multicolumn{1}{|c|@{}}{\shortstack{I\\SD}}
    &\multicolumn{1}{|c|@{}}{\shortstack{E\\IQR}}&\multicolumn{1}{|c|@{}}{\shortstack{E\\SD}}
    
    &\multicolumn{1}{|c|@{}}{PV} &\multicolumn{1}{|c|@{}}{ES} &\multicolumn{1}{|c|@{}}{SW} &\multicolumn{1}{|c|@{}}{SP}
    
    &\multicolumn{1}{|c|@{}}{I} &  \multicolumn{1}{|c|@{}}{E} &\multicolumn{1}{|c|@{}}{\shortstack{PD\\\%}}
    
    &\multicolumn{1}{|c|@{}}{\shortstack{I\\IQR}} &\multicolumn{1}{|c|@{}}{\shortstack{I\\SD}}
    &\multicolumn{1}{|c|@{}}{\shortstack{E\\IQR}}&\multicolumn{1}{|c|@{}}{\shortstack{E\\SD}}

    &\multicolumn{1}{|c|@{}}{PV} &\multicolumn{1}{|c|@{}}{ES}&\multicolumn{1}{|c|@{}}{SW} &\multicolumn{1}{|c|@{}}{SP}
    
     &\multicolumn{1}{|c|@{}}{\shortstack{I\\\%}} &\multicolumn{1}{|c|@{}}{\shortstack{E\\\%}}
     \\
    \hline
    
    Sum Numbers & 17.5 & 44.0 & {$\uparrow$151.43} 
    &19.5&36.0&37.1&27.2 
    & {0.005} & 0.62 & 0.98 & 0.95 & 
    23.5 & 47.5 & {$\uparrow$102.1} 
    &17.7&20.6&45.5&35.7 
    & \textbf{0.003} & 0.56 & 0.97 & 0.58&74.47&92.63\\
    
    Next Prime & 186.0 & 108.0 & {$\downarrow$41.9} 
    &177.2&196.9&271.4&180.4 
    & 0.19 & n/a & 0.97 & 0.55 & 
    191.5 & 111.0 & {$\downarrow$42.0} 
    &161.0&120.2&178.7&170.5 
    & 0.17 & n/a & 0.97 & 0.42&97.11&97.29\\
    
    Highest Grade & 141.5 & 45.6 & {$\downarrow$67.7} 
    &168.5&211.0&96.0&57.3 & \textbf{0.02} & -0.53 & 0.90 & 0.01 & 
    166.0 & 50.0 & {$\downarrow$69.8} 
    &176.2&128.1&113.0&65.5 & \textbf{0.010} & -0.51 & 0.94 & 0.10&85.24&91.20\\
    
    Factorial & 141.0 & 34.0 & {$\downarrow$75.8} 
    &223.6&189.1&101.93&172.5 & \textbf{0.04} & -0.44 & 0.96 & 0.27 & 
    156.0 & 40.0 & {$\downarrow$74.3} 
    &128.2&104.1&61.2&168.5 & \textbf{0.02} & -0.40 & 0.97 & 0.77&90.38&85.00\\
    
    Multiples of Three & 38.5 & 62.5 & {$\uparrow$62.3} 
    &46.5&48.4&51.7&70.0 & 0.33 & n/a & 0.97 & 0.67 & 
    47.5 & 80.5 & {$\uparrow$69.4} 
    &44.2&40.7&65.5&49.5 & 0.26 & n/a & 0.97 & 0.56&81.05&77.64\\
    
    Area of a Square & 4.6 & 11.0 & {$\uparrow$138.8} 
    &3.2&2.5&9.2&15.1 & {0.005} & 0.59 & 0.95 & 0.14 & 
    12.0 & 20.0 & {$\uparrow$66.6} 
    &6.0&5.5&16.5&11.1 & 0.07 & n/a & 0.94 & 0.12&38.33&55.00\\
    
    Check If Even & 8.5 & 20.1 & {$\uparrow$137.1} 
    &5.2&5.0&23.0&21.2 & {0.005} & 0.58 & 0.97 & 0.51 & 
    39.0 & 56.5 & {$\uparrow$44.8} 
    &24.0&17.9&35.5&31.1 & \textbf{0.04} & 0.42 & 0.98 & 0.83&21.79&35.58\\
    
    Number of Digits & 96.4 & 49.0 & {$\downarrow$49.2} 
    &79.2&94.0&71.8&76.2 & 0.19 & n/a & 0.98 & 0.96 & 
    99.0 & 49.0 & {$\downarrow$50.5} 
    &94.2&103.8&64.2&77.0 & 0.06 & n/a & 0.98 & 0.86&97.37&100.00\\
    
    All Programs & 252.0 & 189.5 & {$\downarrow$24.8} &125.7&152.7&68.4&109.2 & 0.26 & n/a & 0.98 & 0.64 & 
    288.0 & 282.5 & {$\downarrow$1.9} &124.5&106.0&76.2&102.3 & 0.24 & n/a & 0.98 & 0.44&87.50&67.07 \\
    \hline
\end{tabular}
}
\end{center}
\label{tab: fixations count}
 \end{minipage}}
\end{table}

\begin{table}[h]
\setlength\tabcolsep{2pt}
\rotatebox{90}{% Rotate the minipage by the desired angle
\begin{minipage}{1\textwidth}
\caption{\newline Results for \textbf{fixation duration in AOI and in Code} (RQ$_{3}$). I = Inline Method; E = Extract Method; PD = percentage difference; PV = \textit{p}-value {after FDR correction}; ES = Cliff's Delta effect size; SW = Shapiro-Wilk; SP = Shapiro \textit{p}-value; IQR = Interquartile Range; SD = Standard Deviation. Columns I and E are based on the median as a measure of central tendency. AOI/Code presents the proportion of fixation duration in AOI divided by the fixation duration in the Code. {The All Programs row compares the median fixation duration across all programs between the Inline and Extract methods.} }
\begin{center}
\resizebox{23cm}{!}{
\begin{tabular}{p{2.5cm}rrrrrrrrrrrrrrrrrrrrrrrr}
    \hline
    \multicolumn{1}{|c|}{\textbf{Tasks}} &\multicolumn{11}{|c|}{\textbf{In AOI}} &  \multicolumn{11}{|c|}{\textbf{In Code}}&\multicolumn{2}{|c|}{\textbf{AOI/Code}}\\
    \hline
    \multicolumn{1}{|c|@{}}{} &\multicolumn{1}{|c|@{}}{\shortstack{I\\sec}} &  \multicolumn{1}{|c|@{}}{\shortstack{E\\sec}} &\multicolumn{1}{|c|@{}}{\shortstack{\\PD\\\%}} 
    
    &\multicolumn{1}{|c|@{}}{\shortstack{I\\IQR}} &\multicolumn{1}{|c|@{}}{\shortstack{I\\SD}}
    &\multicolumn{1}{|c|@{}}{\shortstack{E\\IQR}}&\multicolumn{1}{|c|@{}}{\shortstack{E\\SD}}
    
    &\multicolumn{1}{|c|}{PV} &\multicolumn{1}{|c|@{}}{ES} &\multicolumn{1}{|c|}{SW} &\multicolumn{1}{|c|}{SP}
    
    &\multicolumn{1}{|c|@{}}{\shortstack{I\\sec}} &  \multicolumn{1}{|c|@{}}{\shortstack{E\\sec}} &\multicolumn{1}{|c|@{}}{\shortstack{PD\\\%}}

    &\multicolumn{1}{|c|@{}}{\shortstack{I\\IQR}} &\multicolumn{1}{|c|@{}}{\shortstack{I\\SD}}
    &\multicolumn{1}{|c|@{}}{\shortstack{E\\IQR}}&\multicolumn{1}{|c|@{}}{\shortstack{E\\SD}}

    &\multicolumn{1}{|c|}{PV} &\multicolumn{1}{|c|}{ES} &\multicolumn{1}{|c|}{SW} &\multicolumn{1}{|c|}{SP}
    &\multicolumn{1}{|c|@{}}{\shortstack{I\\\%}} &\multicolumn{1}{|c|@{}}{\shortstack{E\\\%}}
    \\
    \hline
    
    Sum Numbers & 6.2 & 14.2 & {$\uparrow$130.1} 
    &6.4& 5.2 &13.8&9.7 & \textbf{0.01} & 0.60 & 0.87 & 1.26e-03 & 
    8.7 & 18.4 & {$\uparrow$110.5} 
    &7.5& 7.7 & 16.2&11.7 & \textbf{0.03} & 0.53 & 0.86 & 5.76e-04 &0.71&0.77\\
    
    Next Prime & 63.7 & 35.3 & {$\downarrow$44.5} 
    &61.3&51.3&56.0&69.2 & 0.14 & n/a & 0.87 & 1.10e-03 & 
    65.6 & 38.8 & {$\downarrow$40.8} 
    &68.8&54.9&74.2&71.3 & \textbf{0.07} & -0.39 & 0.87 & 9.45e-04 &0.97&0.91\\
    
    Highest Grade & 45.1 & 11.8 & {$\downarrow$73.6} 
    &48.7&42.5&31.5&19.9 & \textbf{0.03} & -0.42 & 0.83 & 5.21e-04 & 
    47.1 & 16.3 & {$\downarrow$65.8} 
    &68.4&59.7&38.0&22.5 & \textbf{0.07} & -0.53 & 0.84 & 3.09e-04 &0.96&0.72\\
    
    Factorial & 40.2 & 8.4 & {$\downarrow$78.9} 
    &54.9&36.1&13.6&21.6 & \textbf{0.01} & -0.64 & 0.83 & 1.57e-04 & 
    46.0 & 11.7 & {$\downarrow$74.5} 
    &61.7&59.9&40.5&74.0 & \textbf{0.07} & -0.51 & 0.86 & 5.35e-04 &0.87&0.72\\
    
    Multiples of Three & 11.7 & 19.9 & {$\uparrow$69.1} 
    &10.0&11.0&19.1&13.2 & 0.38 & n/a & 0.88 & 1.63e-03 & 
    18.1 & 28.9 & {$\uparrow$60.0} 
    &14.8&13.9&26.1&18.7 & 0.30 & n/a & 0.89 & 2.98e-03 &0.65&0.69\\
    
    Area of a Square & 1.4 & 3.0 & {$\uparrow$121.1} 
    &0.8&0.6&3.8&2.4 & \textbf{0.01} & 0.65 & 0.82 & 9.27e-05 & 
    4.1 & 6.5 & {$\uparrow$59.8} 
    &1.5&1.5&6.5&4.0 & 0.12 & n/a & 0.92 & 1.54e-02 &0.34&0.46\\
    
    Check If Even & 2.9 & 5.0 & {$\uparrow$73.1} 
    &2.8&1.7&7.1&5.4 & \textbf{0.06} & 0.46 & 0.81 & 6.55e-05 & 
    14.5 & 19.6 & {$\uparrow$35.1} 
    &10.4&6.8&14.9&11.0 & 0.09 & n/a & 0.94 & 1.07e-01 &0.20&0.26\\
    
    Number of Digits & 21.6 & 14.5 & {$\downarrow$32.7} 
    &31.3&17.7&19.6&27.4 & 0.23 & n/a & 0.83 & 2.75e-04 & 
    40.5 & 17.8 & {$\downarrow$55.9} 
    &44.0&37.1&29.4&33.1 & \textbf{0.07} & -0.38 & 0.88 & 2.14e-03 &0.53&0.81\\
    
    All Programs & 70.9 & 59.2 & {$\downarrow$16.4} 
    &39.1&37.1&19.4&32.0 & 0.21 & n/a & 0.97 & 0.21 & 
    102.4 & 84.5 & {$\downarrow$17.4} 
    &45.9&47.4&27.4&43.0 & 0.12 & n/a & 0.97 & 0.13 &0.69&0.70\\
    \hline
\end{tabular}
}
\end{center}
\label{tab:fixation duration}
 \end{minipage}}
\end{table}

\begin{table}[h]
\setlength\tabcolsep{2pt}
\rotatebox{90}{% Rotate the minipage by the desired angle
\begin{minipage}{1\textwidth}
\caption{\newline Results for \textbf{regressions count in AOI and in Code} (RQ$_{5}$). I = Inline Method; E = Extract Method; PD = percentage difference; PV = \textit{p}-value after FDR correction; ES = Cliff's Delta effect size; SW = Shapiro-Wilk; SP = Shapiro \textit{p}-value; IQR = Interquartile Range; SD = Standard Deviation.  Columns I and E are based on the median as a measure of central tendency. AOI/Code presents the proportion of regressions count in AOI divided by the regressions count in the Code. {The All Programs row compares the median regressions count across all programs between the Inline and Extract methods.} }
\begin{center}
\resizebox{23cm}{!}{
\begin{tabular}{p{2.5cm}rrrrrrrrrrrrrrrrrrrrrrrr}
    \hline
    \multicolumn{1}{|c|}{\textbf{Tasks}} &\multicolumn{11}{|c|}{\textbf{In AOI}} &  \multicolumn{11}{|c|}{\textbf{In Code}}&\multicolumn{2}{|c|}{\textbf{AOI/Code}}\\
    \hline
    \multicolumn{1}{|c|@{}}{} &\multicolumn{1}{|c|@{}}{I} &  \multicolumn{1}{|c|@{}}{E} &\multicolumn{1}{|c|@{}}{\shortstack{\\PD\\\%}} 
    
    &\multicolumn{1}{|c|@{}}{\shortstack{I\\IQR}} &\multicolumn{1}{|c|@{}}{\shortstack{I\\SD}}
    &\multicolumn{1}{|c|@{}}{\shortstack{E\\IQR}}&\multicolumn{1}{|c|@{}}{\shortstack{E\\SD}}
    
    &\multicolumn{1}{|c|@{}}{PV} &\multicolumn{1}{|c|@{}}{ES}&\multicolumn{1}{|c|@{}}{SW} &\multicolumn{1}{|c|@{}}{SP}
    
    &\multicolumn{1}{|c|@{}}{I} &  \multicolumn{1}{|c|@{}}{E} &\multicolumn{1}{|c|@{}}{\shortstack{PD\\\%}}

    &\multicolumn{1}{|c|@{}}{\shortstack{I\\IQR}} &\multicolumn{1}{|c|@{}}{\shortstack{I\\SD}}
    &\multicolumn{1}{|c|@{}}{\shortstack{E\\IQR}}&\multicolumn{1}{|c|@{}}{\shortstack{E\\SD}}

    &\multicolumn{1}{|c|@{}}{PV} &\multicolumn{1}{|c|@{}}{ES}&\multicolumn{1}{|c|@{}}{SW} &\multicolumn{1}{|c|@{}}{SP}
    &\multicolumn{1}{|c|@{}}{\shortstack{I\\\%}} &\multicolumn{1}{|c|@{}}{\shortstack{E\\\%}}
    \\
    \hline
    
    Sum Numbers & 6.0 & 12.5 & {$\uparrow$108.3} 
    &4.7&5.4&9.0&9.6 & \textbf{0.03} & 0.57 & 0.87 & 0.0012 &  
    9.5 & 18.0 & {$\uparrow$89.4} 
    &10.0&10.8&15.0&16.1 & \textbf{0.03} & 0.52 & 0.81 & 7.14e-05 & 63.16 & 69.44\\
    
    Next Prime & 80.5 & 41.0 & {$\downarrow$49.0} 
    &54.2&47.8&62.5&74.3 & 0.12 & n/a & 0.88 & 0.0024 & 
    84.5 & 44.0 & {$\downarrow$47.9} 
    &64.5&55.9&64.5&72.7 & 0.12 & n/a & 0.88 & 0.0024 & 95.27 & 93.18\\
    
    Highest Grade & 43.0 & 11.0 & {$\downarrow$74.4} 
    &54.0&51.6&37.2&23.0 & \textbf{0.03} & -0.50 & 0.80 & 5.59e-05 & 
    75.0 & 19.0 & {$\downarrow$74.6} 
    &68.2&60.0&44.5&29.2 & \textbf{0.03} & -0.53 & 0.84 & 0.0003 & 57.33 & 57.89\\
    
    Factorial & 52.0 & 8.0 & {$\downarrow$84.6} 
    &54.7&34.4&12.0&21.0 & \textbf{0.03} & -0.67 & 0.86 & 0.0017 & 
    70.5 & 15.5 & {$\downarrow$78.0} 
    &53.7&36.1&43.0&36.0 & \textbf{0.03} & -0.49 & 0.90 & 0.0064 & 73.76 & 51.61\\
    
    Multiples of Three & 13.5 & 19.0 & {$\uparrow$40.7} 
    &11.2&13.1&20.7&14.4 & 0.41 & n/a & 0.91 & 0.0088 & 
    21.5 & 28.5 & {$\uparrow$32.5} 
    &18.2&18.5&33.5&24.8 & 0.53 & n/a & 0.95 & 0.1474 & 62.79 & 66.67\\
    
    Area of a Square & 1.0 & 2.0 & {$\uparrow$100.0} 
    &2.0&1.0&2.2&1.6 & 0.11 & n/a & 0.89 & 0.0038 & 
    5.0 & 7.0 & {$\uparrow$40.0} 
    &2.2&2.8&8.5&4.9 & 0.12 & n/a & 0.91 & 0.0177 & 20.00 & 28.57\\
    
    Check If Even & 1.0 & 3.0 & {$\uparrow$200.0} 
    &2.0&1.0&8.2&5.4 & 0.07 & 0.43 & 0.91 & 0.0088 & 
    18.5 & 24.5 & {$\uparrow$32.4} 
    &10.5&7.6&20.5&15.5 & 0.15 & n/a & 0.90 & 0.0056 & 5.41 & 12.24\\
    
    Number of Digits & 21.0 & 13.0 & {$\downarrow$38.0} 
    &26.5&17.5&21.7&28.9 & 0.15 & n/a & 0.71 & 1.45e-06 & 
    46.0 & 21.5 & {$\downarrow$53.2} 
    &39.5&25.5&29.5&37.1 & 0.12 & n/a & 0.84 & 0.0004 & 45.65 & 60.47\\
    
    All Programs & 75.0 & 54.0 & {$\downarrow$28.0} 
    &41.0&39.6&22.2&34.7 & 0.12 & n/a & 0.96 & 0.11 & 
    125.5 & 114.0 & {$\downarrow$9.16} 
    &56.0&44.8&33.5&38.2 & 0.12 & n/a & 0.97 & 0.12 & 59.76 & 47.37\\
    \hline
\end{tabular}
}
\end{center}

\label{tab: regressions count}
 \end{minipage}}
\end{table}

\end{document}